\pdfoutput=1

\documentclass[iop,apj,appfendixfloats]{emulateapj}
\usepackage{apjfonts}
\usepackage{graphicx}
\usepackage{enumerate,enumitem,morefloats}
\usepackage{tabularx}
\usepackage{booktabs}
\usepackage{footnote}
\usepackage{multirow}
\usepackage{array}
\usepackage{amsmath,amssymb}
\usepackage{pifont}
\usepackage{hyperref}

\newcommand{\Sig}[1]{\csname Sig\romannumeral#1\endcsname}

\newcommand{\sigmae}{\ensuremath{\sigma_{\rm e}}}
\newcommand{\Sige}{\ensuremath{\Sigma_{\rm e}}}
\newcommand{\Reff}{\ensuremath{R_{\rm e}}}
\newcommand{\ReffMean}{\ensuremath{\langle R_{\rm e}\rangle}}
\newcommand{\Rvir}{\ensuremath{R_{\rm vir}}}

\newcommand{\spin}{\ensuremath{\langle R_{\rm e}\rangle/R_{\rm vir}}}

\newcommand{\Del}{\ensuremath{\Delta}}
\newcommand{\Mstar}{\ensuremath{M_*}}
\newcommand{\MstarGV}{\ensuremath{M_{*,\rm GV}}}
\newcommand{\Mvir}{\ensuremath{M_{\rm vir}}}
\newcommand{\Mbh}{\ensuremath{M_{\rm BH}}}
\newcommand{\Msol}{\ensuremath{\rm M_{\odot}}}
\newcommand{\Mbul}{\ensuremath{M_{\rm bulge}}}
\newcommand{\Mcrit}{\ensuremath{M_{\rm crit}}}
\newcommand{\Vvir}{\ensuremath{V_{\rm vir}}}
\newcommand{\Tvir}{\ensuremath{T_{\rm vir}}}
\newcommand{\Sersic}{S\'{e}rsic~}
\newcommand{\Ebh}{\ensuremath{E_{\rm BH}}}
\newcommand{\Eheat}{\ensuremath{E_{\rm quench}}}
\newcommand{\Eth}{\ensuremath{E_{\rm bind}}}
\newcommand{\mdotcool}{\ensuremath{\dot{m}_{\rm cool}}}
\newcommand{\mdotheat}{\ensuremath{\dot{m}_{\rm heat}}}
\newcommand{\mdotbh}{\ensuremath{\dot{M}_{\rm BH}}}

\newcommand{\vs}{vs\@.~}
\newcommand{\mdotstar}{\ensuremath{\dot{M}_*}}

\defcitealias{Rodriguez17}{RP17}
\defcitealias{Barro17}{B17}
\defcitealias{Fang13}{F13}
\defcitealias{Fang18}{F18}
\defcitealias{vanDokkum15}{vD15}
\defcitealias{Lilly16}{L16}
\defcitealias{Kormendy13}{KH13}
\defcitealias{vanderWel09}{vdW09}
\defcitealias{vanderWel14}{vdW14}

\lefthead{Chen et al.}
\righthead{Black Holes and Quenching}
\slugcomment{}

\begin{document}

\title{Quenching as a Contest between Galaxy Halos and their Central Black Holes}
\author{
Zhu Chen\altaffilmark{1},
S.~M.~Faber\altaffilmark{2},
David C.~Koo\altaffilmark{2},
Rachel S.~Somerville\altaffilmark{3},
Joel R.~Primack\altaffilmark{4},
Avishai Dekel\altaffilmark{5},
Aldo Rodr\'{i}guez-Puebla\altaffilmark{6},
Yicheng Guo\altaffilmark{7},
Guillermo Barro\altaffilmark{8},
Dale D.~Kocevski\altaffilmark{9},
A.~van der Wel\altaffilmark{10,11},
Joanna Woo\altaffilmark{12,13},
Eric F.~Bell\altaffilmark{14},
Jerome J.~Fang\altaffilmark{15},
Henry C.~Ferguson\altaffilmark{16},
Mauro Giavalisco\altaffilmark{17},
Marc Huertas-Company\altaffilmark{18},
Fangzhou Jiang\altaffilmark{5},
Susan Kassin\altaffilmark{16},
Lin Lin\altaffilmark{19},
F.~S.~Liu\altaffilmark{20},
Yifei Luo\altaffilmark{2},
Zhijian Luo\altaffilmark{1},
Camilla Pacifici\altaffilmark{21},
Viraj Pandya\altaffilmark{2},
Samir Salim\altaffilmark{22},
Chenggang Shu\altaffilmark{1},
Sandro Tacchella\altaffilmark{23},
Bryan A.~Terrazas\altaffilmark{14},
Hassen M.~Yesuf\altaffilmark{24}}

\altaffiltext{1}{Shanghai Key Lab for Astrophysics, Shanghai Normal University, 200234, Shanghai, China}
\altaffiltext{2}{Department of Astronomy and Astrophysics, University of California, Santa Cruz, CA 95064, USA}
\altaffiltext{3}{Center for Computational Astrophysics, Flatiron Institute, New York, NY 10010, USA; Department of Physics and Astronomy, Rutgers University, Piscataway, NJ 08854, USA}
\altaffiltext{4}{Physics Department, University of California, Santa Cruz, CA 95064, USA}
\altaffiltext{5}{Racah Institute of Physics, The Hebrew University, Jerusalem 91904, Israel}
\altaffiltext{6}{Instituto de Astronom\'{i}a, Universidad Nacional Aut\'{o}noma de M\'{e}xico, M\'{e}xico, D.F., M\'{e}xico}
\altaffiltext{7}{Department of Physics and Astronomy, University of Missouri, Columbia, MO 65211, USA}
\altaffiltext{8}{Department of Physics, University of the Pacific, Stockton, CA 95211, USA}
\altaffiltext{9}{Department of Physics and Astronomy, Colby College, Waterville, ME 04901, USA}
\altaffiltext{10}{Max-Planck-Institut f\"{u}r Astronomie, D-69117 Heidelberg, Germany}
\altaffiltext{11}{Sterrenkundig Observatorium, Universiteit Gent, Krijgslaan 281 S9, B-9000 Gent, Belgium}
\altaffiltext{12}{Department of Physics \& Astronomy, University of Victoria, Victoria BC V8P 1A1, Canada}
\altaffiltext{13}{Department of Physics, Simon Fraser University, 8888 University Dr, Burnaby BC, V5A 1S6, Canada}
\altaffiltext{14}{Department of Astronomy, University of Michigan, Ann Arbor, MI 48109-1107, USA}
\altaffiltext{15}{Astronomy Department, Orange Coast College, Costa Mesa, CA, 92626 USA}
\altaffiltext{16}{Space Telescope Science Institute, 3400 San Martin Rd, Baltimore, MD 21218, USA}
\altaffiltext{17}{Department of Astronomy, University of Massachusetts, Amherst, MA 01003, USA}
\altaffiltext{18}{GEPI, Observatoire de Paris, CNRS, Universit\'{e} Paris, 75014 Paris, France}
\altaffiltext{19}{Shanghai Astronomical Observatory of CAS, Shanghai, 200030, China}
\altaffiltext{20}{College of Physical Science and Technology, Shenyang Normal University, Shenyang 110034, People's Republic of China; University of California Observatories and the Department of Astronomy and Astrophysics, University of California, Santa Cruz, CA 95064, USA}
\altaffiltext{21}{Goddard Space Flight Center, Code 665, Greenbelt, MD 20771, USA}
\altaffiltext{22}{Department of Astronomy, University of Indiana, Bloomington, IN 47404, USA}
\altaffiltext{23}{Harvard-Smithsonian Center for Astrophysics, Cambridge, MA 02138, USA}
\altaffiltext{24}{Kavli Institute for Astronomy and Astrophysics, Peking University, Beijing 100871, People's Republic of China}

\begin{abstract}
Existing models of galaxy formation have not yet explained striking correlations between structure and star-formation activity in galaxies, notably the \emph{sloped and moving boundaries} that divide star-forming from quenched galaxies in key structural diagrams. This paper uses these and other relations to ``reverse-engineer'' the quenching process for central galaxies. The basic idea is that star-forming galaxies with larger radii (at a given stellar mass) have lower black-hole masses due to lower central densities. Galaxies cross into the green valley when the cumulative effective energy radiated by their black hole equals $\sim4\times$ their halo-gas binding energy. Since larger-radii galaxies have smaller black holes, one finds they must evolve to higher stellar masses in order to meet this halo-energy criterion, which explains the sloping boundaries. A possible cause of radii differences among star-forming galaxies is halo concentration. The evolutionary tracks of star-forming galaxies are nearly parallel to the green-valley boundaries, and it is mainly the sideways motions of these boundaries with cosmic time that cause galaxies to quench. BH-scaling laws for star-forming, quenched, and green-valley galaxies are different, and most BH mass growth takes place in the green valley.  Implications include: the radii of star-forming galaxies are an important second parameter in shaping their black holes; black holes are connected to their halos but in different ways for star-forming, quenched, and green-valley galaxies; and the same BH-halo quenching mechanism has been in place since $z \sim 3$.  We conclude with a discussion of black hole-galaxy co-evolution, the origin and interpretation of BH scaling laws.
\end{abstract}


\section{Introduction}\label{Sec:Introduction}
How and why star formation comes to an end (``quenches'') in massive galaxies is one of the outstanding puzzles of galaxy evolution.  Though gaseous infall onto halos is decreasing in the current Universe, there is consensus that this effect by itself is not sufficient to curtail star formation to the low levels that are observed in quenched galaxies \citep[see review by][]{Somerville15}, and in any case, quenching began at $z \gtrsim 3$, when the Universe was still quite gas-rich.  The number of massive quenched galaxies at the knee of the quenched mass function has approximately tripled since $z = 1$ \citep{Bell04, Faber07, Huang13, Muzzin13, Ilbert13, Tomczak14}, and the vast majority of that growth is in central galaxies, not satellites \citep{Tinker13}. It is therefore clear that quenching of central galaxies is continuing today, but whether by the same or different mechanisms as at earlier times is not known.\footnote{In this paper, quenched galaxies are defined as objects lying more than $-1.0$~dex below the local star-forming main-sequence ridgeline at their mass, and star-forming galaxies are galaxies that lie above the line lying $-0.45$~dex below the ridgeline. Green valley galaxies lie between these populations from $-0.45$~dex to $-1.0$~dex.}

Many studies have modeled galaxy quenching, from semi-analytic models using parametric prescriptions, to hydrodynamic simulations with more detailed descriptions of gas and feedback physics, and basic data like the mass functions of quenched and star-forming galaxies have been more or less successfully matched.  However, in our opinion, model-makers have not yet focused fully enough on the \emph{full panoply} of physical data exhibited by quenching galaxies \vs mass and time.  In this paper, we assemble a more complete list of relevant observational data and use it to construct a simple picture in which galaxies evolve through a structural parameter space while star-forming (and building black holes) and then encounter a ``quenching boundary'', which is defined in this paper as the entrance to the green valley.  As shown below, this boundary is visible as the dividing line between quenched and star-forming galaxies in various structural diagrams.  Key properties of the boundary (see Section~\ref{SubSec:Review} and  Figure~\ref{Fig:Mstar_Sig1_SMA1}) are its considerable length in stellar mass (more than 1~dex), its sharpness, smooth motion in the diagrams with time, and its slope or tilt) whereby more massive galaxies along the boundary have larger radii, higher central densities, and larger central black holes than smaller galaxies.

If quenching depended only on stellar mass, the boundary lines in structural diagrams would be concentrated at a given stellar mass.  The fact that they are extended \emph{and sloping} implies that at least one more parameter in addition to mass is involved in quenching. We show in this paper that \emph{taking this second parameter to be effective radius and tying black hole mass to it can naturally explain most data.}  This is demonstrated by showing that the observed boundaries are very close to those obtained if galaxies enter the green valley when the total effective energy emitted by their central black holes equals a certain (small) multiple of the binding energy of the hot gas in their dark halos.

Before quenching, the model starts by placing star-forming galaxies in dark halos according to abundance matching, and their stellar masses, radii, and black hole masses are computed from their halo properties using simple rules.  As galaxies enter the green valley, we find that black hole growth accelerates, which leads ultimately to full quenching.  The full history of black hole growth from star-forming to quenched is sketched in the ``cartoon'' shown in Figure~\ref{Fig:Cartoon}.

As of this writing, the exact nature of the mechanism that quenches galaxies is unknown. Two broad classes have been suggested.  One is so-called ``halo quenching'', which involves the fact that halo gas becomes shock-heated as halos develop and is therefore less able to cool and fall onto galaxies \citep[e.g.,][]{Ostriker77,Blumenthal84,Birnboim03,Dekel06,Cattaneo06,Keres09}. The other is ``AGN quenching'' by feedback from growing black holes (BHs), which either ejects gas from galaxies  (``ejective feedback'') \citep[e.g.,][]{Silk98, Springel06, Hopkins06} or heats or disturbs the halo gas in such a way as to prevent it from falling in (``preventive feedback'', ``strangulation'') \citep[e.g.,][]{Croton06,Bower06,Somerville08,Somerville12,Henriques15,Peng15}.\footnote{Feedback is also produced by young stars and supernovae, but the characteristic energy is too small to affect gas in massive galaxies.  The major consequence is to modulate the gas content of lower-mass galaxies while they are still star-forming \citep{Somerville15}.}

Both halo quenching and AGN quenching have considerable observational support.  On the halo side, N-body simulations provide reliable models of halo growth \citep[e.g.,][]{Behroozi13}, the basic physics of halo gas density and temperature within those halos is reasonably well known (see references above), and X-ray data confirm that massive halos are filled with hot gas at the predicted temperatures and densities \citep[e.g.,][]{Arnaud10}. On the AGN side, nearly all quenched galaxies have massive spheroids \citep{Bell08, Bluck14}, and these spheroids harbor massive BHs \citep{Magorrian98, Ferrarese00, Gebhardt00}.  AGN activity peaks just as galaxies are entering quiescence \citep[e.g.,][]{Schawinski07,Schawinski10,Nandra07,Aird19}.  Tight correlations exist between BH mass and host spheroid properties, such as bulge mass ($\Mbh\text{-}\Mbul$: \citealt{Haring04,Kormendy13}) and effective velocity dispersion ($\Mbh\text{-}\sigmae$: \citealt{Gebhardt00, Ferrarese00, Kormendy13,Ho14,vandenBosch16}), signaling some sort of close interplay between spheroid-building and BH growth.  Finally, matter falling onto a black hole may release of order 10\% of its rest mass energy $E = mc^2$, which is comparable in amount to the binding energy of gas in the halo \citep[e.g.,][]{Silk98,Ostriker05,Oppenheimer18}, enough to either heat, blow away, or otherwise significantly disturb the gas and induce quenching.  For reviews of these topics see \citet{Cattaneo09} and \citet[hereafter KH13]{Kormendy13}.

Despite these successes, neither class of model as currently formulated naturally explains the sloping nature of the boundaries that divide quenched from star-forming galaxies above $\Mstar \sim 10^{10}~\Msol$.  Halo quenching depends on only the temperature and density of the halo gas, and thus tends to set in at a characteristic halo mass near $\Mcrit \sim 10^{11\text{-}12}~\Msol$ \citep[e.g.,][]{Birnboim03,Keres05,Dekel06}, and therefore at a corresponding stellar mass because the stellar-mass/halo-mass relation is tight \citep{Wechsler18}. Halo quenching therefore tends to predict a too-sharp boundary in stellar mass near $\Mstar \sim 3\times10^{10}~\Msol$ and also does not naturally explain the systematic variation (sloping boundaries) in other galaxy properties near that mass.

The predicted observational signature of AGN quenching depends on the mode of feeding mass onto the BH.  In ``radiative mode'' (\citealp{Somerville18}; also called ``quasar mode'' \citep{Croton06}, ``bright mode'' \citep{Somerville08}, or ``high mode'' \citep{Pillepich18}), AGN mass-accretion is rapid and is often triggered by major mergers \citep{Hopkins06} and/or disk-gas instabilities \citep{Cattaneo06}. Since these are random events that can occur over a broad range of time in a galaxy's life, it is hard to see why the quenching boundary should be sharp, let alone particularly sloped.  In ``jet mode'' (\citealp{Somerville18}; also called ``radio mode'' \citep{Croton06}, ``kinetic mode'' \citep{Weinberger17}, or ``low mode'' \citep{Pillepich18}), gas drains onto the BH more slowly, creating feedback energy and/or momentum that ejects gas and/or heats the halo.  However, highly collimated jets cannot couple efficiently to the halo gas while the gas is still cool and dense \citep[e.g.,][]{Bower06,Croton06, Cattaneo06}, and so in practice, jet-like feedback also needs to wait until the halo mass exceeds $\Mcrit$.  Prospects may be better in kinetic mode simulations that are more isotropic \citep[e.g., IllustrisTNG,][]{Pillepich18}, but sloping boundaries have not yet been demonstrated.

The above is a brief summary of a rich and complex literature, and variations in points of view and terminology are common.  For example, \citet{Peng10} also identified two quenching mechanisms, which they called ``mass quenching'' and ``environment quenching''.  However, their mass quenching is a more general feedback-driven process that includes both stellar feedback as well as AGN feedback, while their environment quenching refers to satellite galaxies only.  We ignore satellites in this paper owing to their different quenching physics \citep[e.g.,][]{Woo17}, and halo quenching, when it is mentioned, always refers to central galaxies.

The upshot of the preceding is that current models have not yet explained the extent of the quenching boundary in stellar mass or why the radii, central densities, and BH masses of boundary galaxies vary systematically along it.  In this paper, we assume that the quenching process has imprinted these trends, and we attempt to ``reverse-engineer'' the physical nature of the quenching mechanism from them.  This leads us to a picture in which \emph{galaxy quenching is dominated by AGN quenching and becomes perceptible when total effective BH energy output exceeds a certain multiple of the gravitational binding energy of the halo gas.}  The point of perceptibility is defined to be $-0.45$~dex below the star-forming main sequence, where the number of green valley galaxies begins to exceed the lower tail of the star-forming main sequence in CANDELS \citep{Fang18}. For convenience, we will call this the ``quenching boundary" even though galaxies have presumably started to quench before passing this point.

The next section reviews previous works that have also tried to use the structural properties of galaxies as a guide to understanding quenching. It is followed by a section that collects together all of the basic properties of quenching and star-forming galaxies that a satisfactory theory of quenching should explain. The final section in this Introduction sketches the nature of the model and outlines the rest of the paper.

\subsection{Previous Work}
\label{SubSec:PreviousWork}

Many studies have remarked on the striking structural parameter correlations of both quenched and quenching galaxies. Let $\Reff$ be the optical effective radius containing half the light and $\Sige$ be the effective mass surface density given by $\Sige = 0.5\Mstar/\pi\Reff^2$.  \citet{Kauffmann03} showed that quenched SDSS galaxies have higher $\Sige$ than star-forming galaxies at the same mass, and \citet{Franx08} showed the same trend in distant quenched galaxies.  The corresponding offset ridgeline of quenched galaxies to smaller $\Reff$ at fixed $\Mstar$ was shown for SDSS galaxies by \citet{Shen03}, \citet{Omand14}, and \citet{Haines17} and for distant galaxies by  \citet[hereafter vdW09]{vanderWel09}, \citet[hereafter vdW14]{vanderWel14}, and \citet{Haines17}.

\citet{Kauffmann03,Kauffmann06,Kauffmann12} further noted that the boundary between quenched and star-forming galaxies corresponds closely to a line of constant effective surface density: $\Reff \sim \Mstar^{0.5}$.  Separately, \citet{Bell08}, \citet{Kauffmann12}, and \citet{Bluck14} showed that local quenched galaxies have high \Sersic indices characteristic of bulges, and the same was shown for distant galaxies by \citet{Bell12} and \citet{Wuyts12}. Collectively, these studies established that quenched galaxies are both smaller (more ``compact'') than star-forming galaxies at the same stellar mass, and also that quenched galaxies have prominent bulges.

\citet{Cheung12} introduced the use of \emph{central} properties to predict quenching. After testing numerous quantities, they concluded that high stellar-mass surface density within a projected radius of 1~kpc ($\Sig1$) is the best predictor of quenching at $z \sim 0.7$. \citet[hereafter F13]{Fang13} found a similar correlation between $\Sig1$ and stellar mass for SDSS galaxies, with quenched and green valley galaxies occupying a narrow sloping ridgeline at the top of the $\Sig1\text{-}\Mstar$ distribution.  With an rms scatter of only 0.16~dex \citepalias{Fang13}, the $\Sig1\text{-}\Mstar$ ridgeline is among the tightest scaling relations known for quenched galaxies.  \citetalias{Fang13} further showed that specific star-formation rate (SSFR) declines sharply near this ridgeline and suggested that the residual $\Delta\Sig1$ could be a \emph{marker} for quenching, possibly signaling a threshold in BH feedback. \citet{Wake12} (also \citealp{Bluck16,Teimoorinia16}) found a similar result for SDSS galaxies using central velocity dispersion $\sigma_1$ within 1~kpc (which \citetalias{Fang13} showed correlates closely with $\Sig1$). Finally, \citet{Luo19} showed that high $\Delta\Sig1$ correlates with high bulge-to-total ratio ($B/T$), consistent with the known quenching-spheroid connection.

\citet{vanDokkum14F} demonstrated that the threshold relationship between $\Sig1$ and SSFR found by \citetalias{Fang13} obtains throughout cosmic time for high-mass galaxies.  Lower-mass galaxies were added by \citet{Tacchella15}, \citet{Barro17}, \citet{Whitaker17}, \citet{Lee18}, and \citet{Mosleh18}, who found that the  $\Sig1\text{-}\Mstar$ relations for quenched and star-forming galaxies at $z = 0$ are replicated almost unchanged out to $z = 3$ -- both slopes are constant with time, and both zero points march downward \emph{in unison} by $\sim0.3$~dex from $z = 3$ to now.\footnote{We note parenthetically that, while $\Sig1$ and $\Sige$ are both surface densities, their behaviors \vs mass and time are different \citep[hereafter B17]{Barro17}. The outer envelopes of quenched galaxies grow due to dry mergers after quenching, which increase $\Reff$ and decrease $\Sige$ while leaving $\Sig1$ roughly unchanged \citep[e.g.,][]{Daddi05,Trujillo06,vanDokkum08,Bezanson09,Naab09,vanDokkum10,Oser12,Szomoru12}.  The quenched-galaxy relation is therefore broader in $\Sige\text{-}\Mstar$ than in $\Sig1\text{-}\Mstar$, and its zero point declines with time, gradually encroaching on the domain of star-forming galaxies. The division between star-forming and quenched galaxies is therefore not as clean in $\Sige\text{-}\Mstar$ as in $\Sig1\text{-}\Mstar$ \citepalias{Barro17}.  $\Sig1$ and $\Sige$ are also different for star-forming galaxies, and the difference makes $\Sig1$ more suitable for predicting black-hole mass.  We show below that $\Sig1 \sim  \Mstar/\Reff$ (Section~\ref{SubSec:InputPowerLaws}), whereas $\Sige \sim \Mstar/\Reff^2$.  Because BH mass scales approximately as $\sigmae^4$  (e.g., \citetalias{Kormendy13}, \citealt{vandenBosch16}) and $\sigmae^4 \sim (\Mstar/\Reff)^2 \sim \Sig1^2$, a close power-law relation is predicted between $\Mbh$ and $\Sig1$, as borne out below. In contrast, no simple power-law relation is predicted between $\Mbh$ and $\Sige$.}

Various quenching models have tried to reproduce the structural differences between quenched and star-forming galaxies.  \citet[hereafter vdW09]{vanderWel09F} introduced a model in which galaxies evolve in $\Reff\text{-}\Mstar$ and quench when they reach a boundary set by their effective velocity dispersion, $\sigmae$.  Since quenched galaxies with higher $\sigmae$ today are older at fixed $\Mstar$ \citep{Bernardi05, Graves09}, \citetalias{vanderWel09} reasoned that the threshold $\sigmae$ for quenching must have been higher in the past. These facts motivated them to adopt a \emph{moving boundary} that shifts upwards with time in $\Reff\text{-}\Mstar$ (and downwards in $\sigmae\text{-}\Mstar$). This  moving boundary was actually detected by \citet{Haines17} in $\Reff\text{-}\Mstar$.  However, neither of these works specified how galaxies approach the boundary before quenching.  This was added by \citet[hereafter vD15]{vanDokkum15}, who envisioned that galaxies \emph{evolve on parallel tracks} through $\Reff\text{-}\Mstar$ while star-forming and then quench when meeting a moving boundary in $\sigmae$. A similar picture for the motion of galaxies in $\Reff\text{-}\Mstar$ and the origin of quenched galaxies was presented by \citet{Cappellari13,Cappellari16}.

\citetalias{Barro17} and \citet{Tacchella15} considered the evolution of galaxies in $\Sig1\text{-}\Mstar$.  They concluded from the narrowness of the star-forming locus that galaxies must evolve approximately along it while star-forming. In their picture, galaxies reach a threshold value of $\Sig1$ for their mass, quench, and then pile up along the $\Sig1\text{-}\Mstar$ ridgeline.  Because the ridgeline in $\Sig1\text{-}\Mstar$ moves down with time, this quenching boundary also moves down.  Thus, like \citetalias{vanderWel09} and \citetalias{vanDokkum15}, \citetalias{Barro17} envisioned a moving boundary but modeled it as a power law in $\Sig1\text{-}\Mstar$, not as a line at constant $\sigmae$.  None of these papers offered any physical explanation for their boundaries -- they were simply set empirically to match the data.

\citet{Omand14} introduced yet another aspect of quenching by focusing on $\Reff\text{-}\Mstar$. They observed that star-forming SDSS galaxies populate a shallow ridgeline in $\Reff\text{-}\Mstar$ with rms scatter $\sigma_{\rm Re} \sim 0.20$~dex at fixed mass, while quenched galaxies occupy a narrower and more steeply sloped ridgeline at the bottom edge of this distribution (\citealp{Omand14}; see also \citealp{Shen03}, \citetalias{vanderWel09}, \citetalias{vanderWel14}, \citealp{Lange16}). Contours of equal quenched fraction are sloping stripes running parallel to the quenched ridgeline, and quenched galaxies therefore \emph{co-exist with star-forming galaxies at the same $\Mstar$}. This fact had been noticed before \citep[e.g.,][]{Kauffmann03}, but Omand et al.~seized on it as evidence that quenching is controlled by at least one more parameter than mass and suggested that galaxies are a \emph{two-parameter family} whose quenching is governed by $\Reff$ as well as $\Mstar$.

Further evidence that both $\Reff$ and $\Mstar$ play a role came from the work of \citet{vandenBosch16} on local black holes.  Starting with the $\Mbh\text{-}\sigmae$ relation, he noted that $\sigmae^2 \sim \Mstar/\Reff$ from the virial theorem and proposed that BHs populate a ``fundamental plane'' in  $\Mbh\text{-}\Reff\text{-}\Mstar$.  In words, $\Reff$ can be thought of as cleaning up the scatter in $\Mbh$ \vs $\Mstar$ that appears when star-forming galaxies are added to quenched galaxies \citepalias{Kormendy13} in the sense that \emph{smaller-radii galaxies must have larger black holes}.  Similar arguments were made by \citet{Krajnovic18}. Both works stressed that $\Reff$ \vs $\Mstar$ is a map of black hole mass, an insight that will prove basic to our model.

Finally, \citet[hereafter L16]{Lilly16} presented yet another quenching model. Galaxies evolve along parallel log-log tracks in $\Reff\text{-}\Mstar$, as in \citetalias{vanderWel14} and \citetalias{vanDokkum15}, but quenching depends on $\Mstar$ probabilistically, increasing with $\Mstar$ until all massive galaxies eventually quench.  Given that earlier-quenching galaxies tend to be smaller and that effective radii tend to shrink a lot on quenching in their model, \citetalias{Lilly16} showed that the contours of equal quenched fraction would actually be tilted despite the fact that quenching formally depends on $\Mstar$ only.  This model is discussed further in Section~\ref{SubSec:L16}.

\subsection{Collected Properties of Quenched \vs Star-forming Galaxies}
\label{SubSec:Review}

The previous section has introduced several different types of data that a successful quenching model must explain.  For convenience, these and other data yet to be mentioned are collected together with basic references in Table~\ref{Tb_1}.  The first two lines of this table describe the green valley, which in this paper is taken to be the main channel through which \emph{high-mass central galaxies} evolve from star-forming to quenched.  Data on the mass function of quenched galaxies indicate that the green valley consists of two separate components: a high-mass channel that appears early at $z \ge 2.5$ with a roughly log-normal mass function, and a later channel after $ z \sim 1.5$ that appears as a gradually rising low-mass tail \citep[e.g.,][]{Ilbert10,Huang13,Muzzin13}\footnote{Note that the earlier-cited increase in the number of massive quenched galaxies with time referred to the high-mass channel near the knee of the quenched mass function, not the low-mass tail.}.  This tail is generally due to satellite quenching, not AGN quenching \citep{Tinker13,Fossati17,Guo17} and is therefore not treated in this paper.  The high-mass channel is the object of our model, and its two striking features are the near constancy of its peak mass near log $\Mstar/\Msol \simeq 10.7$ from $z = 3$ to now\footnote{\citet{Bundy06} and \citet{Haines17} both claim to see a decrease in the characteristic quenching mass with time.  However, their definition of quenching mass is sensitive to the relative numbers of quenched and star-forming galaxies and tends to fall as galaxies quench.  We focus on the typical green-valley mass itself, which is constant, as seen in e.g., \citetalias{Barro17}, \citet{Straatman16}, \citet{Haines17}, \citet{Pan16}, and \citet{Fang12}.} and its substantial width, approaching 1 full dex in stellar mass. Line 3 of Table~\ref{Tb_1} also notes the continuing steady flow of galaxies though the high-mass channel down to recent times (e.g., \citealt{Straatman15}, F18).

\begin{table*}[!htbp]
\caption{\label{Tb_1}Observations To Be Explained$^{\rm a}$}
\renewcommand\arraystretch{1.5} %adjust table height
{\centering
\begin{tabularx}{\textwidth}{cXl}
\toprule
Item No.$^{\rm b}$ & Observation & References\\
\midrule
1 & The stellar mass distribution in the high-mass green valley peaks at approximately the same value  & \citet{Ilbert10}\\
& $\sim 10^{10.7}~\Msol$ from $z=3$ to now. & \citet{Huang13}\\
& & \citet{Barro17}\\
2 & The high-mass green valley is $\sim 1$~dex wide in stellar mass from $z = 3$ to now. & \citet{Ilbert10}\\
& & \citet{Huang13}\\
& & \citet{Barro17}\\
3 & There is a steady flow of high-mass central galaxies across the quenching boundary from $z = 3$ to now. & \citet{Bell04}\\
& &\citet{Faber07}\\
& &\citet{Tinker13}\\
& & \citet{Tomczak14}\\
4 & The ridgeline of star-forming galaxies in $\Reff$ \vs $\Mstar$ has constant slope $\Reff \sim \Mstar^{0.21}$ from $z = 3$ to now. & \citet{vanderWel14}\\
5 & The zero point of the star-forming ridgeline in $\Reff$ \vs $\Mstar$ has increased by $\sim 0.4$~dex from $z = 3$ to now. & \citet{vanderWel14}\\
$[6]$ & The ridgeline of quenched galaxies in $\Reff$ \vs $\Mstar$ has constant slope $\Reff \sim \Mstar^{0.75}$ from $z = 3$ to now. & \citet{vanderWel14}\\
$[7]$ & The zero point of the quenched ridgeline in $\Reff$ \vs $\Mstar$ has increased by $\sim 0.75$ dex from $z = 3$ to now. & \citet{vanderWel14}\\
8 & The ridgeline of star-forming galaxies in $\Sig1$ \vs $\Mstar$ has constant slope $\Sig1 \sim \Mstar^{0.88}$ from  $z = 2.5$ to 0.5. & \citet{Barro17}\\
9 & The zero point of the star-forming ridgeline in $\Sig1$ \vs $\Mstar$ has decreased by $\sim 0.30$~dex from $z = 2.5$ to 0.5. & \citet{Barro17}\\
$[10]$ & The ridgeline of quenched galaxies in $\Sig1$ \vs $\Mstar$ has constant slope $\Sig1 \sim \Mstar^{0.66}$ from $z = 2.5$ to 0.5. & \citet{Barro17}\\
$[11]$ & The zero point of the quenched ridgeline in $\Sig1$ \vs $\Mstar$ has decreased by $\sim 0.30$~dex from $z = 2.5$ to 0.5. & \citet{Barro17}\\
$[12]$ & The zero points of the quenched and AGN galaxy ridgelines in $\Mbh$ \vs $\Mstar$ appear to be decreasing with time & \citet{Bennert11}\\
& by similar amounts.  & \citet{Ding17}\\
13 & $\Mbh$ can be expressed as a power law of either $\Mstar/\Reff$ or $\sigmae$, but the slopes of quenched galaxies are  & \citet{Kormendy13}\\
&  shallower than the slopes of all galaxies taken together. & \citet{vandenBosch16}\\
14 & The ridgelines of quenched galaxies are narrow in $\Sig1$ \vs $\Mstar$ and $\Mbh$ \vs $\Mstar$ but not so narrow in $\Reff$ \vs $\Mstar$. & \citet{Fang13}\\
& & \citet{Barro17}\\
& & \citet{Kormendy13}\\
& & \citet{vanderWel14}\\
15 & The quenched ridgelines in $\Sig1$ \vs $\Mstar$ and $\Mbh$ \vs $\Mstar$ show age stratification, with older galaxies  & \citet{Tacchella17}\\
& having higher values of $\Sig1$ and $\Mbh$ at fixed $\Mstar$. \\
& & \citet{Luo19}\\
$[16]$ & The zero points of quenched and star-forming galaxies have evolved together in lockstep in $\Sig1$ \vs $\Mstar$ since $z = 3$. & \citet{Barro17}\\
$[17]$ & The zero points of quenched and star-forming galaxies have evolved differently in $\Reff$ \vs $\Mstar$ since $z = 3$. & \citet{vanderWel14}\\
18 & Star-forming galaxies mix with quenched galaxies on the quenched ridgeline in $\Sig1$ \vs $\Mstar$ today but lie    & \citet{Fang13}\\
& well below the quenched ridgeline in $\Mbh$ \vs $\Mstar$. & \citet{Barro17}\\
& & \citet{Terrazas17}\\
\bottomrule
\end{tabularx}}

\footnotesize{$^{\rm a}$ Referring to central galaxies only.\\}
\footnotesize{$^{\rm b}$ Items without brackets are thought to be fully explained by the model; items with brackets are only partly explained.  The model being scored is the binding-energy model (blue lines in Figure~\ref{Fig:Mstar_Sig1_SMA1}), for which there is only one adjustable parameter, $f_{\rm B} = 4$ at entry into the green valley, which has been used to match the zero point of the blue lines to that of the red lines.  All of the other successes listed here are new.  See further discussion in Section~\ref{SubSec:Successes}.}
\end{table*}

Lines 4-7 in Table~\ref{Tb_1} describe the properties of quenched and star-forming ridgelines in the $\Reff\text{-}\Mstar$ plane.  Star-forming galaxies need accurate treatment in the model because they feed the quenching process and their properties determine what quenched galaxies ultimately will look like.  Important facts about the $\Reff\text{-}\Mstar$ plane are the constancy of the ridgeline slopes for both quenched and star-forming galaxies over time and the increase in both ridgeline zero points since $z = 3$.  Lines~8-11 summarize similar properties of quenched and star-forming ridgelines in $\Sig1\text{-}\Mstar$.  Important facts are again the constancy of both ridgeline slopes over time, accompanied (in this case) by \emph{decreasing} zero points.

Line~12 of Table~\ref{Tb_1} says that the ridgeline zero point of quenched galaxies in $\Mbh\text{-}\Mstar$ also decreases with time. Since less is known about the history of this ridgeline, we do not cite data on its slope back in time, but its general behavior seems similar to the motion of the quenched ridgeline in $\Sig1\text{-}\Mstar$. The distribution in $\Mbh\text{-}\Mstar$ of star-forming AGN galaxies has been studied locally by \citet{Ho14} and \citet{Reines15} and in distant galaxies by \citet{Ding17,Ding20}.  This ridgeline seems to move down by a similar amount to the quenched ridgeline, which keeps the two populations in lockstep.  To the extent that AGNs can be taken to represent all star-forming galaxies, this suggests that the entire $\Mbh\text{-}\Mstar$ distribution for all galaxies maintains the same shape and moves down in bulk, just like that in $\Sig1\text{-}\Mstar$.  Line~13 points out that there seem to exist various scaling laws linking $\Mbh$ to both $\Mstar/\Reff$ and $\sigmae$ and that the slopes of these laws differ depending on whether all galaxies or only quenched ridgeline galaxies are included (\citetalias{Kormendy13};\citealp{vandenBosch16}).

Line~14 refers to the width of the quenched ridgelines, which are quite narrow in $\Sig1\text{-}\Mstar$ and $\Mbh\text{-}\Mstar$ at fixed $\Mstar$ but not so narrow in $\Reff\text{-}\Mstar$, especially at early times. Our model says that these quenched ridgelines are not excesses due to sighting along a plane edge-on (like the Fundamental Plane for ellipticals or the Tully-Fisher relation for spirals) but are actual pile-ups of galaxies that stop evolving and come to a halt along particular loci.  The slopes and zero points of these loci therefore contain important information about quenching, which is basic data for the model.  However, the quenched ridgeline in $\Reff\text{-}\Mstar$ is somewhat broader than the other two due to variable amounts of galaxy-wide dissipation during quenching \citep[e.g.,][]{Zolotov15,Tacchella16a} and to variable amounts of dry merging after quenching (\citealp[e.g.,][]{Naab09}; \citetalias{vanderWel09}), both of which tend to decouple global $\Reff$ from central $\Sig1$ and $\Mbh$.  The narrowness of the quenched ridgelines in $\Sig1\text{-}\Mstar$ and $\Mbh\text{-}\Mstar$ therefore suggests that those relations more closely preserve the memory of quenching; significantly, both are properties that relate to the \emph{centers} of galaxies, not to their outer parts.  However, line~15 points out that recent data are beginning to reveal \emph{age stratification} even within both of these ridgelines, the quenched stellar populations above the ridgelines being older (\citealp{Tacchella17,Martin18,Luo19}).  The authors attribute this to ``progenitor bias'' \citep{vanDokkum96,Carollo13, Poggianti13} caused by the moving boundary, which evidently leaves early-quenched objects stranded above the boundary at late times.

Lines~16 and 17 highlight the different zero-point behaviors of quenched \vs star-forming galaxies in the different spaces.  In $\Sig1\text{-}\Mstar$, ridgelines of both populations move down together in lock-step so that the relative positions of quenched \vs star-forming galaxies are preserved and the $\Sig1\text{-}\Mstar$ diagram \emph{looks the same at all times} except for this bulk downward shift (cf.~Figure~2 in \citetalias{Barro17}).  However, in $\Reff\text{-}\Mstar$, the ridgeline of quenched galaxies moves up \emph{faster} than the ridgeline of star-forming galaxies and the two populations begin to overlap, as in \citetalias{vanderWel14}.

\begin{figure*}[htbp]
\centering
\includegraphics[scale=0.53]{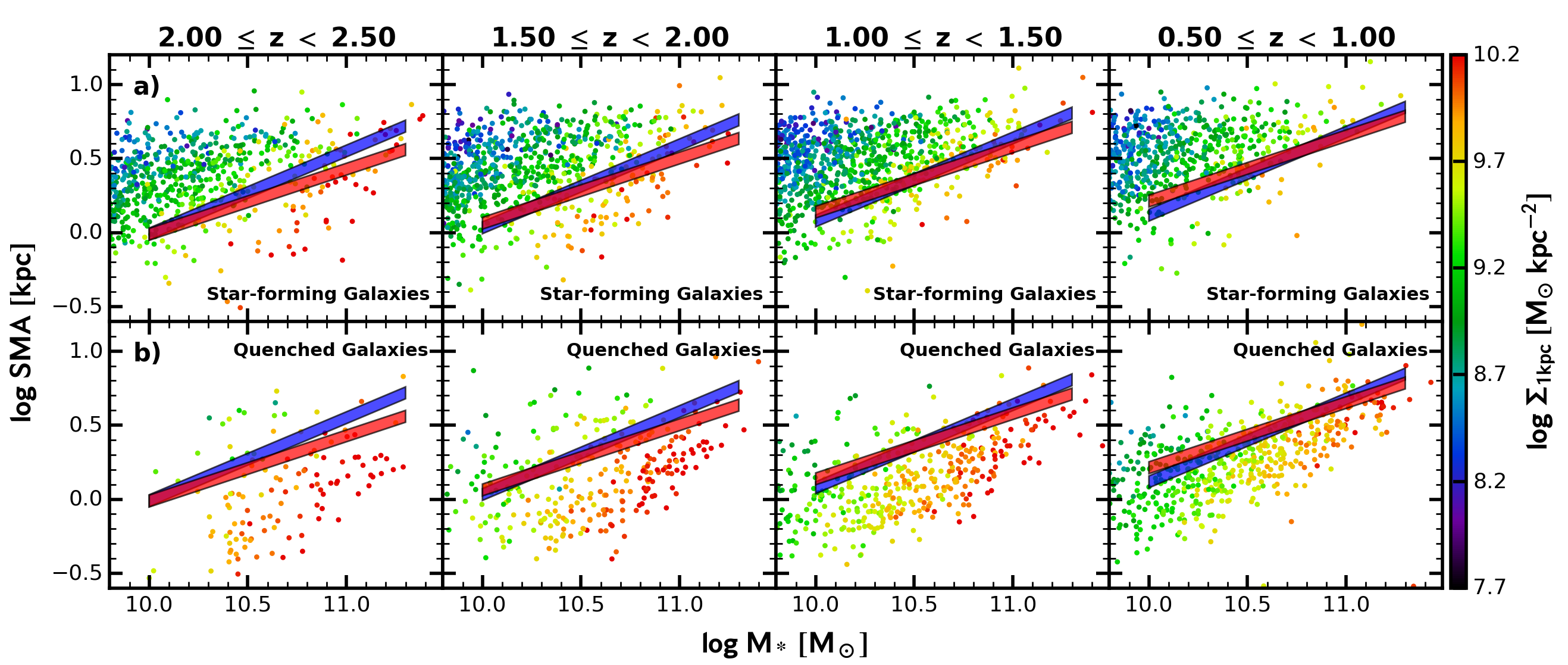}
\includegraphics[scale=0.53]{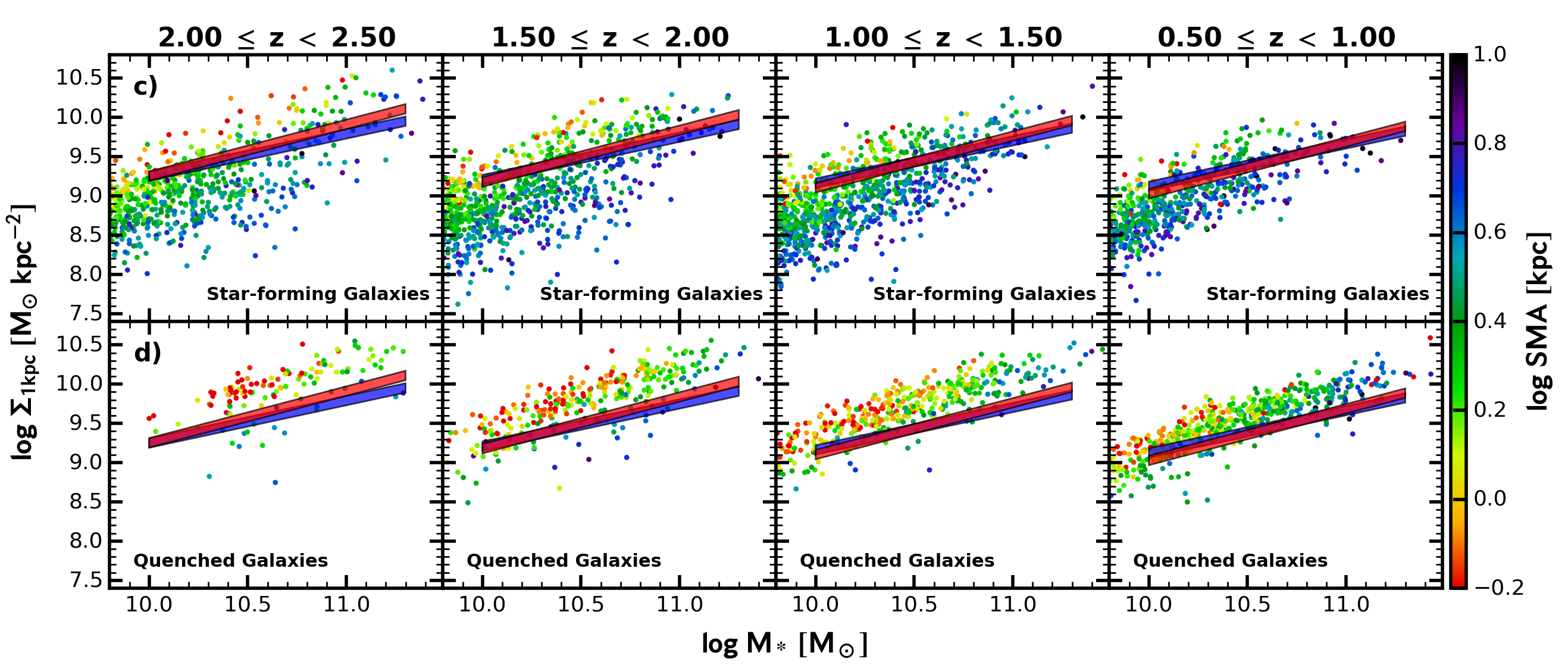}
\caption{\label{Fig:Mstar_Sig1_SMA1} Data from all five CANDELS fields divided by redshift, patterned after similar figures in \citetalias{vanderWel09} and \citetalias{Barro17}.  The data and sample are described in Section~\ref{Sec:Data}; only face-on galaxies with $b/a \geq 0.5$ are used.  Satellites and centrals are included even though the model technically refers to centrals only.  Panel a) (top row):  Effective radius $\Reff$ along the semi-major axis (SMA) \vs stellar mass, colored by $\Sig1$ for star-forming galaxies.  Panel b) (second row):  The same as panel a) but for quenched galaxies. Panel c) (third row): $\Sig1$ \vs stellar mass colored by $\Reff$ for star-forming galaxies. Panel d) (bottom row): Same as panel c) but for quenched galaxies. Star-forming galaxies are defined as lying above the line running $-0.45$~dex below the main sequence ridgeline, and quenched galaxies lie more than $-1.0$~dex below it.  Green valley galaxies (between these two limits) are not shown. The figure illustrates the dramatic difference in structure between star-forming and quenched galaxies and the sharp boundaries between them.  The red lines are the empirical quenching boundaries (tilted boundaries) that are developed from the data in Section~\ref{Sec:PowerLawModel}. The blue lines are predicted by the halo gas binding-energy model in Section~\ref{Sec:ThermalModel}, which says that entry to the green valley occurs when total emitted BH energy ($\Ebh$) equals $\times4$ the binding energy of hot halo gas. The agreement between the red and blue lines is the major result of this paper.}
\end{figure*}

The first column in Table~\ref{Tb_1} numbers the items for future reference.  Item numbers without brackets indicate phenomena that we think are fairly well explained by the new model, while item numbers with brackets indicate phenomena that are only partially explained. We stress again that the model only predicts $\Reff$ for star-forming galaxies before they reach the quenching boundary and that $\Reff$ is not modeled later, when dissipation and dry merging become important. Since lines~6, 7, and 17 all involve quenched galaxies, their values of $\Reff$ are not accurately known, which is why they are given brackets.  Other brackets are described in Section~\ref{SubSec:Successes}.

Figure~\ref{Fig:Mstar_Sig1_SMA1} illustrates many of the properties listed in Table~\ref{Tb_1}. This figure borrows heavily from similar figures in \citetalias{vanderWel14} and \citetalias{Barro17} using CANDELS data. The upper two panels compare star-forming and quenched galaxies in the space of $\Reff$ \vs $\Mstar$, and the lower two panels compare the same samples in the space of $\Sig1$ \vs $\Mstar$ (in order to sharpen differences, green valley galaxies are not plotted).  Looking ahead, the red lines represent empirically-determined  quenching boundaries that are derived in Section~\ref{Sec:PowerLawModel}, and the blue lines represent the quenching boundaries of the ``binding-energy" model of this paper, in which the threshold quenching BH energy is equal to a multiple of the binding energy of halo gas.

Focusing first on the lower panels with $\Sig1$, we see that virtually all quenched galaxies lie on a ridgeline above the red quenching boundary and their numbers are growing with time, while the vast majority of star-forming galaxies lie below the same line.  Galaxies approach the quenching boundary from below-left in this plane, as shown in Figure~\ref{Fig:ThreePanel} below.  A smattering of star-forming galaxies also lies above the line, which is the so-called ``elbow effect'' pointed out by \citetalias{Fang13} and \citetalias{Barro17} in which star-forming and quenched galaxies co-exist at nearly the same values of $\Sig1$ (cf.~also \citealp{Whitaker17,Lee18}).  This is mentioned in line~18 of Table~\ref{Tb_1} and signifies a highly non-linear relation between $\Sig1$ and the decline of star-formation in the GV.  Line~18 also says that the downward trend in star-formation rate \vs $\Mbh$ is more gradual (straighter) \citep[cf.~Figure 3,][]{Terrazas17} so that star-forming galaxies \emph{never} lie on the quenched ridgeline in $\Mbh\text{-}\Mstar$ but always lie well below it \citep[Figure 2,][]{Terrazas16}.  These different behaviors of $\Sig1$ \vs $\Mstar$ and $\Mbh$ \vs $\Mstar$ near quenching are illustrated in Figures~\ref{Fig:ThreePanel} and \ref{Fig:Elbow} below, and matching them will prove to be a major clue to black-hole growth rates in the green valley.  Note that the quenched ridgelines, and hence also the red lines, maintain constant slope with time in $\Sig1\text{-}\Mstar$ but move down in zero point.  The ridgelines of star-forming galaxies also maintain constant slope and move down in lockstep with the quenched ridgeline in the same diagram.  Collectively these are Lines~ 8-11, 14, and 16 in Table~\ref{Tb_1}.

The upper two panels in Figure~\ref{Fig:Mstar_Sig1_SMA1} show the same objects in $\Reff\text{-}\Mstar$, where the pattern is reversed -- virtually all quenched galaxies now fall \emph{below} the red quenching boundary while the vast majority of star-forming galaxies fall above it.  The picture here is that galaxies approach the quenching boundary from above-left, as shown in Figure~\ref{Fig:ThreePanel}.  Note that the quenched ridgelines, and hence also the red lines, again maintain constant slope but now move \emph{up} in zero point with time.  The ridgelines of the star-forming galaxies also maintain constant slope and move up.  However, in this case, the zero point of the quenched galaxies moves up faster than the zero point of the star-forming galaxies, and the two populations are not in lock-step the way they are in $\Sig1\text{-}\Mstar$. Collectively these are properties 4-7, 14, and 17 in Table~\ref{Tb_1}.

Three more features of Figure~\ref{Fig:Mstar_Sig1_SMA1} are important.  First, one sees that quenched galaxies cover roughly a full dex in mass that is centered at about $10^{10.7}~\Msol$ at all redshifts.  These are lines~1 and 2 in Table~\ref{Tb_1}.  Second, there is a dearth of quenched galaxies near and below $10^{10}~\Msol$ at $z > 2$ that is gradually filled in at lower redshift. This is the growing tail of low-mass quenched galaxies that was referred to above and that we  attribute to satellite quenching; it is not treated in the current paper.  The third point is that the trends seen Figure~\ref{Fig:Mstar_Sig1_SMA1} are ancient, going all the way back to $z \sim 3$.  The enduring nature of these relations seems to imply that the quenching mechanism for high-mass galaxies above $10^{10}~\Msol$ has remained basically the same over all these billions of years despite major declines in star formation-rates, gas contents, and merger rates.  To our knowledge, no SAM or hydro model has as yet reproduced the full collection of properties listed in Table~\ref{Tb_1} and illustrated in Figure~\ref{Fig:Mstar_Sig1_SMA1}.

\subsection{Outline for the Remainder of the Paper}
\label{SubSec:Outline}

The first step in this paper, in Section~\ref{Sec:PowerLawModel}, is an empirical parametrization of the quenching boundary that is fitted to real data. Specifically, we assume that the effects of quenching become apparent on galaxy star-formation rates when the total effective energy emitted by their BH equals some \emph{total energy quota} needed by the halo to stop cooling: $\Ebh = \Eheat$, and we derive an empirical power-law expression for $\Eheat$ as a function of $\Mvir$ and redshift from the observed (red) boundaries in Figure~\ref{Fig:Mstar_Sig1_SMA1}. Starting with this empirical formulation for $\Eheat$ rather than a theoretical expression provides a simple form that can be experimentally manipulated and also shows intuitively how the real halo energy quota must scale if quenching is in fact a balance between cumulative BH energy output and halo mass.

The second major step, in Section~\ref{Sec:ThermalModel}, compares this empirical halo-energy quota to the gas binding energies of real halos.  This comparison is shown by the blue lines in Figure~\ref{Fig:Mstar_Sig1_SMA1}, which are seen to agree fairly well with the red lines with no adjustment in slope and a plausible one-time choice of zero point.  The blue lines are constructed by assuming that quenching sets in when the total effective energy emitted by the BH equals some constant, $f_{\rm B}$, times the binding energy content of the halo gas.  The precise quenching mechanism is unspecified -- it could be gas heating and ejection, a physical deflection of infalling primordial gas, or actual gas unbinding -- and the cumulative energy input from BHs is in principle large enough to accomplish any of these things. Equating the onset of quenching to the tipping point between total BH energy and halo-gas binding energy has also been proposed by others \citep[e.g.,][]{Silk98,Ostriker05,Booth10,Booth11,Bower17,Oppenheimer18,Davies19a,Davies19b,Oppenheimer19}.  Our major new result is that this simple model explains the observed quenching boundaries in Figure~\ref{Fig:Mstar_Sig1_SMA1} quite well, along with many other data in Table~\ref{Tb_1}.

Four other aspects of the model deserve mention.  First, our knowledge of halos and the halo-galaxy connection is used to model how halos and their central galaxies evolve together \emph{before quenching}. A second key assumption is the rule for putting BHs into star-forming galaxies, which assumes that BH mass scales as a power law of $\Sig1$ during the star-forming phase, which is suggested by empirical data. This rule implies that black holes grow modestly but steadily (averaging over possible short-term high-accretion episodes) while galaxies form stars.\footnote{Black-hole growth as a power of $\Sig1$ during the star-forming phase is the default model for most of the paper.  However, the Discussion section considers a complementary picture in which black holes remain smaller than this throughout most of the star-forming phase and only begin to grow rapidly later (see Section~\ref{SubSec:BHGrowth}).}  The third feature replaces this early, modest power law with a steeper power law (\vs $\Sig1$) in the green valley (see Figure~\ref{Fig:Cartoon}).  This also agrees with a variety of data and can also reconcile the different values of BH scaling-law slopes found in different works. The fourth and final feature, following \citet{Omand14,vandenBosch16,Krajnovic18}, is to model star-forming galaxies as a \emph{two-parameter family} labeled by $\Reff$ and $\Mstar$. This says that BH mass depends on galaxy radius as well as $\Mstar$ -- larger galaxies have smaller BHs because they have lower central $\Sig1$-- which means they have to evolve to higher mass before they overcome their halos and quench.  This last is the key factor that explains the sloped quenching boundaries seen in Figures~\ref{Fig:Mstar_Sig1_SMA1} and \ref{Fig:ThreePanel}.

Section~\ref{Sec:Variants} explores four variants of the basic model. All fail, but each one teaches an important lesson, as discussed in Appendix~\ref{App:A4}.  Section~\ref{Sec:Discussion} is the Discussion.  Topics include the model's success in matching the data goals in Table~\ref{Tb_1}, the origin of scatter in effective radii, how quenching happens in the binding-energy model compared to other models, compaction, the connection between black holes and dark halos, black hole scaling laws, and what it means to say that black holes  and galaxies ``co-evolve''.  Finally, Section~\ref{Sec:Summary} reviews the basic findings and important take-away messages, and four appendices present various formulae, summarize important power laws, and describe the variant models in more detail.

We end this Introduction with a list of limitations and phenomena that are not covered by the model. All scaling laws are assumed to be perfectly tight with zero scatter.  Mergers and satellite quenching are ignored.  It is assumed that entry into the green valley is a one-way street and that rejuvenation does not occur \citep[cf.][]{Chauke19}.  Because of its use of $\Sig1$ to track black hole growth, the model cannot be used before $z \sim 3$, where many galaxies fit entirely within 1~kpc.  Likewise, no description is provided for the microscopic state of gas in the halo or how the BH energy interacts with it.  Finally, the data matched are mainly global structural relations, and no attempt is made to study the pattern of star formation or quenching within galaxies; predict mass functions or other structure functions; or predict AGN frequencies. These topics are left for future work.

All data used in this paper are consistent with $H_0 = 70~{\rm km~s^{-1}~Mpc^{-1}}$, $\Omega_{\rm m} = 0.3$ and $\Omega_\Lambda = 0.7$.

\section{Data}\label{Sec:Data}

The data used in this paper come from the multi-wavelength and ancillary datasets produced by the Cosmic Assembly Near-Infrared Deep Extragalactic Legacy Survey on the \emph{Hubble Space Telescope} \citep[CANDELS; ][]{Grogin11,Koekemoer11}. All five fields are used (GOODS-S, UDS, COSMOS, EGS, and GOODS-N) with $H \leq 24.5$~mag.  AB magnitudes, spectroscopic/photometric redshifts, stellar masses, star-formation rates, and $A_{\rm V}$ attenuation values are taken from the following catalogs: \citet{Guo13,Galametz13,Nayyeri17,Stefanon17,Barro19}. Methodologies for the stellar masses and $A_{\rm V}$ values are described in \citet{Santini15}. $\Reff$ values and axis ratios are measured using GALFIT by \citet{vanderWel12}. The selection and completeness of the sample are discussed in Appendix~\ref{App:A1}.

An important parameter in this paper is central stellar density, $\Sig1$, whose measurement
is explained in \citetalias{Barro17}.  Briefly, mass profiles were constructed from multiband \emph{HST} images, and raw values of \Sig1\ were measured based on the projected stellar mass within 1 kpc. A PSF correction was then applied from a grid of high-resolution model images having the same  values of \Reff\ and \Sersic index.   Details are given in Appendix C of \citetalias{Barro17}.
Two separate sets of values were computed using slightly different techniques by Barro and Chen, and they agree well. All data including \Sig1\ are available at the \emph{Rainbow} database \citep{Barro11}\footnote{http://arcoiris.ucsc.edu//Rainbow\_navigator\_public/}.  The values of \Sig1 used here are the Chen values.

CANDELS galaxies have not yet been divided into centrals and satellites even though the model is nominally for central galaxies only.  Environments for CANDELS galaxies are being measured, and this distinction is left for future work.

\section{Power-Laws in the Empirical Model}\label{Sec:PowerLawModel}

\subsection{Input Power Laws}\label{SubSec:InputPowerLaws}

This section and the next construct an empirical model that describes the evolutionary tracks that galaxies take while star-forming, and then quenches them as they pass over the quenching boundary. Several key relations are power laws.  The present section derives their slopes, and the next section derives their zero points and how they move with time. Full versions of all laws, including the zero points, are given in Appendix~\ref{App:A3}.

Figure~\ref{Fig:ThreePanel} plots three structural diagrams that illustrate the basic landscape of the model.  Star-forming galaxies evolve along straight, power-law trajectories in the four-dimensional space comprised of $\Mstar$, $\Reff$, $\Sig1$, and $\Mbh$.  The three panels in Figure~\ref{Fig:ThreePanel} show the three 2-D projections of this four-dimensional space onto $\Mstar$.  Eventually galaxies encounter a boundary in this 4-D space, as shown by the dashed lines, where they start to quench.  Crossing the boundary is equated with \emph{entering the green valley}, i.e., the point where AGN feedback has a just-perceptible effect on the star-formation rate.  The zero point of the boundary is defined using a well-defined feature in the $\Sig1\text{-}\Mstar$ distribution of SDSS galaxies from \citet{Luo19} and is then mapped into the other two projections using the power law mappings developed in this section. The blue points are star-forming galaxies, and the quenching boundaries mark their maximum extent along the arrows.  In panels a and b, the red points are fully quenched galaxies only, and in panel c they are both quenched and green-valley galaxies. Panels a and b show CANDELS galaxies while panel c shows a selection of local galaxies with measured black holes from \citet{Terrazas16}.

\begin{figure*}[htbp]
\centering
\includegraphics[scale=0.56]{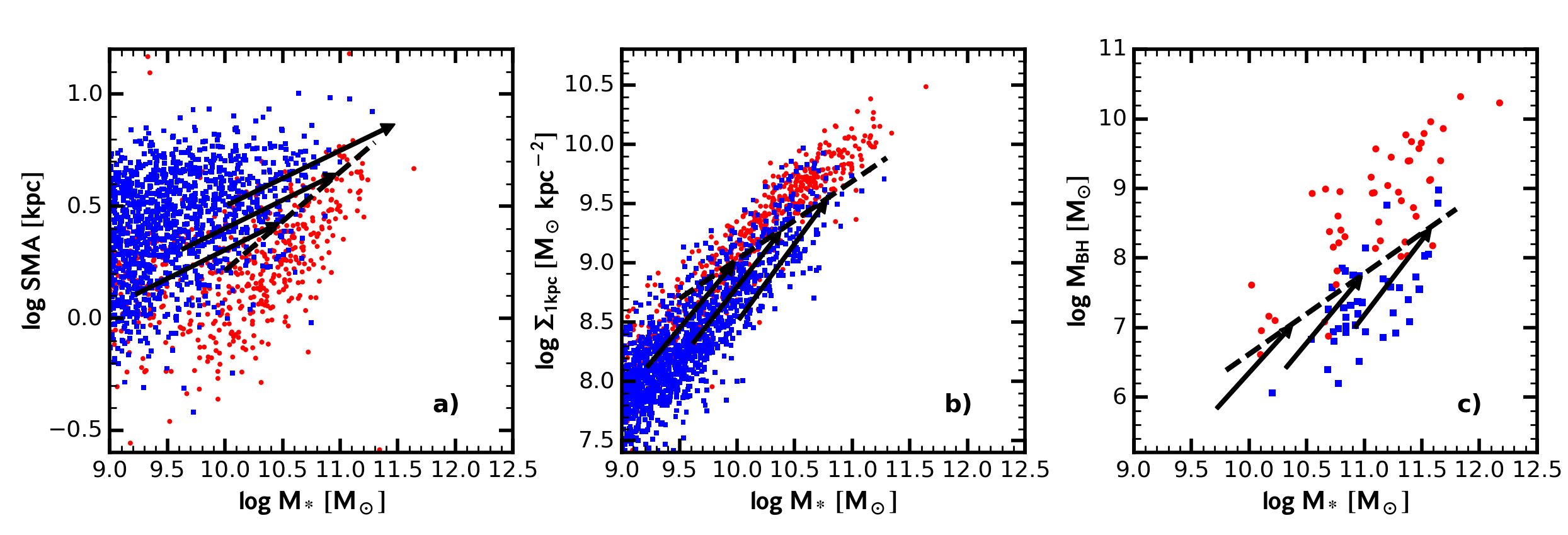}
\caption{\label{Fig:ThreePanel} Schematic illustration of how star-forming galaxies evolve in the empirical power-law model.  Galaxies evolve along approximately power-law tracks in a four-dimensional space comprised of $\Mstar$, $\Reff$, $\Sig1$, and $\Mbh$. The panels show projections onto $\Mstar$ of each of the other three quantities.  The arrows show the evolutionary tracks of individual galaxies, and the sloping (tilted) dashed lines show the quenching boundaries at which galaxies enter the green valley. The fact that the dashed lines are not vertical indicates that at least one other parameter beyond \Mstar\ is involved in quenching. Panel a: $\Reff$ \vs $\Mstar$.  The points are face-on CANDELS galaxies with $b/a > 0.5$ in the redshift range $z = 0.6\text{-}0.9$. Blue points are star-forming galaxies on the SFMS with $\Delta$logSSFR above $-0.45$~dex; red points are fully quenched galaxies with  $\Delta$logSSFR below $-1.0$~dex.  Green valley galaxies are not shown.  Panel b: $\Sig1$ \vs $\Mstar$.  The objects are the same as in panel a.  Panel c: $\Mbh$ \vs $\Mstar$.  This diagram reproduces Figure~2 from \citet{Terrazas16}, which plots black-hole masses for 90 nearby galaxies.  Blue points are star-forming galaxies with the same selection criteria as in \citet{Terrazas16}; red points are now both quenched and green valley galaxies.  Note the strong bias to high masses in this panel compared to panels a and b.  All three panels of the figure are schematic: the slopes of the dashed boundary lines are accurate, but the evolutionary track slopes are illustrative only, and the dashed boundaries \emph{move with time}.}
\end{figure*}

\begin{figure*}[htbp]
\centering
\includegraphics[scale=0.53]{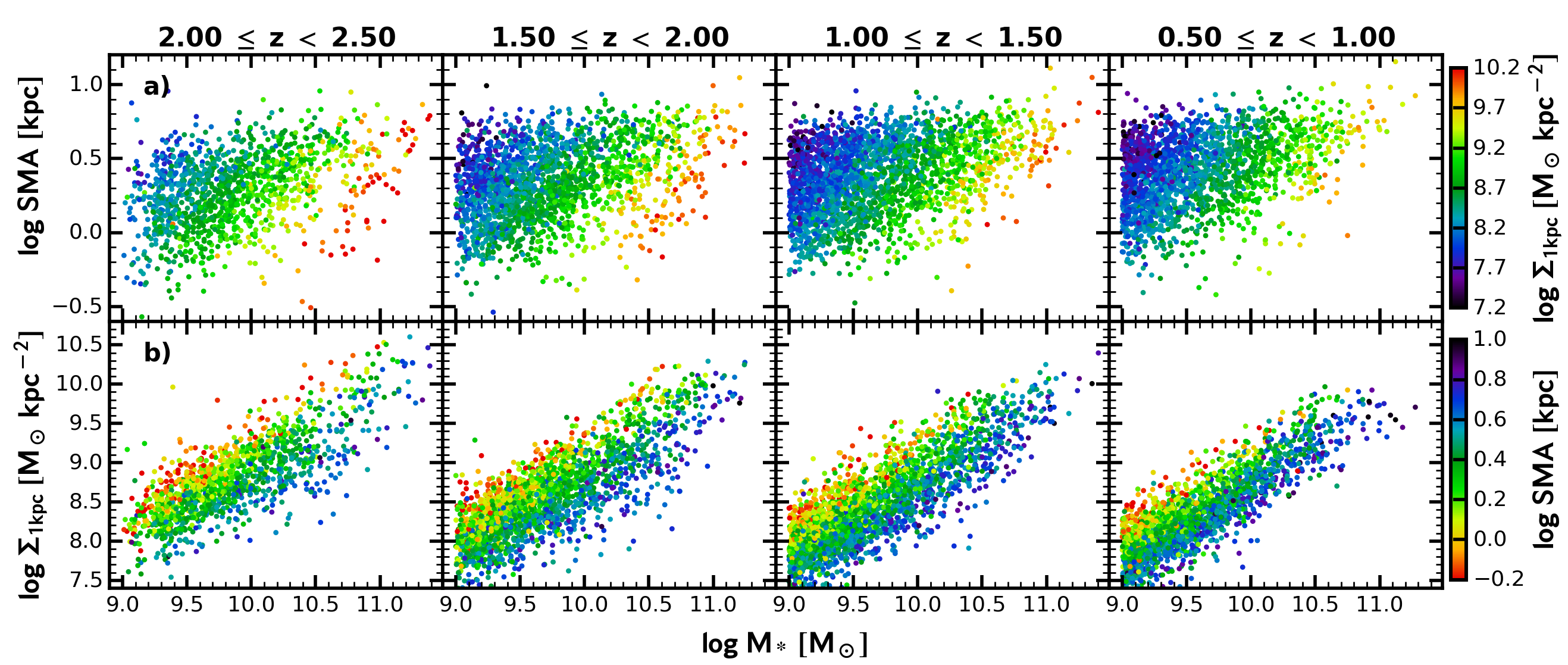}
\caption{\label{Fig:Mstar_Sig1_SMA2} $\Reff$ and $\Sig1$ \vs $\Mstar$ from all five CANDELS fields for star-forming galaxies only.  This figure differs from Figure~\ref{Fig:Mstar_Sig1_SMA1} in showing only star-forming galaxies and to a lower mass limit.  Galaxies satisfy $H \leq 24.5$ mag, axis ratio $b/a \geq 0.5$, and are classed as star-forming using the same criterion as in Figure~\ref{Fig:Mstar_Sig1_SMA1}.  Upper panel: Effective radius $\Reff$ along the semi-major axis (SMA) \vs stellar mass, colored by $\Sig1$.  Lower panel: $\Sig1$ \vs stellar mass colored by $\Reff$.  Loci of constant $\Sig1$ (above) and constant $\Reff$ (below) are visible as colored stripes.  These indicate the presence of a plane in $\Reff\text{-}\Sig1\text{-}\Mstar$ space for star-forming galaxies, and the two plots are different projections of it. $\Sig1 \sim \Mstar^q \Reff^{-p} \sim \Mstar^{1.1}\Reff^{-1}$  is the best mapping of this plane from one projection to another (Appendix~\ref{App:A1}).  The red points in the plots of SMA at high redshift are compact star-forming galaxies that have been identified as the precursors of compact quenched ``red nuggets'' at this redshift (\citealp{Barro13,Barro14,Nelson14,Williams14}; \citetalias{vanDokkum15}).}
\end{figure*}

The use of power laws for boundaries and scaling laws is motivated by the fact that the ridgelines of quenched galaxies are approximately power laws.  As Section~\ref{Sec:Variants} shows, this all but requires that the equations of quenching also be power laws in basic galaxy/halo properties, which places important constraints on quenching physics (see Section~\ref{Sec:Variants}).

The power laws used in this paper are in several categories:

$\bullet$ A \emph{structural mapping} relates one set of structural variables to another.  The main example is the mapping $\Sig1 \sim \Mstar^q\Reff^{-p}$, which maps the $\Reff \text{-} \Mstar$ plane onto the $\Sig1 \text{-} \Mstar$ plane.  The variables $p$ and $q$ depend on \Sersic $n$ \citepalias{Barro17}, and hence a given $p$,$q$ mapping applies only to galaxies with similar \Sersic indices.  The needed mapping is for star-forming galaxies, which generally have $n$ in the range 1-2 \citepalias{Barro17}.

$\bullet$ A \emph{quenched ridgeline} is the fit at a given redshift to the locus of galaxies that have quenched.  It is narrowest and best defined in $\Sig1\text{-}\Mstar$, where it has the slope 0.66 at all redshifts \citepalias{Barro17}.  Quenched galaxies, as we have noted, are defined as objects that lie more than 1.0~dex below the star-forming main sequence at their mass and redshift.

$\bullet$ A \emph{star-forming ridgeline} is the fit at a given redshift to the locus of galaxies that are still star-forming.  It is conventionally shown in the $\Reff\text{-}\Mstar$ plane (e.g., \citetalias{vanderWel14}), but it exists also in the $\Sig1\text{-}\Mstar$ plane, being mapped there by the $p$ and $q$ values for star-forming galaxies.  Star-forming galaxies are defined as all objects that lie above the line lying $-0.45$~dex below the star-forming main sequence at their mass and redshift.

$\bullet$ A \emph{scaling law} relates one variable to the power of another, usually with some implied physics.  Examples are our assumed scaling laws that BH mass scales as $\Mbh \sim \Sig1^v$ in star-forming galaxies and that total effective energy radiated during the course of BH mass growth varies as $\Ebh \sim \eta \Mbh c^2$.

$\bullet$ An \emph{evolutionary track} is a locus that represents the evolution of one variable versus another over time.  An example is the stellar-mass/halo-mass (SMHM) relation, which we take to have the form $\Mstar \sim \Mvir^s$. Since we assume no change in slope or zero point with redshift, it also is an evolutionary track.  Hence, some scaling laws, like the SMHM relation, can also be evolutionary tracks.

$\bullet$ A \emph{quenching boundary} is a power-law locus in the space of some set of structural variables at a given redshift that signals the beginning of the end of star formation when galaxies cross over it.  There are three projected boundaries, in $\Sig1\text{-}\Mstar$, $\Reff\text{-}\Mstar$, and $\Mbh\text{-}\Mstar$, which together form a single line in the master 4-D space.

The following five power laws are combined to derive expressions for the slopes of the projected quenching  boundaries.  The first two of these are derived from abundance matching:

\begin{enumerate}
\item \label{item_BasicPowerLaw:1} \emph{SMHM relation for star-forming galaxies:} This evolutionary track is parameterized as
    \begin{equation}\label{Eq:MstarMvir}
    \Mstar = {\rm const.}\times\Mvir^s,
    \end{equation}
where $\Mstar$ is total stellar mass and $\Mvir$ is halo virial mass.  $\Mstar$ is defined as the current stellar mass including stellar remnants but does not include the mass lost in stellar evolution; the definitions of halo virial mass and virial radii are given in Section~\ref{SubSec:EvolTracks}.  In the basic version of the model that we are describing now, the SMHM relation is assumed to be redshift-independent based on the nearly constant zero point obtained by \citet[hereafter RP17]{Rodriguez17} from abundance matching (see Figure~\ref{Fig:HMSM}).  Other works give slightly different results, and Variant~2 in Section~\ref{Sec:Variants} explores the effect of varying the SMHM zero point with redshift.  The exponent $s$ is set to 1.75, which is the slope below the knee from \citetalias{Rodriguez17}.  We note that this is very close to the value $s \sim 5/3$ predicted from the energetics of SNae feedback \vs halo potential well depth \citep{Dekel86,Dekel03}.  The SMHM law below the knee applies to star-forming galaxies only, and since no scatter is assumed, it is simply referred to as the SMHM relation.  Above the knee, the SMHM or its inverse, the HMSM, is used as appropriate.\footnote{The knee occurs at approximately \(\Mstar \sim 10^{10.7} ~\Msol\), which is not quite massive enough to encompass all star-forming galaxies.  This problem is addressed in Variant~1 in Section~\ref{Sec:Variants}.}  The model does not use star-formation rates explicitly, but using a particular SMHM relation for star-forming galaxies implicitly assumes a matching star-formation rate to make it true.

\item \label{item_BasicPowerLaw:2} \emph{Median galaxy effective radius, $\ReffMean$, \vs halo virial radius, $\Rvir$, for star-forming galaxies:} This evolutionary track assumes that the median $\ReffMean$ of star-forming galaxies at a given stellar mass is the same fixed fraction of the halo virial radius, $\Rvir$, at all masses and redshifts ($\Reff$ is the observed half-light semi-axis major from \citet{vanderWel12}, called SMA in Figures~\ref{Fig:Mstar_Sig1_SMA1}-\ref{Fig:Mstar_Sig1_SMA2}).  Thus:
    \begin{equation}\label{Eq:ReRvir}
    \ReffMean = {\rm const.}\times\Rvir.
    \end{equation}
This assumption is consistent with the fact that the observed median radii of star-forming galaxies appear to equal about the same fixed fraction of model halo virial radii back to $z = 3$ to within approximately 20\% (\citealp{Kravtsov13,Shibuya15,Huang17}; \citetalias{Rodriguez17}; \citealp{Somerville18}).\footnote{All observed radii in this paper are projected 2-D semi-major-axis half-light radii.  These can differ significantly and systematically from 3-D half-mass stellar radii, as shown in \citet{Somerville18}.  Eq.~\ref{Eq:ReRvir} is obeyed more accurately by 3-D half-mass radii, but the differences are unimportant given the approximate nature of our treatment.}
\end{enumerate}

The next  three power laws feature the central role of $\Sig1$:

\begin{enumerate} [resume]
\item \label{item_BasicPowerLaw:3} \emph{$\Sig1$ \vs $\Reff$ and $\Mstar$ for star-forming galaxies:} This relation is an empirical structural mapping of the form
    \begin{equation}\label{Eq:Sig1MstarRe}
    \Sig1 = {\rm const.}\times\Mstar^q\Reff^{-p}.
    \end{equation}
    where $\Sig1$ is the projected stellar mass within 1~kpc of the center of the galaxy.  Figure~\ref{Fig:Mstar_Sig1_SMA2}a plots $\Reff$ \vs $\Mstar$ for CANDELS star-forming galaxies color-coded by $\Sig1$, and Figure~\ref{Fig:Mstar_Sig1_SMA2}b plots $\Sig1$ \vs $\Mstar$ color-coded by $\Reff$.  The slanting colored stripes in both panels indicate the existence of a plane in $\Sig1\text{-}\Reff\text{-}\Mstar$ space (a similar plane was shown for VIPERS galaxies at $z = 0.5$-1.0 by \citet{Haines17}). Fitted values for the exponents of this plane are $p = 1.00\pm0.01$, $q = 1.10\pm0.01$, which are essentially constant with redshift, as is the zero point.  Comparisons of data to predictions from this relation are shown in Figure~\ref{Fig:A1} in Appendix~\ref{App:A1}, where rms residuals in $\Reff$ and $\Sig1$ are both $\sim 0.2$~dex.  This $p$, $q$ mapping applies to star-forming galaxies only and is valid because their \Sersic indices are all in the range $n = 1\text{-}2$ \citepalias{Barro17}.  (The analogous mapping for quenched galaxies has $p = 0.50$, $q = 1.10$ and is less tight owing to the more variable \Sersic indices of quenched galaxies.)

\item \label{item_BasicPowerLaw:4} \emph{$\Sig1$ \vs central velocity dispersion, $\sigma_1$:} \citetalias{Fang13} empirically demonstrated the scaling law that
    \begin{equation}\label{Eq:Sig1sig1}
    \Sig1 = {\rm const.}\times\sigma_1^{2.0\pm0.2},
    \end{equation}
    for SDSS galaxies, where $\sigma_1$ is the average velocity dispersion along the line of sight within 1~kpc.  Because SDSS does not resolve rotational motion close to the centers of galaxies, $\sigma_1^2$ is a measure of the total kinetic energy at the center of the galaxy, including both rotation and dispersion. The relation emerges from the virial mass estimator $\Mstar \sim R \sigma^2/G$ applied to the mass inside a sphere if we note that
    \begin{equation*}
    \Mstar \sim \Sig1\times\pi R^2
    \end{equation*}
    and evaluate all quantities at $R = 1$~kpc.  It then follows that  $\Sig1 \sim \sigma_1^2$ except for (evidently small) geometric factors due to light and mass profile differences, variable mass projected in front of and behind the central sphere, variable orbital geometries, and different inclinations.  Eq.~\ref{Eq:Sig1sig1} is assumed to be valid for star-forming and quenched galaxies with the same constant at all redshifts.  The main use of Eq.~\ref{Eq:Sig1sig1} is not for actual computations but to help motivate what comes next.
\end{enumerate}

\begin{figure*}[htbp]
\centering
\includegraphics[scale=1.0]{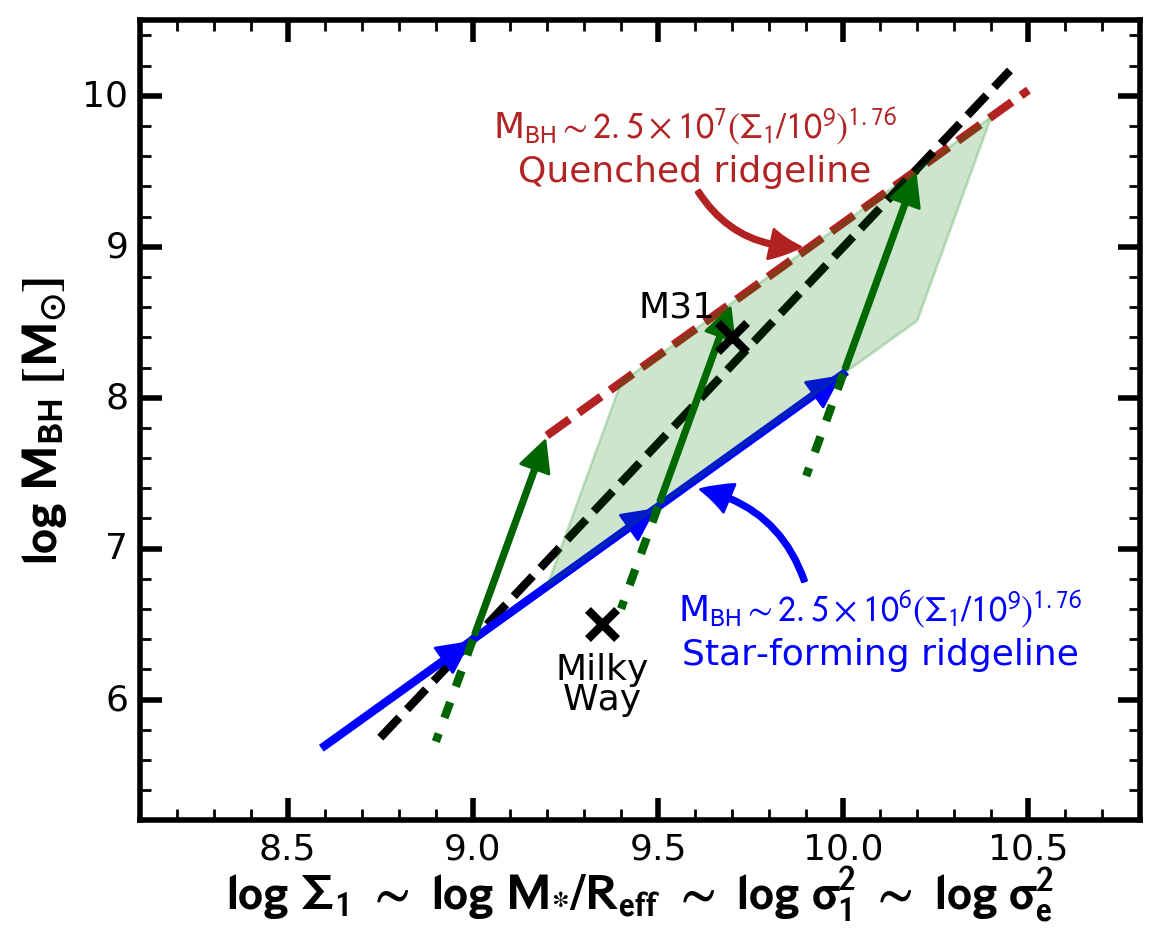}
\caption{\label{Fig:Cartoon} Cartoon showing scaling laws for $\Mbh$ \vs $\Sig1$.  In the default model, star-forming galaxies enter along the blue track at lower left, turn off into the green valley at some point along the steep green tracks, and come to rest at the top of the tracks along the quenched ridgeline (red dashed line).  The quenching boundaries shown as the red lines in Figure~\ref{Fig:Mstar_Sig1_SMA1} represent entry into the green valley and correspond to the bends in the tracks.  The three green vectors correspond to galaxies with the same starting values of $\Mstar$ at a given epoch but different values of $\Reff/\ReffMean$: small-radii galaxies turn off sooner and produce quenched galaxies with smaller $\Mstar$, $\Mbh$, $\Reff$, and $\Sig1$, etc.  The effect of the quenching boundary moving in time is illustrated by the green parallelogram, which shows the location of the green valley at an earlier time, when quenching values of $\Sig1$ and $\Mbh$ were larger; this parallelogram slides down along the scaling laws with time.  Except for this, no other aspects of the figure change with time.  Section~\ref{SubSec:BoundaryCrossing} cites evidence that the green-valley portions of the tracks (the green vectors) have the same slope and length independent of mass and time.  If so, the star-forming and quenched galaxy ridgelines will be parallel, and the observed slope of the upper relation can be used to determine the slope of the lower relation.  This is the assumption employed in Item~\ref{item_BasicPowerLaw:5} (this section).  The derivations of the slopes and zero points of the quenched and star-forming relations are fully described in Appendix~\ref{App:A3}.  Our fitted values of $p$, $q$ say that $\Sig1 \approx \Mstar/\Reff$ for star-forming galaxies, which is also roughly $\sigmae^2$ from \citet{vandenBosch16}, and thus also roughly $\sigma_1^2$ if $\sigma_1 \approx \sigmae$.  We have therefore used these approximate relations to label the X-axis as the log of all of these quantities (ignoring the fact that $\Sig1 \approx \Mstar/\Reff$ is not quite true for quenched galaxies).  The black dashed line shows schematically what happens if one attempts to fit the BH masses of all galaxies with a single power law, as in \citet{vandenBosch16}.  The slope of the black dashed line (in the same units) is $\sim 2.6$.  The value of $\Sig1$ for M31 is based on the enclosed mass at 1~kpc in the M31 mass model by \citet{Blanadiaz18}, and the value of $\Sig1$ for the Milky Way is based on a rotation speed of $175~\rm{km~sec^{-1}}$ at 1~kpc \citep{Courteau14,Blandhawthorn16}.  The dashed green lines below the blue track show alternative growth histories for BHs in star-forming galaxies, as discussed in Section~\ref{SubSec:BHGrowth}.}
\end{figure*}

The last step is to specify rules for BH growth.  The basic assumption is that BH mass scales as a power law of  $\Sig1$, but with potentially different slopes and zero points for star-forming, quenched, and green valley galaxies. The adopted laws are shown in Figure~\ref{Fig:Cartoon}.  Star-forming galaxies enter along the blue track at the bottom.  At some point they cross the quenching boundary and then turn off into the green valley along the steep (green) vectors.  This happens at different values of $\Sig1$ for different-$\Reff$ galaxies at different times (see below), which creates multiple tracks through the GV regime.  Eventually BHs stop growing, and galaxies come to a halt along the fully quenched ridgeline (dashed red line).  Further evolution is not included in the model.

\begin{figure*}[htbp]
\centering
\includegraphics[scale=0.53]{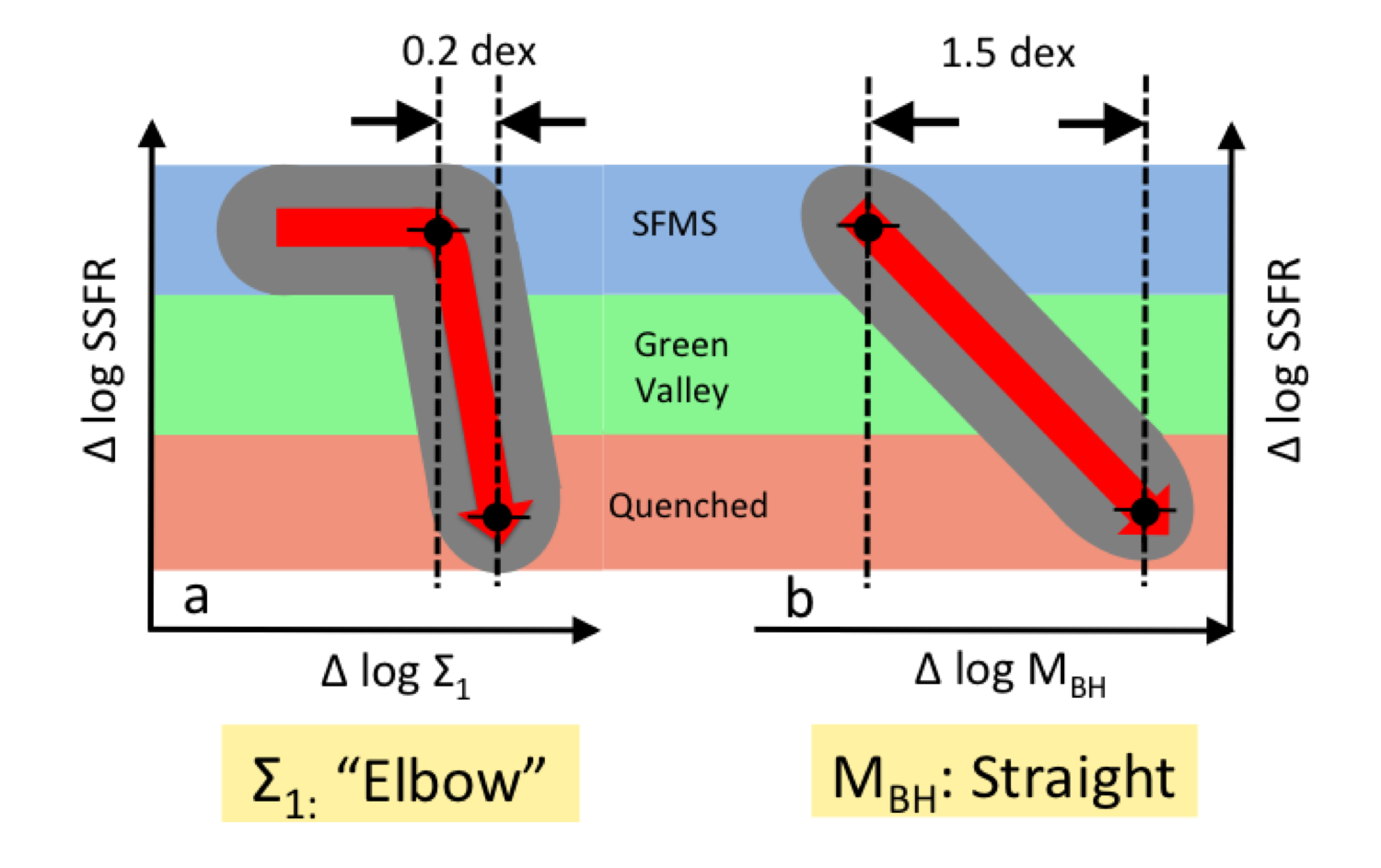}
\caption{\label{Fig:Elbow} Cartoon illustrating the different evolution of $\Sig1$ and $\Mbh$ through the green valley.  Panel a: $\Delta \log$\Sig1\ is the residual of $\log \Sig1$ with respect to the quenched ridgeline in $\Sig1\text{-}\Mstar$, and $\Delta \log SSFR$ is the residual of $\log SSFR$ with respect to the star-forming main sequence.  $\Sig1$ shows a characteristic ``elbow'' shape whereby $SSFR$ falls sharply but $\Sig1$ grows only slightly \citepalias{Fang13, Barro17} through the green valley.  Panel b: The analogous plot for $\Delta \log \Mbh$ based on data from  \citet[][Figure 3]{Terrazas17}.  In contrast to $\Sig1$ in panel a, black hole mass grows considerably in the green valley.  The delta-quantities above the figures are used to plot the green-valley vectors in Figure~\ref{Fig:Cartoon}.}
\end{figure*}

The series of broken power-law tracks in Figure~\ref{Fig:Cartoon} is motivated by the following logic:  1) Black holes in local quenched galaxies obey a power-law relation $\Mbh \sim \sigmae^{u}$ \citepalias{Kormendy13}.  We assume $\sigma_1 \sim \sigmae$, in which case $\Mbh \sim \Sig1^{v}$, where $v = u/2.0$ from Item~\ref{item_BasicPowerLaw:4} above.  The value of $v$ is the slope of the upper, quenched ridgeline in Figure~\ref{Fig:Cartoon}.  2) A steep track in the GV is then indicated by the fact that star-forming and quenched galaxies can have similar $\Sig1$ values today at the same stellar mass but that $\Mbh$ differs strongly for these same two populations (line~18 of Table~\ref{Tb_1}).  This phenomenon is visible by comparing panels b and c of Figure~\ref{Fig:ThreePanel} and is illustrated more graphically in Figure~\ref{Fig:Elbow}.  Together these imply that $\Sig1$ largely stops increasing in the GV (hence the ``elbow") whereas $\Mbh$ keeps growing. 3) Based on a feature seen in SDSS galaxies \citep[Figure 5b,][]{Luo19}, the entry to the GV (i.e., the quenching boundary) is taken to be parallel to the quenched ridgeline in $\Sig1\text{-}\Mstar$ but is displaced downward from it by a fixed 0.2~dex at all stellar masses.  This motivates the increase of 0.2~dex that has been applied to $\Sig1$ across the GV for all galaxies in Figure~\ref{Fig:Cartoon} (green vectors; see also Figure~\ref{Fig:Elbow}a).  4) Likewise, the separation between the quenched and star-forming ridgeline in $\Mbh\text{-}\Mstar$ is observed to be $\sim 1.5$~dex at all stellar masses \citep[Figure 3]{Terrazas17}, which we take to be the amount of BH mass growth across the GV (see illustration in Figure~\ref{Fig:Elbow}b).  5) Assumptions 3) and 4) applied to all galaxies then imply that the lengths and slopes of the green GV tracks are independent of galaxy mass today and thus are all parallel.  6) Finally, point 5) means that the slope of the lower track for star-forming galaxies is the same as the slope of the upper scaling law for quenched galaxies, and hence can be determined from it.\footnote{Three further aspects of the cartoon model in Figure~\ref{Fig:Cartoon} merit comment: 1) The evolution in BH mass is steep \vs $\Sig1$, but this does not mean that it is necessarily steep in time.  The total time needed to cross the GV is not specified and could be long.  2) Similarly, since AGN luminosities depend on BH mass accretion \emph{rates}, they also cannot be calculated without additional assumptions about evolutionary times.  3) The fractional BH mass growth in the GV is taken to be the offset in BH mass between the star-forming and quenched populations in \citet[Figure 3]{Terrazas17}, which are separated by about $\sim$1.5 dex.  However, this snapshot of mass differences now does not equal the BH growth factor \emph{for an individual galaxy} if boundaries and ridgelines have been moving over time.  The sense is that 1.5~dex would be an overestimate of the growth factor, but the correction cannot be more than a few tenths of a dex (Section~\ref{SubSec:BoundaryCrossing}) and has not been applied in Figure~\ref{Fig:Cartoon}. It is discussed further in Section~\ref{Sec:Discussion}.}

The main justification for the above framework is the empirical observation that $\Mbh \sim \sigma_e^{4\text{-}5}$ in many types of galaxies (\citetalias{Kormendy13}; \citealt{Ho14}), which translates to a power law in $\Sig1$ from Eq.~\ref{Eq:Sig1sig1}.  Many papers have tried to derive the $\sigma^4$ law analytically \citep[e.g.,][]{Silk98,Ostriker00,Ciotti01,Adams01,Colgate03,King03,Cen07,Cen15,King15}, but there is as yet no generally accepted explanation.  As time passes, evidence is actually growing for two \emph{different} laws with different zero points, one for quenched galaxies \citepalias{Kormendy13} and one for star-forming galaxies \citep{Ho14,Reines15,Savorgnan16}, separated by the green valley.  The slope of the quenched law is better determined, which is why we have used it as our starting point, but the basic idea of two laws is consistent with our scheme in Figure~\ref{Fig:Cartoon}.  More discussion of the adopted laws and their derivation is given in Section~\ref{Sec:Discussion}.

The preceding paragraphs have motivated the law that $\Mbh = \rm{const.} \times \Sig1^v$ for star-forming galaxies and that BHs grow more steeply relative to $\Sig1$ in the GV.  We now extrapolate those rules and apply them to star-forming and GV galaxies at all redshifts. The main argument in favor of this assumption is the similar appearance of the $\Reff\text{-}\Mstar$, $\Sig1\text{-}\Mstar$, and  $\Mbh\text{-}\Mstar$ scaling laws at all redshifts, which suggests that the same basic physics has operated at all epochs since $z \sim 3$.  Thus, if bulges build BHs today, we assume that they did so in the same way at early times, and we model the behavior of early galaxies on current galaxies.

Before leaving this topic, we note that the cartoon in Figure~\ref{Fig:Cartoon} interprets the blue BH scaling law for star-forming galaxies \emph{as an evolutionary track} that galaxies evolve along.  Indeed, that is how it was introduced above.  Considerable evidence suggests that star-forming galaxies in fact spend most of their time below the blue relation and build their BHs only later as they enter the green valley.  This means that they would cross the blue line from below when passing into the green valley.  This alternate scenario is discussed in Section~\ref{SubSec:BHGrowth}.

Returning to the main thread and combining the above logic we now have:

\begin{enumerate} [resume]
\item \label{item_BasicPowerLaw:5} \emph{BH mass \vs $\Sig1$ for star-forming galaxies:}
    \begin{equation}\label{Eq:MbhSig1}
    \Mbh = {\rm const.}\times\Sig1^v,
    \end{equation}
    where $u$ is the slope of the quenched ridgeline, and $v = u/2.0$.  This is the blue line in Figure~\ref{Fig:Cartoon}, which is assumed to hold for all star-forming galaxies at all redshifts $z<3$.  Since $\Mbh$ is uniquely specified by $\Sig1$, this means that $\Mbh\text{-}\Mstar$ maps onto $\Sig1\text{-}\Mstar$ and that $\Mbh$ is now also joined to the space of  $\Sig1\text{-}\Reff\text{-}\Mstar$ for star-forming galaxies.
\end{enumerate}

To this point, star-forming galaxies in the model have only one degree of freedom, which is $\Mvir$ today (or alternatively $\Mstar$ today), and all galaxies of a given $\Mstar$ at a given redshift have the same $\Reff = \ReffMean$.  It remains now to introduce the crucial radius variations about $\ReffMean$ as a second parameter, without which there would be no variations in $\Mbh$ at a given stellar mass. Measurable scatter is observed in $\Reff$ at fixed $\Mstar$ for star-forming galaxies, with $\sigma_{\Reff} \sim 0.20$~dex at all masses \citepalias{vanderWel14}.  At this point, the introduction of scatter is purely empirical; discussion of possible origins is deferred to Section~\ref{SubSec:RadiusScatter}.

Adding radius scatter makes star-forming galaxies a two-parameter family labeled by $\Mstar$ and $\Del \log R_{\rm e}$.  Looking ahead to the model with zero points in Section~\ref{Sec:EvolPLModel}, we find it convenient to re-express these numbers in terms of the values for a galaxy and its halo at a starting redshift, $z_{\rm s}$, when the stellar mass $\Mstar = 10^9~\Msol$.  Thus, the initial parameters at $z_{\rm s}$ are $M_{\rm *,s} = 10^9~\Msol$, $\Mvir = M_{\rm vir,s}$, $\Reff = R_{\rm e,s}$, and $\Rvir = R_{\rm vir,s}$.  How these parameters are set is described in  Section~\ref{SubSec:GalIniCon}.

To model the scatter, we multiply each starting radius by a scatter factor $R_{\rm e,s}/\langle R_{\rm e}\rangle$, where $\langle R_{\rm e}\rangle$ is the mean value of $\Reff$ for $10^9~\Msol$ galaxies at $z_{\rm s}$.  This scatter factor is assumed to be constant throughout the star-forming phase, and its logarithm is assumed to be Gaussianly distributed with $\sigma_{\Reff} = 0.2$~dex.   Every galaxy is now labeled by two numbers, the redshift $z_{\rm s}$ at which $\Mstar = 10^9~\Msol$ and its value $\Del \log R_{\rm e,s}$ at that redshift.  The latter is also equivalent to the scatter in $\Mbh$ and $\Sig1$ since the previous equations imply:
\begin{equation}\label{Eq:Sig1Re}
\Delta \log \Sigma_{\rm 1,s} = -p\Delta \log R_{\rm e,s},
\end{equation}
and
\begin{equation}\label{Eq:Sig1Mbh}
\Delta \log M_{\rm BH,s} = v\Delta \log \Sigma_{\rm 1,s},
\end{equation}
and
\begin{equation}\label{Eq:ReMbh}
\Delta \log M_{\rm BH,s} = -vp\Delta \log R_{\rm e,s}.
\end{equation}
Note that galaxies with larger $\Reff$ at fixed $\Mstar$ have smaller BHs.

The next step is to evolve galaxies forward in time.  To do this, another crucial assumption is made, namely, that \emph{the star-formation rate at a given mass is independent of $\Reff$}.  This fact is now established from several studies on both the SDSS \citep{Omand14,Lin19} and CANDELS/3D-HST (\citealp{Whitaker17}; \citetalias{Fang18}; \citealp{Lin19}).  Galaxies with the same starting $\Mstar$ but different starting radii therefore grow in $\Mstar$ at the same rate, and each starting mass cohort stays together.  Since all relationships are power laws, it follows that $\Del \log R_{\rm e,s}$, $\Del \log \Sigma_{\rm 1,s}$, and $\Del \log M_{\rm BH,s}$ are constant through time. Hence, galaxies at a given $\Mstar$ that start out with large radii always have smaller-than-average black holes, and vice versa.  If quenching requires reaching some threshold BH mass that does not vary too rapidly with $\Mstar$, large galaxies will tend to do this later and attain a larger stellar mass before quenching. The fact that starting radius confers an $\Mbh$ enhancement to some galaxies and a penalty to others is what creates the \emph{sloped quenching boundaries} in $\Reff$\text{-}$\Mstar$, $\Sig1$\text{-}$\Mstar$, and $\Mbh$\text{-}$\Mstar$ in Figure~\ref{Fig:ThreePanel}.

The final step is to specify the equation of the quenching boundary.  For this, our first-stage, empirical power-law model assumes that the \emph{total effective energy} emitted by a BH during its lifetime is equal to some \emph{total energy quota} that must be transferred to the halo gas in order to inhibit further infall onto the galaxy.  We call the latter the \emph{halo quenching-energy quota}, and in the empirical model we parameterize it as a simple power of halo mass.  Thus
\begin{equation}
\Ebh = \Eheat,
\end{equation}
or
\begin{equation}\label{Eq:MbhMvir}
\eta \Mbh c^2 = {\rm const.}\times\Mvir^t,
\end{equation}
where $\eta$ is an efficiency factor for converting BH accretion energy into useful feedback energy.  We reiterate that crossing over the boundary corresponds to \emph{entry} into the GV, and thus that BHs are just starting to affect their halos and galaxies at that point.  The use of $\Eheat$ to denote the critical halo energy is meant to be generic and does not mean that the energy transferred must be thermal; it could be mechanical, cosmic rays, or any other mode of energy/momentum transfer that can stop halo gas from falling onto the galaxy.  This expression for necessary BH feedback as a power law in $\Mvir$ is needed only during the star-forming phase in order to calculate the location of the quenching boundary and is not needed (or assumed to be valid) after galaxies pass over the quenching boundary.  The value of $t$ and the (evolving) constant are adjusted to match the adopted quenching boundary, which was taken to be offset by 0.2~dex in $\Sig1$ from the quenched ridgeline above.  We will find in the next section that the empirical value of $t$ must be close to 2.0 in order to match the observed \emph{slopes} of the quenched ridgelines in $\Sig1\text{-}\Mstar$ and $\Mbh\text{-}\Mstar$, and the constant \emph{must vary with time} in order to match the moving boundaries in those same spaces.

\subsection{Quenching Boundary Power Laws}\label{SubSec:QuenchSlope}

We now show that the above assumptions suffice to compute the slopes of the quenching boundaries in our various spaces as a function of $s$, $t$, $v$, $p$, and $q$.  Galaxies on the boundary at the same epoch obey
\begin{equation*}
\Mbh \sim \Mvir^t.
\end{equation*}
by Eq.~\ref{Eq:MbhMvir}.  But
\begin{equation*}
\Mbh \sim \Sig1^v,
\end{equation*}
from Eq.~\ref{Eq:MbhSig1}, and
\begin{equation*}
\Mvir \sim \Mstar^{1/s}.
\end{equation*}
from Eq.~\ref{Eq:MstarMvir}.  Substituting, we have
\begin{equation}\label{Eq:Sig1MstarSlope}
\Sig1 \sim \Mstar^{t/sv},
\end{equation}
which gives the slope, $t/sv$, of the instantaneous boundary in $\Sig1\text{-}\Mstar$.  Now apply the mapping
\begin{equation*}
\Sig1 \sim \Mstar^q\Reff^{-p}
\end{equation*}
from Eq.~\ref{Eq:Sig1MstarRe} to obtain
\begin{equation}\label{Eq:ReffMstarSlope}
\Reff \sim \Mstar^{q/p - t/svp},
\end{equation}
which gives the slope, $q/p - t/svp$, of the instantaneous boundary in $\Reff\text{-}\Mstar$.
Finally, $\Mbh \sim \Sig1^v$ gives
\begin{equation}\label{Eq:MbhMstarSlope}
\Mbh \sim \Mstar^{t/s},
\end{equation}
which is the slope, $t/s$, of the instantaneous boundary in $\Mbh\text{-}\Mstar$.

Of these five exponents, $p$ and $q$ are set by the $p$, $q$ mapping while $s$, $t$, and $v$ are set by three additional observational constraints:

\begin{enumerate}
\item \label{item_QuenchSlope:1} \emph{Slope of the quenching boundary in $\Sig1$ \vs $\Mstar$:} Even though the spaces $\Sig1\text{-}\Mstar$, $\Reff\text{-}\Mstar$, and $\Mbh\text{-}\Mstar$ are all closely related (for star-forming galaxies), the quenched ridgeline is narrowest and best determined in $\Sig1\text{-}\Mstar$, and so we start there.  We have further assumed that the boundary is offset by a constant distance below the ridgeline by 0.2~dex at all masses (based on a feature in the local SDSS $\Sig1\text{-}\Mstar$ distribution found by \citealp{Luo19}) and therefore has the same slope as the ridgeline.  \citetalias{Barro17} found the slope of this ridgeline to be nearly constant at 0.66 from $z = 3$ to now.  Eq.~\ref{Eq:Sig1MstarSlope} now gives: \footnote{\citetalias{Barro17} set the cut for quenched galaxies at $-0.7$~dex below the star-forming ridgeline, which differs from our definition using $-1.0$~dex in Figure~\ref{Fig:Mstar_Sig1_SMA1}.  However, the ridgeline slope is insensitive to this definition since green valley galaxies and fully quenched galaxies occupy nearly the same regions in $\Sig1\text{-}\Mstar$ \citepalias{Fang13,Barro17}.}
    \begin{equation}\label{Eq:Sig1Mstar}
    \frac{\Delta \log \Sig1}{\Delta \log \Mstar} = \frac{t}{vs} = 0.66\pm0.02,
    \end{equation}
where the error is based on the scatter in four independent measurements at four different redshifts in \citetalias{Barro17}.  This is the first constraint.\footnote{For simplicity, the shape of the quenched ridgeline in $\Sig1\text{-}\Mstar$ is assumed here to be a pure power law, as found by \citetalias{Barro17}.  Recent work by \citet{Luo19} for SDSS galaxies indicates some downward curvature in $\Sig1\text{-}\Mstar$ at high masses, which may actually improve quenching predictions for high-mass galaxies, as discussed under Variant~1 in Section~\ref{SubSec:BrokenPLModel}.}

As discussed below, we do not use the quenched ridgeline slope in $\Reff\text{-}\Mstar$ as an observational constraint because of the sensitivity of effective radii to dry merging and other post-quenching processes.  However, the slope of the \emph{quenching boundary} in $\Reff\text{-}\Mstar$ will be needed in future.  Eq.~\ref{Eq:ReffMstarSlope} now gives:
    \begin{eqnarray}\label{Eq:ReffMstar}
    \frac{\Delta \log \Reff}{\Delta \log \Mstar} & = & \frac{q}{p}  - \frac{t}{vps} \nonumber\\
                                         & = & 1.10 - \frac{t}{vs}  = 0.44\pm0.02,
    \end{eqnarray}
    where the second line follows from inserting $t/vs = 0.66\pm0.02$ and $p = 1.0\pm0.01$, $q = 1.1\pm0.01$ from Appendix~\ref{App:A1}.

    Note that, although the quenching boundary and quenched ridgeline are assumed to be parallel in $\Sig1\text{-}\Mstar$, they are not parallel when mapped into $\Reff\text{-}\Mstar$ because quenched galaxies have different \Sersic indices and $\Reff$ due to dry merging.  This is a major reason why the quenched ridgeline in $\Reff\text{-}\Mstar$ has the steep slope of 0.75 \citepalias{vanderWel14} whereas the quenching boundary slope in Eq.~\ref{Eq:ReffMstar} is only 0.44.

  \item \label{item_QuenchSlope:2} \emph{Slope of the quenching boundary in $\Mbh$ \vs $\Mstar$:} As shown in Figure~\ref{Fig:ThreePanel}c here and more clearly in Figure~3 of \citet{Terrazas17}, the separation between fully quenched and star-forming galaxies in $\Mbh\text{-}\Mstar$ is a fixed $\sim 1.5$~dex today for galaxies above $10^{10}~\Msol$, and we assume that this is true at all redshifts.  This implies that the quenching boundary is parallel to the quenched ridgeline in $\Mbh\text{-}\Mstar$, as it is in $\Sig1\text{-}\Mstar$, and the quenched slope in $\Mbh\text{-}\Mstar$ can therefore be used as another observational constraint on the unknown exponents $s$, $t$, and $v$. The local observed $\Mbh\text{-}\Mstar$ relation consists of a narrow ridgeline of fully quenched galaxies with star-forming galaxies scattering below it (\citetalias{Kormendy13}; \citealp{vandenBosch16, Terrazas16, Krajnovic18}).  We equate the observed quenched ridgeline slope to that of local ellipticals and bulges, which is 1.16 from \citetalias{Kormendy13}.  Thus:
    \begin{equation}\label{Eq:MbhMstar}
    \frac{\Delta \log \Mbh}{\Delta \log \Mstar} = \frac{t}{s} = 1.16\pm0.08,
    \end{equation}
    where the error comes from \citetalias{Kormendy13}.  Since classical bulges and ellipticals follow the same law, this slope is insensitive to the precise definition of the quenched sample.  This is the second constraint.

\item \label{item_QuenchSlope:3} Finally, the slope of the SMHM relation, $s$, is observed to be $1.75\pm0.17$ based on the abundance matching results of \citetalias{Rodriguez17} and \citet{Behroozi19}.  This is the third constraint.
\end{enumerate}

Constraints are now available from Items~\ref{item_QuenchSlope:1}, \ref{item_QuenchSlope:2}, and \ref{item_QuenchSlope:3}, and the solution is uniquely determined.  The combination $s = 1.75\pm0.17$, $t/s = 1.16\pm0.08$, and $t/vs = 0.66\pm0.02$ implies $t = 1.16 \times 1.75 = 2.03\pm0.24$, and $v = 1.16/0.66 = 1.76\pm0.13$, and from this last we derive $u = 2.0v = 3.52\pm0.26$, as was employed in Eq.~\ref{Eq:MbhSig1} above.

We pause here to consider the uncertainties in these values.  The formal errors are small, but the important unknowns are whether the cartoon model in Figure~\ref{Fig:Cartoon} is actually correct and whether the same $\Mbh\text{-}\Sig1$ scaling relations obtain at all redshifts.  The general constancy of the cartoon relations is supported by the fixed power-law appearance of the $\Sig1\text{-}\Mstar$ and $\Mbh\text{-}\Mstar$ distributions back in time (Lines 12 and 16 in Table~\ref{Tb_1}), and by correct predictions for both evolving zero points when combined with the moving boundary hypothesis, as discussed in Section~\ref{SubSec:Successes}.  In short, the constant slopes and offsets of quenched and star-forming galaxies in $\Sig1\text{-}\Mstar$ and $\Mbh\text{-}\Mstar$ over time impose strict constraints on the constancy of the quenching process, which is why these relations are so important.  If further data confirm these diagrams, something like the cartoon model \emph{must} be taking place.

Another concern is that the slope of the quenched ridgeline has been used to set the slope of the star-forming galaxies in Figure~\ref{Fig:Cartoon}, but the quenched slope may have evolved after quenching due to mergers.  For example, to match the scaling laws of quenched galaxies today requires $\Reff \sim \Mstar^2$ along the merging trajectory (\citealp[e.g.,][]{Naab09}; \citetalias{vanderWel09}), and thus $\sigmae \sim \Mstar^{-0.5}$, i.e., $\sigmae$ actually declines as galaxies merge!  Both of these vectors (in $\Reff$ \vs $\Mstar$ and $\sigmae$ \vs $\Mstar$) point at angles to their respective ridgelines, and evolution along these vectors might therefore change the original slopes.  Mergers of clusters of galaxies add further orbital energy to central galaxies, puff them up, and reduce $\sigmae$ still more \citep{Boylan06, Lauer07}.  Some evidence of steepening in $\Mbh$ \vs $\sigmae$ at high masses has been noted (\citealt{Lauer07, Gultekin09, Krajnovic18}; \citetalias{Kormendy13}).  To summarize, effective quantities like $\Reff$ and $\sigmae$ are strongly impacted by dry merging, and their expected evolutionary vectors are at angles to today's ridgelines, with unknown consequences for the ridgeline slopes over time.  This makes effective quantities unsuitable as slope constraints, which is why we have not used them.

By contrast, merging is probably not as important for $\Sig1\text{-}\Mstar$, since centers are more immune to merging than envelopes and the slope of $\Sig1\text{-}\Mstar$ is actually measured to be constant over time \citepalias{Barro17}.  In $\Mbh\text{-}\Mstar$, if pre-merger BHs merge along with their host galaxies and if BH mass is conserved, galaxies would tend to move \emph{along} this ridgeline since its slope is near unity.  This slope would then also be invariant.  Hence, it is plausible that both our calibrating slopes are stable over time.

A final matter is the low value of $u$ that we have derived, which implies that $\Mbh \sim \sigma_1^{3.52\pm0.26}$. This value stems from $v = 1.76$, since $u$ is $2.0 \times v$.  But $v$ is simply the ratio of the slopes of the quenched ridgelines in $\Sig1\text{-}\Mstar$ and  $\Mbh\text{-}\Mstar$, both of which are well determined.  Alternative values for $u$ are $4.40\pm0.30$ for quenched ridgeline galaxies from \citetalias{Kormendy13} and $5.35\pm0.23$ for all galaxies together from \citet{vandenBosch16}.  At first glance, these seem incompatible with our value of 3.52 and also with each other. (These laws use $\sigmae$ while we use $\sigma_1$, but the difference is probably not large.)  However, as described in Section~\ref{SubSec:PredictionsTests} below, the Figure~\ref{Fig:Cartoon} cartoon model can explain the differences as well as the low value that we get for $v$ (and $u$) for star-forming galaxies alone.

To summarize, the empirical power-law model predicts a series of slopes and relations that can be fitted to observed quantities.  The following three exponents are the basic unknowns: $\Mbh \sim \Mvir^t$ (the halo energy quota needed to halt gas cooling), $\Mstar \sim \Mvir^s$ (the SMHM relation), and $\Mbh \sim \Sig1^v$ (the assumed relation between BH mass and central stellar density for star-forming galaxies). These three exponents, $t$, $s$, and $v$, predict the slopes of the two quenching boundaries in $\Sig1\text{-}\Mstar$ and $\Mbh\text{-}\Mstar$, which are observed, and $s$ is separately measured from abundance matching.  This makes three constraints, which together determine the three exponents.

\section{The Evolving Empirical Power-law Model}\label{Sec:EvolPLModel}

\subsection{Model for Galaxy Radii; Evolutionary Tracks}\label{SubSec:EvolTracks}

The power-law exponents $s$, $t$, and $v$ are now specified. The present section adds moving zero points and evolves star-forming galaxies along evolutionary tracks to create an empirical  \emph{evolving} power-law model, so-called because the halo energy quota, $\Eheat$, is represented as a power law in $\Mvir$ with exponent $t$ set to 2.0, and a moving boundary is introduced to match the moving (red) quenching boundaries in Figure~\ref{Fig:Mstar_Sig1_SMA1}.  In what follows, this name is shortened to the ``empirical power-law model''.

\begin{figure}[htbp]
\centering
\includegraphics[scale=0.44]{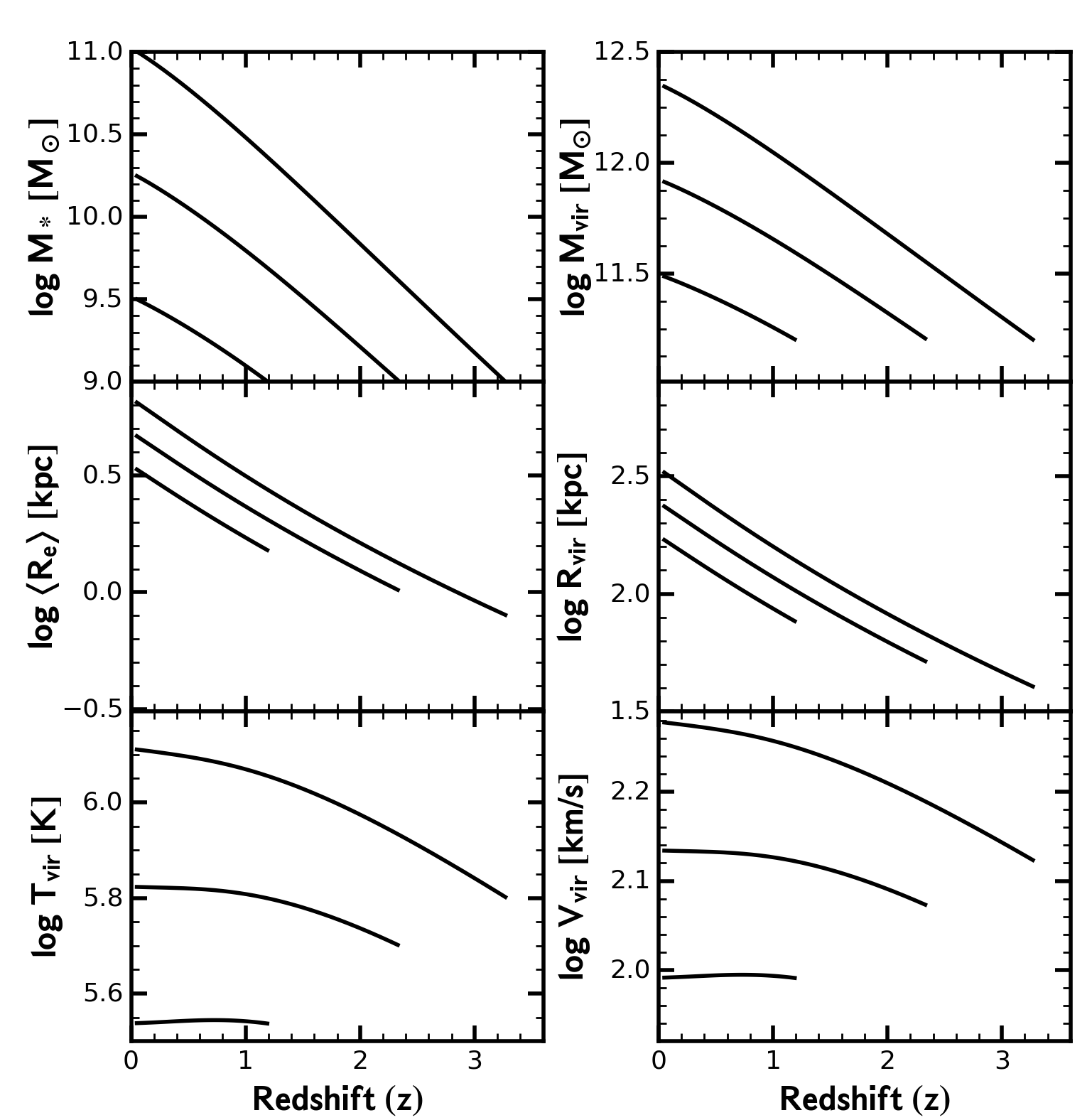}
\caption{\label{Fig:Trajectoryz} Evolutionary tracks for $\Mstar$, $\Mvir$, $\ReffMean$, $\Rvir$, $\Tvir$, and $\Vvir$ for three sample star-forming central galaxies \vs redshift.  $\Mvir$ and $\Rvir$ are based on the evolutionary model for dark halos summarized in \citetalias{Rodriguez17}.  $\Mstar$ is derived from $\Mvir$ using the SMHM relation $\Mstar \sim \Mvir^s$ with $s = 1.75$ from Section~\ref{SubSec:InputPowerLaws}.  The constant value $\ReffMean/\Rvir = 0.02$ is taken from \citetalias{Rodriguez17} (see Item~\ref{item_BasicPowerLaw:2} in Section~\ref{SubSec:InputPowerLaws} and Figure~\ref{Fig:HMSM}).  Tracks start at the starting redshift, $z_{\rm s}$, which is the redshift where stellar mass $\Mstar = 10^9~\Msol$.  The quantity $\Vvir$ is used in Section~\ref{Sec:ThermalModel}, which discusses the halo binding-energy interpretation of the halo energy quota.  The quantity $\Tvir$ is used to compute heating and cooling rates in Variants~3 and 4 in Appendix~\ref{SubSec:AppRatesorTotals}.}
\end{figure}

The model starts by using our knowledge of how halos evolve in $\Rvir$ \vs $\Mvir$ to derive evolutionary tracks for \emph{central star-forming galaxies} in $\Reff\text{-}\Mstar$, $\Sig1\text{-}\Mstar$, and $\Mbh\text{-}\Mstar$.  Ingredients and assumptions in the model are listed in Table~\ref{Tb_2}.  Accurate evolutionary tracks are necessary during the star-forming phase in order to have galaxies cross the quenching boundaries in the right place.  To achieve this, a halo-based model is used to predict the evolution of $\Mstar$ and $\Reff$ for star-forming galaxies.  The starting point is the formulae for halo mass growth and related galaxy parameters of central star-forming galaxies from \citet{Rodriguez15} and \citetalias{Rodriguez17}.  Figure~\ref{Fig:Trajectoryz} plots the resulting stellar and virial parameters for three representative different-mass halos and their associated galaxies \vs time.  The adopted constant value for the ratio $\spin = 0.02$ comes from the abundance matching results in \citetalias{Rodriguez17} for star-forming galaxies below $\Mstar = 10^{10.5}~\Msol$ (see Item~\ref{item_BasicPowerLaw:2} in Section~\ref{SubSec:InputPowerLaws} and Figure~\ref{Fig:HMSM} in Appendix~\ref{App:A4}).  See also the discussion of radii evolution in \citet{Somerville18}, which justifies a constant value.

\begin{table*}[!htbp]
\renewcommand\arraystretch{1.5} %adjust table height
{\centering
\caption{\label{Tb_2}Model Ingredients and Assumptions}
\begin{tabularx}{\textwidth}{cXl}
\toprule
No. & Item & References\\
\midrule
1 & Halo masses grow according to mean trajectories versus time derived from N-body simulations. & \citet{Rodriguez17}\\
2 & Central galaxies evolve along a power-law SMHM relation while star-forming, with constant slope $s = 1.75$, constant zeropoint, and no scatter. & \citet{Rodriguez17}\\
3 & The median effective radii $\ReffMean$ of star-forming galaxies are $0.02 \times \Rvir$ at all masses and redshifts. & \citet{Huang17}\\
& & \citet{Somerville18}\\
4 & There is significant real scatter in both $\Reff$ and $\Sig1$ among star-forming galaxies at fixed $\Mstar$ and $\Rvir$. & \citet{Omand14}\\
& & \citet{vanderWel14}\\
& & \citet{Fang13}\\
& & \citet{Barro17}\\
5 & A galaxy's offset in radius $\Del\log \Reff$ from the median $\ReffMean$ is constant during the star-forming phase. SF galaxies evolve along parallel tracks in $\Reff\text{-}\Mstar$. & Assumed, Section~\ref{SubSec:EvolTracks}\\
6 & For star-forming galaxies, $\Sig1 = {\rm const.}~\Mstar^q \times \Reff^{-p}$, where $q = 1.1$ and $p = 1.0$. & Fitted to CANDELS data in Appendix~\ref{App:A1}. \\
7 & The star-formation rate of galaxies on the SF main sequence is independent of $\Reff$ and $\Sig1$ at fixed $\Mstar$. & \citet{Omand14}\\
& & \citet{Whitaker17}\\
& & \citet{Fang18}\\
& & \citet{Lin19}\\
8 & $\Delta \log \Sig1 = 0.2$~dex and $\Delta \log \Mbh = 1.5$~dex across the green valley, independent of stellar mass and   & \citet{Luo19}\\
& redshift. & \citet{Terrazas17}\\
& & Assumed in Section~\ref{SubSec:InputPowerLaws}\\
& & and discussed in Section~\ref{SubSec:BoundaryCrossing}\\
9 & Black-hole masses in star-forming galaxies grow as $\Mbh = {\rm const.}~\Sig1^{1.76}$ with the same zero point and slope at all redshifts and no scatter.$^{\rm a}$ & Assumed, Section~\ref{SubSec:InputPowerLaws}\\
10 & The effective energy radiated from the black hole is $\Ebh = \eta \Mbh c^2$, where $\eta = 0.01$. & Assumed, Section~\ref{SubSec:EvoTracksQuenchBound}\\
11 & The energy quota that must be absorbed by the halo gas to initiate entry into the green valley in the empirical power-law model is $\Eheat = k_{\rm e}(z) \left(\frac{\Mvir}{10^{12}~\Msol}\right)^t$ with no scatter. & Assumed, Section~\ref{SubSec:EvoTracksQuenchBound}\\
12 & The energy quota that must be absorbed by the halo gas to initiate entry into the green valley in the binding-energy model is $4 \times \Eth = 4 \times 1/2~f_{bary}\Mvir\Vvir^2$ with no scatter. & Assumed, Section~\ref{Sec:ThermalModel} \\
13 & Mergers during the star-forming and GV phases are ignored, and galaxies do not rejuvenate in the GV. & Assumed throughout\\
\bottomrule
\end{tabularx}}

\footnotesize{$^{\rm a}$ This is the default assumption in both the empirical power-law and binding-energy models.  An alternative version in which black holes are smaller than this in star-forming galaxies and \emph{cross over} this scaling law as they enter the green valley is discussed in Section~\ref{SubSec:BHGrowth}.}
\end{table*}

The conventions we are using for $\Mvir$ and $\Rvir$ are based on the definition of halo overdensity $\Delta_{\rm vir}$ by \citet{Bryan98} for an isothermal sphere mass profile.  The formula relating virial mass and radius is:
\begin{equation}\label{Eq:RvirMvir}
\Mvir = \frac{4\pi}{3}\Delta_{\rm vir}\rho_{\rm m}\Rvir^3,
\end{equation}
where $\rho_{\rm m}$ is the mean matter density $\Omega_{\rm m}\rho_{\rm c}$ \citep{Rodriguez16_2} and $\Delta_{\rm vir}$ is given by \citet{Bryan98}.  The quantity $\Vvir$ is defined such that $\Mvir \equiv \Vvir^2 \Rvir/G$.

\begin{figure}[htbp]
\centering
\includegraphics[scale=0.44]{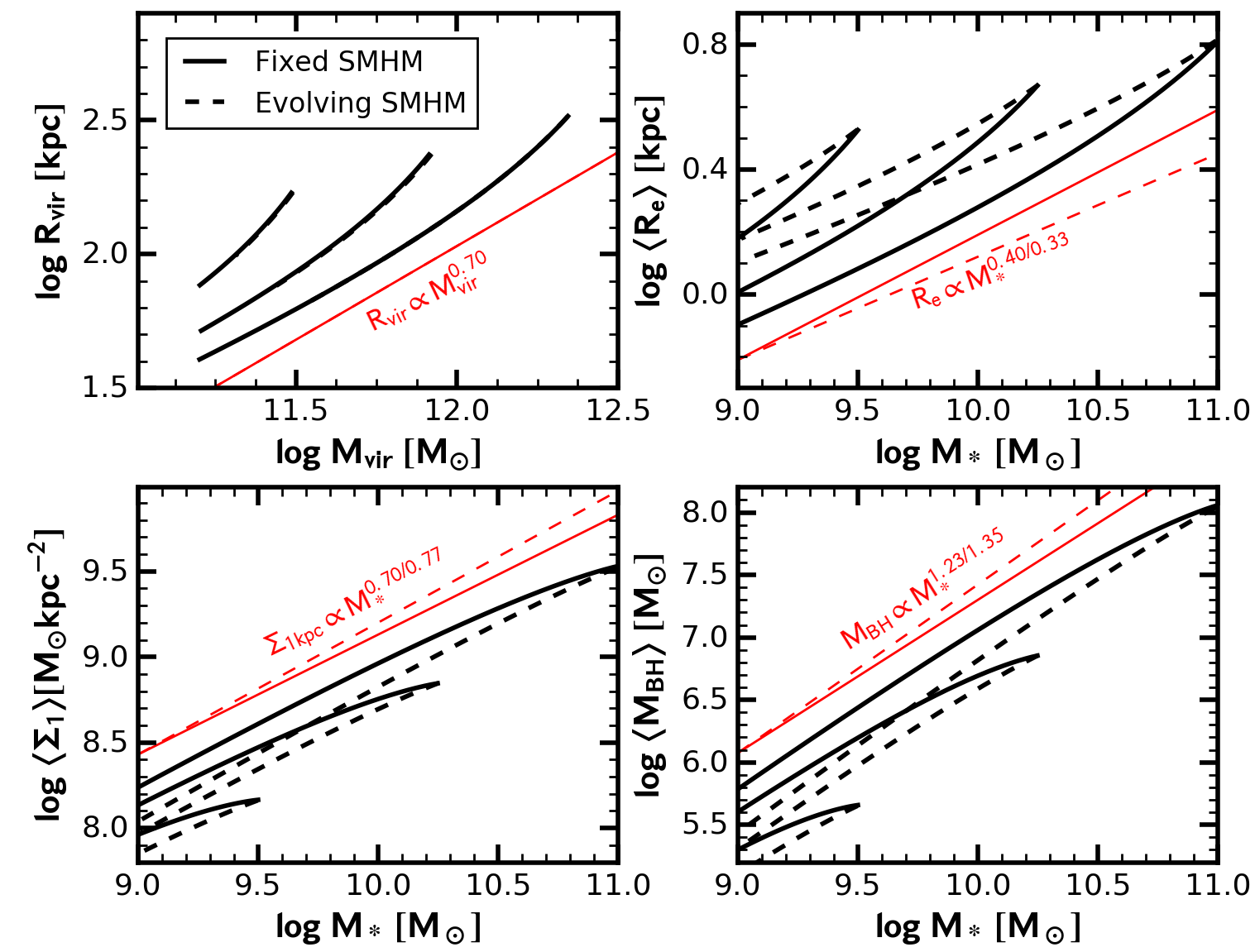}
\caption{\label{Fig:TrajectoryM} The same evolutionary tracks from Figure~\ref{Fig:Trajectoryz} for central star-forming galaxies are replotted \vs mass, with $\langle\Sig1\rangle$ and $\langle\Mbh\rangle$ added.  The black solid lines are for the empirical power-law model with fixed SMHM zeropoint.  The black dashed lines are analogous tracks for the Variant~2 model, which has an evolving zeropoint in the SMHM relation (Section~\ref{SubSec:EvolvingSMHMZP}).  The corresponding red lines show approximate power-law slopes (with arbitrary zero points) that are valid for galaxies in the range $\Mvir = 10^{11.5\text{-}12.5}~\Msol$ and $z = 3$ to 0.5.  The first exponent refers to the empirical power-law model with fixed SMHM zero point; the second exponent refers to the Variant~2 model with the evolving SMHM zero point.  These slopes are useful for rough estimates; the real curves are used when computing model predictions.}
\end{figure}

The tracks in Figure~\ref{Fig:Trajectoryz} are used to evolve central star-forming galaxies in the $\Reff\text{-}\Mstar$ plane through time.  In order to check them and to develop handy power-law approximations, Figure~\ref{Fig:TrajectoryM} replots them \vs $\Mvir$ and $\Mstar$, where the relations in Eq.~\ref{Eq:Sig1MstarRe} and Eq.~\ref{Eq:MbhSig1} are used to derive $\langle\Sig1\rangle$ and $\langle\Mbh\rangle$ from $\ReffMean$ and $\Mstar$.  The solid lines are for the standard empirical power-law model, which has a fixed zero point in the SMHM relation.  The dashed lines refer to the Variant~2 model, which puts the evolving zero point into the SMHM relation instead of in the halo quenching condition  as in Eq.~\ref{Eq:EheatEbh} and is discussed in Section~\ref{Sec:Variants} and Appendix~\ref{App:A4}.

The predicted tracks for $\ReffMean$ in Figure~\ref{Fig:TrajectoryM} show the motions of single galaxies through time, whereas plots of observed $\Reff$ \vs $\Mstar$ are snapshots of different galaxies at the same time.  Such snapshots are seen to populate a broad range in $\Reff$, have a shallower slope than the individual evolutionary tracks, and have a ridgeline zero point that evolves upwards in time \citepalias{vanderWel14}.  By definition, the peak of the ridgeline is the locus of galaxies with $\Reff = \ReffMean$.  The properties of a ridgeline can therefore be calculated from the formula for $\ReffMean$ versus time from \citetalias{Rodriguez17}.  From Eq.~\ref{Eq:MstarMvir}, Eq.~\ref{Eq:ReRvir}, and Eq.~\ref{Eq:RvirMvir}, we get:
\begin{eqnarray}\label{Eq:ReMstar}
\ReffMean & \sim & \Rvir \nonumber \\
          & \sim & \Mvir^{1/3}\Delta_{\rm vir}^{-1/3}\rho_{\rm m}^{-1/3} \nonumber \\
          & \sim & \Mstar^{1/(3s)}\Delta_{\rm vir}^{-1/3}\rho_{\rm m}^{-1/3}.
\end{eqnarray}
Thus the predicted ridgeline slope $\Delta \log \ReffMean/\Delta \log \Mstar = 1/(3\times1.75) = 0.19$, and the zero point evolves as $\Delta_{\rm vir}^{-1/3}\rho_{\rm m}^{-1/3}$, which is approximately equal to $h(z)^{-0.74}$ from $z = 0.5$ to $z = 3$, where $h(z) \equiv H(z)/H(0)$ and the \citet{Bryan98} prescription for $\Delta_{\rm vir}$ is used.  Both trends agree well with observed data as noted by \citetalias{vanderWel14}, who found an average ridgeline slope of 0.22 and a zero point evolution of $h(z)^{-0.66}$.  Further support for this radius model comes from \citetalias{vanDokkum15}, who reviewed observational and theoretical data on evolutionary tracks for star-forming galaxies in $\Reff\text{-}\Mstar$.  They found an average evolutionary track slope $\Delta \log \Reff/\Delta \log \Mstar = 0.3$, in reasonable agreement with our average track slope of 0.40 in Figure~\ref{Fig:TrajectoryM}.

A final consistency check can be made by comparing to slopes in $\Sig1\text{-}\Mstar$ using the $p,q$ mapping.  A given log slope $m$ in $\Reff\text{-}\Mstar$ corresponds to slope $q-pm$  in $\Sig1\text{-}\Mstar$.  For $p = 1$, $q = 1.1$, the predicted ridgeline slope for star-forming galaxies is then $1.1 - 1.0\times0.19 = 0.91$.  This compares well to the fitted ridgeline slope in \citetalias{Barro17}, which averages 0.88 at all redshifts.  The corresponding evolutionary track slope is $1.1 - 1.0\times0.40 = 0.70$, just slightly steeper than the quenched ridgeline slope of 0.66.  An additional check on the evolutionary track slope can be made by identifying a median star-forming galaxy at $10^{8.9}~\Msol$ at $z = 2.5$ in \citetalias{Barro17}, computing a stellar mass increase to $10^{10.2}~\Msol$ at $z = 0.75$ \citepalias{Rodriguez17}, and placing it in the middle of the star-forming cloud in $\Sig1\text{-}\Mstar$ of \citetalias{Barro17} at the new redshift and mass.  The resulting changes are 1.30~dex in $\Mstar$ and 0.90~dex in $\Sig1$, for a net evolutionary track slope of 0.69, in excellent agreement with the predicted slope (inferred from $\Reff\text{-}\Mstar$) of 0.70.

Finally, star-forming galaxies are assumed to evolve along the power-law SMHM relation (Eq.~\ref{Eq:MstarMvir}) until they enter the green valley.

In summary, a simple halo-based model for $\Mstar$ and $\Reff$ appears to match fairly well the observed average radii of star-forming galaxies, their evolution in time after $z = 3$, and estimates of how individual galaxies move in $\Reff\text{-}\Mstar$ and $\Sig1\text{-}\Mstar$.

\subsection{Starting Conditions for Each Galaxy}\label{SubSec:GalIniCon}

In order to ensure that galaxies cross the quenching boundary in the right place, star-forming galaxies must be properly positioned in $\Sig1\text{-}\Mstar$ and $\Reff\text{-}\Mstar$ to start with. Power-law evolutionary tracks are now in place, but each model galaxy needs to be provided with appropriate starting conditions for entry onto its track. We start by choosing a galaxy stellar mass today and use the observed halo-mass-to-stellar-mass (HMSM) relation at $z = 0$ from \citetalias{Rodriguez17} to calculate a corresponding halo mass.\footnote{The HMSM relation is the inverse of the SMHM relation, being the mean value of $\Mvir$ observed at fixed $\Mstar$ today.  It is taken from \citetalias{Rodriguez17} and is not a power law.}  This halo mass is then propagated backwards in time using the \citetalias{Rodriguez17} halo mass evolution formula together with the power-law SMHM relation, $\Mstar = k_{\rm m} \Mvir^s$, until $\Mstar = 10^9~\Msol$, which sets the starting redshift, $z_{\rm s}$.  This halo mass, stellar mass, and virial radius are then evolved forward using the halo evolution model of \citetalias{Rodriguez17} and the same SMHM relation.  For the empirical power-law model, which we are describing here, the zero point of the SMHM relation is kept fixed, and $k_{\rm m}$ is a constant.

It remains to scatter the starting effective radii.  A mean starting $\langle R_{\rm e,s} \rangle$ at $z_{\rm s}$ for galaxies with $\Mstar = 10^9~\Msol$ is computed from the $\Rvir$ formula in Eq.~\ref{Eq:ReRvir}, and galaxies are scattered above and below this value with a Gaussian distribution having $\sigma \log (\Reff/\ReffMean) = 0.20$~dex, which is the observed rms scatter in the radii of star-forming galaxies (\citetalias{vanderWel14}; \citealp{Somerville18}).  An important simplification, noted above, is that the star-formation rates of ridgeline galaxies on the star-forming main sequence are observed to be independent of $\Reff$ at all redshifts (\citealp{Omand14,Whitaker17}; \citetalias{Fang18}; \citealp{Lin19}) so that all galaxies in a given $\Mstar$ cohort stay together in mass regardless of their starting $\Reff$.

\subsection{Evolutionary Tracks in Relation to the Quenching Boundary}\label{SubSec:EvoTracksQuenchBound}

When the above evolutionary tracks were computed, a surprising fact emerged.  We had anticipated that galaxies would evolve through the $\Reff\text{-}\Mstar$ and $\Sig1\text{-}\Mstar$ planes toward the sloping boundary, where specific star formation rates fall -0.45 dex below the star-forming main sequence value.  Galaxies would move toward this boundary because their evolutionary tracks were angled relative to it, and eventually the two would intersect.  However, we saw above that the observed quenching boundary in $\Sig1\text{-}\Mstar$ has slope 0.66 \citepalias{Barro17} whereas the estimated evolutionary track has slope 0.70, i.e., nearly the same. Likewise, the quenching boundary in $\Reff\text{-}\Mstar$ has slope 0.44 whereas the evolutionary track has slope 0.40, again nearly the same.  In other words, the evolutionary tracks of star-forming galaxies make very acute ``angles of attack'' relative to the quenching boundaries. The concern then arises that many galaxies would never quench, i.e., that galaxies initially located far from the boundary would never cross it.

The solution to this problem turned out to be the \emph{moving  boundary}. A downward motion of $-0.3$~dex in the $\Sig1\text{-}\Mstar$ quenched ridgeline zero point was measured by \citetalias{Barro17} from $z = 2.5$ to 0.5, and the corresponding upward motion in $\Reff$ is $+0.3$~dex.  Our quenching boundaries are tied to these ridgelines.  This motion, which seems small at first sight, is actually the \emph{key ingredient in the empirical power-law model that causes galaxies to quench}. Individual galaxies evolve roughly parallel to the boundary, but meanwhile the boundary sweeps over them, from above in $\Sig1$ and from below in $\Reff$.

To make further progress, this motion needs to be inserted into the expression for the quenching boundary.  It was shown above in Eq.~\ref{Eq:MbhMvir} that equating the total energy emitted by the BH to a power of halo mass matches the \emph{slope} of the quenching boundary if $t = 2.0$.  We now take that power-law relation and add a moving zero point:
\begin{equation}\label{Eq:EheatEbh}
  \Ebh = \Eheat
\end{equation}
or
\begin{equation}\label{Eq:EbhEhalo}
  \eta \frac{\Mbh}{10^8~\Msol} c^2 = k_{\rm e}(z) \left(\frac{\Mvir}{10^{12}~\Msol}\right)^t,
\end{equation}
where $k_{\rm e}(z)$ varies with time and we have taken $\eta = 0.01$ \citep[cf.][]{Oppenheimer19}.  The left side is the effective emitted BH energy, $\Ebh$, and the right side is the halo energy quota for quenching, $\Eheat$.  From $\Mbh \sim \Sig1^v$ (Eq.~\ref{Eq:Sig1Mbh}), we find that the change $\Delta \log k_{\rm e}(z)$ with time is equal to $v$ times the change in the zero point of the quiescent $\Sig1$ ridgeline in \citetalias{Barro17} at fixed $\Mstar$. A fit of this zero point \vs redshift, called $B^Q$ in \citetalias{Barro17}, is shown in Figure~\ref{Fig:A2} in Appendix~\ref{App:A2}.  It yields $\log B^Q = 9.44 + 0.66 \log (H(z)/H(0))$, and thus $k_{\rm e}(z) = k_{\rm e}(0) \times h(z)^{0.66v}$.

The constant $k_{\rm e}(0)$ at $z = 0$ is used as a free parameter to match the absolute location of the quenching boundary at $z = 0$, assuming $\eta = 0.01$.  Rather than match to the quenched ridgeline itself, however, we have offset entry into the green valley by 0.2~dex below the quenched ridgeline in $\Sig1\text{-}\Mstar$ based on the data on SDSS green valley galaxies of \citet{Luo19}, and we apply that same shift at all redshifts.  The value of $k_{\rm e}(0)$ is then found to be $10^{17.66}$ km$^2$ sec$^{-2}$.  Finally, the moving boundary in  $\Sig1\text{-}\Mstar$ is mapped into $\Reff\text{-}\Mstar$ using the $p,q$ mapping.  This prescription for $k_{\rm e}(z)$ (and $t$) perfectly matches the red lines in Figure~\ref{Fig:Mstar_Sig1_SMA1}.

We pause here to make two points.  The first is that the moving zero point in Eq.~\ref{Eq:EbhEhalo} is needed to match the empirically determined quenching boundaries (red lines in Figure~\ref{Fig:Mstar_Sig1_SMA1}), which move with time.  However, a more physical model for quenching based on real halo properties is coming up in Section~\ref{Sec:ThermalModel}, which also naturally produces a moving boundary.  The moving boundary thus has two independent justifications: the need to match observational data on the one hand and the changing physical properties of gaseous halos on the other.   The second point is that the presentation and labeling of quantities in Eq.~\ref{Eq:EheatEbh} explicitly evoke the notion of a BH-halo energy balance, with BH energy on the left side and a halo quenching energy quota on the right side.  We think that this presentation is helpful because it points the way to the more physically-based model in Section~\ref{Sec:ThermalModel}.  On the other hand, one could omit the labels and view Eq.~\ref{Eq:EbhEhalo} simply as an empirical relation between $\Mbh$ and a power of $\Mvir$ (plus a moving zero point) and save the physical interpretation until later.  That is equally valid and, depending on one's preferences, might be regarded as a ``cleaner'' presentation.

\begin{figure}[htbp]
\centering
\includegraphics[scale=0.6]{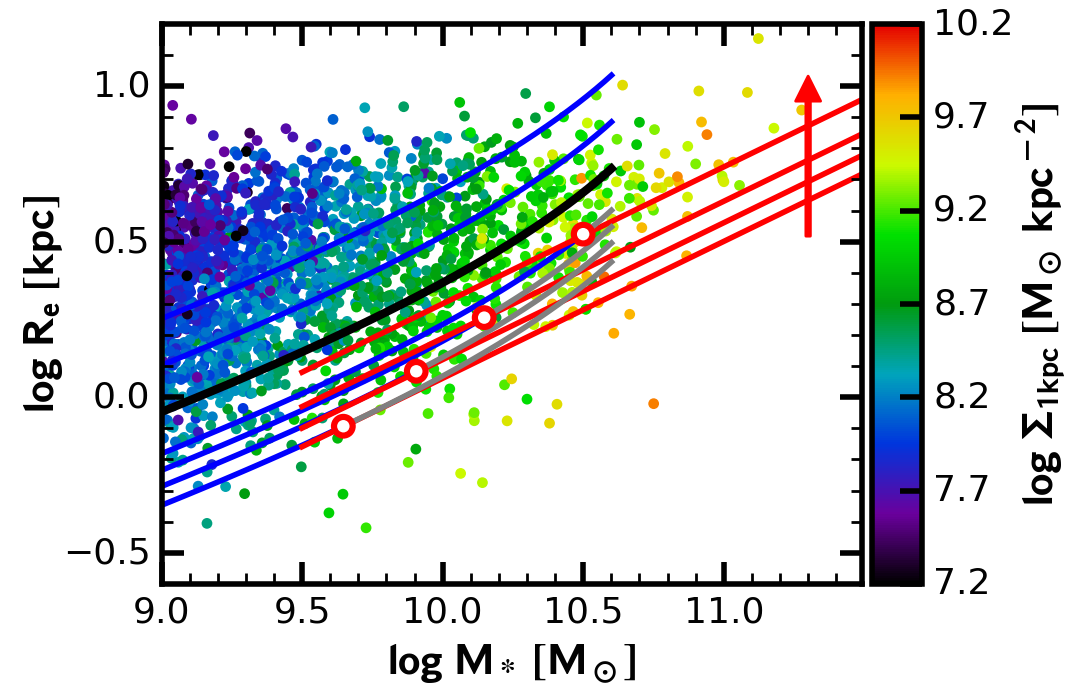}
\caption{\label{Fig:TrajectoryRM} Sample evolutionary paths for a selection of galaxies of a given stellar mass in $\Reff$ \vs $\Mstar$ in the empirical power-law model.  The black line is for an average galaxy with radius $\Reff = \ReffMean$ and $\Mstar = 10^{10.6}~\Msol$ today (if it does not quench first).  The blue tracks above and below that line show galaxies of the same starting mass but scattered up and down in $\Reff$ by $\pm0.75~\sigma_{\rm Re}$ and $\pm 1.5 \sigma_{\rm Re}$ (two extra values are added on the low side).  Tracks with lower values of $\Reff$ have higher BH masses.  The observed distribution of galaxies is shown for comparison at $z = 0.5\text{-}1.0$, colored by $\Sig1$.  The diagonal red lines show the location of the quenching boundary, which moves upward with time.  Entry to the green valley occurs when a galaxy crosses the instantaneous boundary; such locations are shown by the red/white dots, which correspond to $z = 1.76$, 1.36, 0.90, and 0.43, from the bottom.  The evolution of galaxies in $\Reff\text{-}\Mstar$ after entering the green valley is not covered by the model and is not shown.}
\end{figure}

Evolutionary paths and quenching points of typical Milky Way-mass galaxies in $\Reff$ \vs $\Mstar$ are illustrated in Figure~\ref{Fig:TrajectoryRM}.  These tracks use the actual trajectories of halos, not the  approximate power-laws.  The black line is for an average galaxy with radius $\Reff = \ReffMean$ and $\Mstar = 10^{10.6}~\Msol$ today.  The blue tracks above and below this line are galaxies of the same starting mass but scattered up and down in $\Reff$ by $\pm0.75\sigma_{\rm Re}$ and $\pm1.5\sigma_{\rm Re}$, where $\sigma_{\rm Re} = 0.2$~dex (two extra values are added on the low side).  Tracks with lower values of $\Reff$ have higher BH masses and vice versa.  The observed distribution of galaxies at $z = 0.5\text{-}1.0$ is shown for comparison in the background.  The diagonal red lines show the location of the quenching boundary at various times.  Entry into the green valley occurs when a galaxy crosses the instantaneous boundary (red and white dots).  It is seen that the tracks are inherently quite parallel to the boundary, and it is really the upward motion of the boundary \emph{zero point} that causes galaxies to cross it.

To summarize, a galaxy is identified as quenching perceptibly (i.e., entering the green valley) when the quenching condition in Eq.~\ref{Eq:EbhEhalo} is satisfied, and the redshift when this happens and its location are marked on the tracks. If the model is working, the locus of perceptibly quenching galaxies at a given redshift should match the adopted quenching boundary in $\Sig1\text{-}\Mstar$ at that redshift, which is taken to be 0.2~dex below the observed quenched ridgeline.

All the power-law expressions used in the model and their slopes and zero points are collected and explained for reference in Appendix~\ref{App:A3}.

\subsection{Results for the Empirical Power-Law Model}\label{SubSec:EvolModelResults}

\begin{figure*}[htbp]
\centering
\includegraphics[scale=0.66]{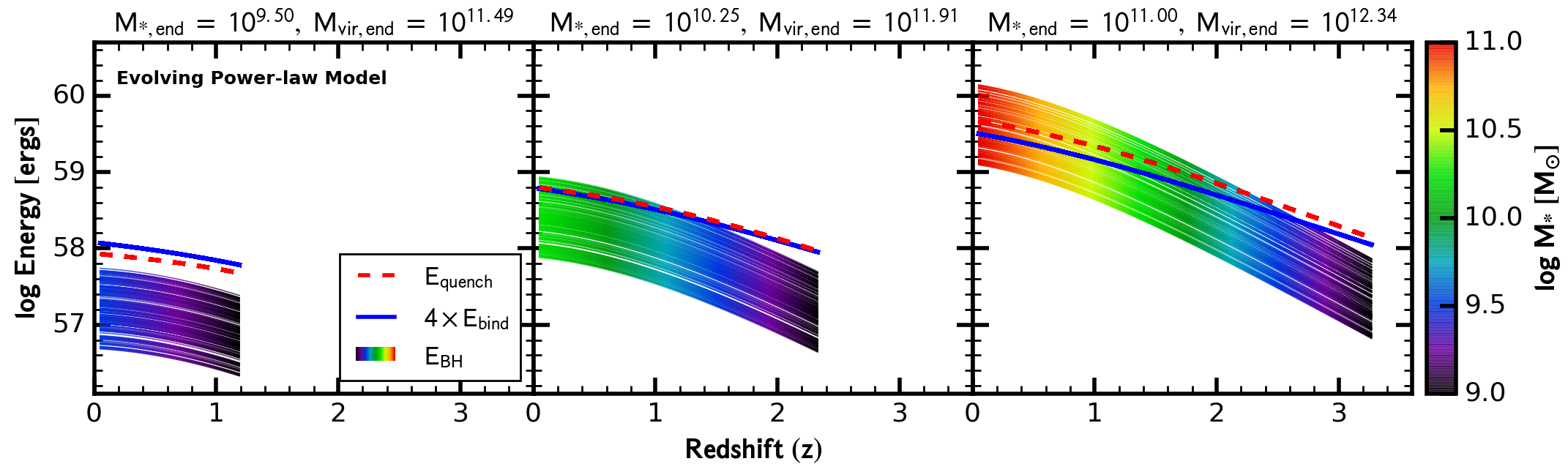}
\caption{\label{Fig:BHEnergy} The BH energy output ($\Ebh$, rainbow bands) and the halo quenching-energy quota ($\Eheat$, red dashed lines) are shown for star-forming galaxies \vs redshift in the evolving power-law model.  Three representative $z = 0$ stellar-mass/halo-mass combinations are shown.  $\Ebh$ is the total effective energy, $\eta\Mbh c^2$, emitted by the black hole to that epoch, and $\Eheat$ is the total energy quota that needs to be absorbed by the halo as galaxies enter the green valley (Eq.~\ref{Eq:EbhEhalo}).  Galaxies below the red dashed lines have large radii, small BHs, and are still star-forming, while galaxies above the red dashed lines have small radii, large BHs, and have entered the green valley. (Note: since galaxies above the lines are in the green valley, their real values of $\Mbh$ and $\Ebh$ would be different than shown.  Star-forming galaxies do not actually populate these locations; see text.)  The width of the rainbow bands corresponds to scattering the initial values of $R_{\rm e,s}$ by $ \pm 1.5\sigma_{\rm Re} = 0.30$~dex, which scatters $\Mbh$ by $\pm 0.52$~dex from Eq.~\ref{Eq:ReMbh}.   Galaxies with smaller $\Reff$'s (and bigger BHs) at a given $\Mstar$ quench sooner.  Notice the acute ``angle of attack'' between $\Ebh$ and $\Eheat$ and the tendency for a larger fraction of massive galaxies to quench.  These trends fall out naturally from setting $t = 2.0$ and the evolving zero point of $\Eheat$ to match the observed quenching boundary.  The blue solid lines show the halo quenching-energy quota, 4$\times$\Eth, from the binding-energy model in Section~\ref{Sec:ThermalModel}.  It is seen that $\Eheat$ follows $4 \times \Eth$ very closely, showing that the empirical power-law energy quota we have used to quench a halo (Eq.~\ref{Eq:EbhEhalo}) is closely equal to the actual gas binding energies of halos. The red dashed lines and blue solid lines here correspond to the red and blue quenching boundaries in Figure~\ref{Fig:Mstar_Sig1_SMA1}.}
\end{figure*}

\begin{figure*}[hbtp]
\centering
\includegraphics[scale=0.58]{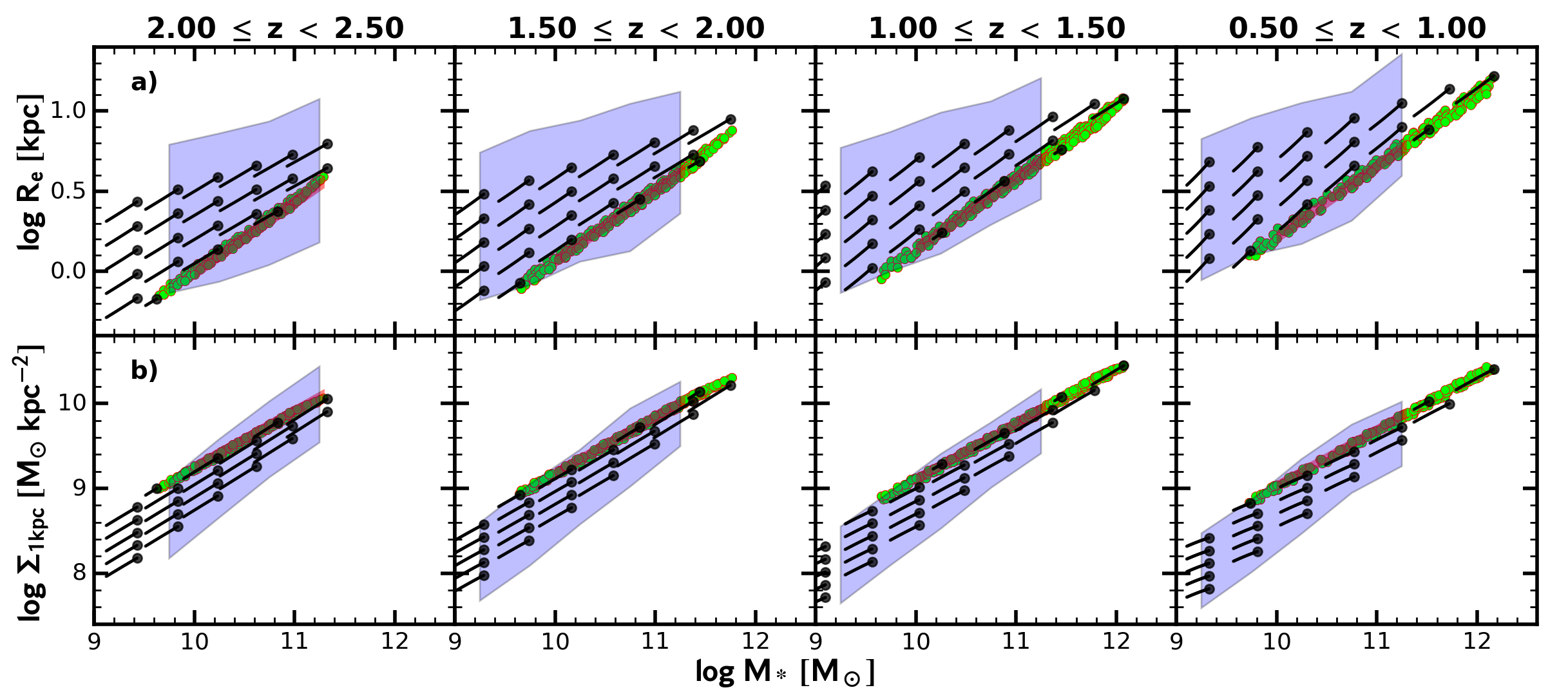}
\caption{\label{Fig:SimuCartoon}  Model star-forming galaxies from the empirical power-law model in $\Reff$ \vs $\Mstar$ (panel a) and in $\Sig1$ \vs $\Mstar$ (panel b).  Seven masses and four redshift ranges are shown.  Each dot is a galaxy, and the short line attached to it is its recent evolutionary track.  The pale green circles show the loci of galaxies that are entering the quenching boundary at that redshift, and the red lines underneath them are the observed quenching boundaries from Figure~\ref{Fig:Mstar_Sig1_SMA1} (red lines).  A perfect match has been attained by choosing the right values of $s$, $t$, and $v$ and by allowing the zero point in Eq.~\ref{Eq:EbhEhalo} to evolve appropriately.  The model galaxy points have been scattered by a total of $\pm 1.5\sigma_{\rm Re}$ in $R_{\rm e,s}$, and the shaded regions outline the corresponding $\pm 1.5\sigma_{\rm Re}$ areas that are covered by star-forming galaxies in \citetalias{vanderWel14}.  The similarity of the two distributions indicates that our method for choosing starting values of $R_{\rm e,s}$ (Section~\ref{SubSec:GalIniCon}) has achieved a reasonable match between the regions populated by the model star-forming galaxies \vs observed galaxies.  Notice that the individual evolutionary tracks are almost parallel to the quenching boundary and that galaxies cross the boundary because the boundary moves down in $\Sig1$ (or up in $\Reff$) to intercept them.  Notice also the ``tongue'' of star-forming galaxies at late times that extends to higher masses than observed.  This is discussed in Variant~1 in Section \ref{Sec:Variants}.}
\end{figure*}

Results for the empirical  power-law model are shown in Figures~\ref{Fig:BHEnergy} and \ref{Fig:SimuCartoon}.  Figure~\ref{Fig:BHEnergy} plots the BH heating energy, $\Ebh$, and the halo heating-energy quota, $\Eheat$, \vs redshift for three different halo-galaxy combinations.  The curves are labeled with ending stellar and halo mass today.  The red dashed lines show $\Eheat$ from  Eq.~\ref{Eq:EbhEhalo}, which depends only on halo mass and is therefore shown as a single line for each $\Mstar$.  The rainbow-colored bands show $\Ebh$; their spread reflects the range of BH masses coming from a $\pm 1.5\sigma$ scatter in $\Reff$ (smaller $\Reff$ and larger BHs are at the upper edges).  A given galaxy quenches when its $\Ebh$ line crosses above $\Eheat$.  We stress that $\Eheat$ is not necessarily thermal energy but rather expresses some sort of generic energy (or momentum) transfer from the BH to the halo gas that stops cooling.

We note that the $\Ebh$ trajectories have been plotted ignoring the fact that galaxies perceptibly quench when they cross over the $\Eheat$ lines and their subsequent $\Mstar$ and $\Mbh$ evolution switches to the green valley vectors in Figure~\ref{Fig:Cartoon}.  Star-forming galaxies therefore do not actually exist in these locations, but we have left them in place in order to better convey the fraction of galaxies at each mass and redshift that have quenched.

Several features are apparent.  First, the $\Eheat$ curves bend over with time for a given stellar mass despite the fact that the halo mass is always growing.  In the empirical  power-law model, this bend comes from the declining value of $k_{\rm e}(z)$ in the $\Eheat$ zero point (Eq.~\ref{Eq:EbhEhalo}).  Without this bend, $\Eheat$ would continue to rise with $\Mvir$, and galaxies would quench very late or not at all. Note that small-$\Reff$ galaxies with bigger BHs quench sooner (upper envelopes).  Finally one sees that a critical parameter is the ``angle of attack'' that a given $\Ebh$ trajectory makes with $\Eheat$. We mentioned this angle before in the $\Sig1\text{-}\Mstar$ and $\Reff\text{-}\Mstar$ diagrams; the concept here is similar -- if the angle of attack is large, all galaxies of a given starting $\Mstar$ will quench in a short space of time, and thus at nearly the same mass.  This means that the length of the quenching locus (and the width of the green valley) for that-mass galaxy will be small.  On the other hand, if the angle is acute, quenching takes a long time from smallest to largest $\Reff$, and the extent of the quenching locus in $\Mstar$ will be large. Matching the observed large width of the green valley requires acute angles of attack, as discussed further below.\footnote{The need for acute angles of attack to match the wide width of the green valley is exaggerated here by the fact that all of the assumed power-law relations in the model are scatter-free.  Scatter around the scaling relations would help to broaden the green valley naturally and should be included in future versions of the model.}

Figure~\ref{Fig:SimuCartoon} shows corresponding snapshots of galaxies in the $\Reff\text{-}\Mstar$ and $\Sig1\text{-}\Mstar$ planes at four redshifts.  Each dot is a model star-forming galaxy of a given mass, and attached to each point is a short line showing its recent evolutionary track.  Five model galaxies are plotted in each mass cohort, corresponding to galaxies scattered in radius by $\pm0.75\sigma_{\rm Re}$ and $\pm 1.5\sigma_{\rm Re}$. They are superimposed on a shaded region denoting the observed $\pm 1.5 \sigma_{\rm Re}$ range of observed star-forming galaxies at that redshift from \citetalias{vanderWel14}.  The model galaxy points populate the observed range of galaxies fairly well, which validates our scheme for starting galaxies at $10^9~\Msol$ with $\Reff = 0.02 \times \Rvir$ and growing $\Reff$ in proportion to $\Rvir$ thereafter.

Galaxies that are entering the green valley at each redshift are shown by the pale green points, which are populated more densely in starting mass in order to show their locus clearly.  The straight red lines underneath them are the quenched ridgelines of \citetalias{Barro17} displaced downward by 0.2~dex in $\Sig1$ to represent the adopted quenching boundary (entry into the green valley). They are the same as the red lines shown in Figure~\ref{Fig:Mstar_Sig1_SMA1}.  The good match between the red lines and the green points verifies that the zero point and redshift evolution of the quenching expression in Eq.~\ref{Eq:EbhEhalo} have been set correctly. The predicted slopes also match, indicating that the adopted values of $t/vs$, $p$, and $q$ are also proper.  They are taken from Section~\ref{Sec:PowerLawModel} and are $t = 2.03$, $v = 1.76$, $s = 1.75$, $p = 1.0$, and $q = 1.1$.

\section{Interpretation of $\Eheat$ in Terms of Halo-Gas Binding Energy}\label{Sec:ThermalModel}

Galaxies enter the green valley in the empirical power-law model when the total effective energy radiated by their accreting BHs equals the halo heating-energy quota $\Eheat = k_{\rm e}(z) \left(\frac{\Mvir}{10^{12}~\Msol}\right)^t$ in Eq.~\ref{Eq:EbhEhalo}.\footnote{In adopting this criterion, we have neglected both radiative losses in the AGN feedback and possible energy transfer to the halo by supernovae \citep[e.g.][]{Dekel86} and by gravitational heating \citep[e.g.,][]{Dekel08,Johansson09}.  Recent hydro simulations give some sense of the importance of the first two factors.   In both EAGLE \citep{Bower17} and TNG50 \citep{Nelson19}, successful AGN winds tend to conserve energy, which justifies the use of cumulative BH energy in Eqs.~\ref{Eq:EbhEth} and \ref{Eq:EbhEbind}.  In contrast, supernovae winds are prone to large radiative losses \citep[e.g.][]{Somerville15}, and adding them to the current framework would require a major energy-loss correction factor. In TNG50, the relative roles of supernovae \vs BHs are hard to separate near \Mstar = $10^{10.5}~\Msol$, but AGN winds clearly dominate above this \citep{Nelson19}.}  The formulation of $\Eheat$ as a power law of $\Mvir$ was adopted to match the approximate power-law nature of the quenching boundary.  However, there was little physical motivation, and it does not quite work -- an \emph{ad hoc} evolving zero-point correction had to be added.

An additional obvious energy to be considered is the binding energy of the halo gas.  It is already known that $\Ebh$ is of the same order of magnitude as the halo-gas binding energy and thus that actual BHs emit enough power to influence their halos \citep{Silk98,Ostriker05,Bower17,Oppenheimer18, Davies19a,Oppenheimer19,Nelson19}.  If in addition we can show that halo-gas binding energy varies similarly to $\Eheat$ as a function of $\Mvir$ and time, we will have additional evidence that BH energy beating against halo-gas binding energy is the gatekeeper for quenching.

Our adopted expression for halo-gas binding energy is:
 \begin{equation}\label{Eq:GasBindingEnergy}
 \Eth = \frac{1}{2} f_{\rm bary} \Mvir V_{\rm vir}^2,
 \end{equation}
 where $f_{\rm bary}$ is the fraction of $\Mvir$ in the form of hot gas, set equal to 0.1 throughout. Inserting this into Eq.~\ref{Eq:EbhEhalo} we now have a new quenching condition:
 \begin{equation}\label{Eq:EbhEth}
  \Ebh =  f_{\rm B} \times \Eth
\end{equation}
or
\begin{equation}\label{Eq:EbhEbind}
  \eta \frac{\Mbh}{10^8~\Msol} c^2 = f_{\rm B} \times \frac{1}{2} f_{\rm bary} \Mvir V_{\rm vir}^2,
\end{equation}
where $f_{\rm B}$ is the multiple of halo-gas binding energy that needs to be supplied by the black hole at the quenching boundary.

The quantity $ f_{\rm B} \times \Eth$ is plotted as the solid blue lines for the three typical halo masses in Figure~\ref{Fig:BHEnergy}, and the equivalent quenching boundaries for this new model are the blue lines in Figure~\ref{Fig:Mstar_Sig1_SMA1}.  Comparing these with the observed quenching boundaries (red lines), we see fairly good agreement in slope, average zero point, and zero-point evolution.  The agreement in average zero point has been optimized by choosing $f_{\rm B} = 4$, which is the best value given $\eta = 0.01$.

In contrast to the average zero point, the agreements in slope and zero-point evolution are unforced, as can be seen by considering the ratio of $\Eheat$ to $\Eth$.  At fixed $z$, $\Eheat$ varies as $\Mvir^2$, while $\Eth$ varies as $\Mvir\Vvir^2 \sim \Mvir^{1.67}$, which is similar but slightly shallower.  This slight difference accounts for the shallower slope of the blue lines in Figure~\ref{Fig:Mstar_Sig1_SMA1} and also explains why the blue line lies above the red dashed line for small galaxies but below it for massive galaxies in Figure~\ref{Fig:BHEnergy}.\footnote{We note that the slope of the blue lines could be increased in Figure~\ref{Fig:Mstar_Sig1_SMA1} to match that of the red lines if the slope of the SMHM relation, $s$, were reduced from 1.75 to 1.44.  This change is larger than the quoted error of 0.17 but is perhaps not outside the realm of possibility.}

For a given galaxy over time, it may also be shown using the above equations plus $\Mvir \sim \Vvir^3/h(z)^{1.11}$ that $\Eth/\Eheat$ evolves as $\Mvir^{-0.33}h(z)^{0.74-0.66v} \sim \Mvir^{-0.33}h(z)^{-0.42}$.  But $\Mvir$\ grows approximately as $h(z)^{-1.50}$ over this redshift range, and so $\Eth/\Eheat \sim h(z)^{0.08}$.  In other words, $\Mvir$ is rising while $h(z)$ is falling, and the two almost cancel along a given evolutionary track.  This is why the red dashed lines and the blue solid lines are parallel in Figure~\ref{Fig:BHEnergy}.  An additional insight comes from noting that $\Eheat$ varies as $k_{\rm e}(z)\Mvir^{2.0}$ while $\Eth$ varies as $\Mvir^{2.0}\Rvir^{-1}$.  Thus, the evolving term $k_{\rm e}(z)$ in the empirical power-law model is needed to mimic the evolution of $\Rvir^{-1}$ over time.  The question of the zero-point evolution of the \emph{boundary} over time is different from the evolution of an individual galaxy and is considered in Section~\ref{SubSec:Successes}.

This is a good place to make a point about the uniformity of quenching.  If $f_{\rm B}$ varies from halo to halo, say, with rms scatter $\delta_{\rm {fB}}$ dex at a given epoch, this will induce the same amount of scatter in $\Mbh$ and a scatter of $\delta_{\rm {fB}}/v$ in $\Sig1$.  The $\Sig1$ ridgeline has rms scatter equal to 0.16~dex today \citepalias{Fang13}, which implies that $\delta_{\rm {fB}} = 0.28$ dex.  However, this is an upper limit since age stratification has been detected \citep{Tacchella17,Luo19}, so the real scatter in $\delta_{\rm {fB}}$ must be less.  The point is that the halo binding-energy criterion in Eq.~\ref{Eq:EbhEbind}, represented by $f_{\rm B}$, must be \emph{highly uniform from halo to halo} in order to create the tight ridgelines in $\Sig1\text{-}\Mstar$ that are seen.

To summarize, in agreement with previous work we have shown that the energy available in the course of BH growth is the same order of magnitude or greater than the binding energy of the halo gas, and thus that BHs potentially produce enough energy to modify the state of their halo gas and retard gas infall.  In addition, we have shown that halo-gas binding energy has the right general behavior \vs $\Mvir$ and time to match the observed sloping quenching boundaries in plots of galaxy structural parameters, and thus to explain important structural scaling laws of quenching galaxies.  The notion that halo-gas binding energy controls BH growth, and through that, quenching, was modeled by \citet{Booth10,Booth11} and has been seen and analyzed in the EAGLE simulations \citep{Davies19a,Oppenheimer19} and in IllustrisTNG (\citealp{Terrazas19}).  Comparisons to these and other models are discussed in Section~\ref{Sec:Discussion}.

\section{Variants of the Empirical Power-Law Model}\label{Sec:Variants}

This section describes four variations on the empirical power-law model.  Summaries are presented here, and details are given in Appendix~\ref{App:A4}.  Each variant fails in some respect, but the reasons are instructive, yielding important insights into the sensitivities and basic physics of the problem. Readers interested in final conclusions can skip directly to the Discussion in Section~\ref{Sec:Discussion}.

\subsection{Variant~1: A Broken Power-Law for the SMHM Relation} \label{SubSec:BrokenPLModel}

A feature of the empirical power-law model that does not match observed data is the long ``tongue'' of star-forming galaxies extending up to $\Mstar > 10^{12}~\Msol$ at late redshifts in Figure~\ref{Fig:SimuCartoon}.  This tongue is due to the fact that we have used the curved HMSM relation from \citetalias{Rodriguez17} at $z = 0$ to find the starting redshifts, $z_{\rm s}$, but have calculated stellar masses going forward using the power-law SMHM relation of Eq.~\ref{Eq:MstarMvir}, $\Mstar = k_{\rm m} \Mvir^s$.  Since the latter keeps climbing with $\Mvir$, massive galaxies that start near the knee of the real SMHM relation today can wind up with higher stellar mass than their assumed initial mass.  Galaxies with $\Mstar > 10^{10.7}~\Msol$ and large $\Reff$ (and small BHs) are the objects that populate the massive tongue.

To attempt to cure this problem, the Variant~1 broken-power-law model uses the same power-law SMHM relation for star-forming galaxies up to $10^{10.7}~\Msol$ but switches to a flatter power-law for SMHM above that. Results are shown in Figures~\ref{Fig:HMSM} and \ref{MEnergy_V1} in Appendix~\ref{SubSec:AppBrokenPLModel}.  With the new relation, massive galaxies now largely stop growing but are permanently stuck in a star-forming state because their BHs remain small. An excess of massive star-forming galaxies should therefore accumulate over time, which is not seen.

However, there may be another solution, based on recent information about the shape of the quenching boundary from a large sample of SDSS galaxies \citep{Luo19}.  The quenched ridgeline in $\Sig1\text{-}\Mstar$ in this new sample is seen to bend over to lower $\Sig1$ at high stellar mass, which would create a matching bend in the quenching boundary.  Our assumed linear power law lies about 0.2~dex above the new SDSS ridgeline at $10^{11.4}~\Msol$, and it is seen from Figure~\ref{Fig:SimuCartoon} that bending the ridgeline down by this amount would remove the problematic massive galaxies.

The lesson of Variant~1 is that the fraction of quenched \vs star-forming galaxies at high mass is very sensitive to the exact shape of the quenching boundary, and a deep physical understanding of the boundary will be needed to predict the fraction  accurately.  A corollary is that the shape and amplitude of the quenched mass function of quenched galaxies at high values of $\Mstar$ is difficult to predict.

\subsection{Variant~2: Move the Evolving Zero Point from $\Eheat$ to the SMHM Relation}\label{SubSec:EvolvingSMHMZP}

We have noted that the quenching boundary evolves in $\Sig1\text{-}\Mstar$, and the empirical  power-law model matches this by evolving the zero point, $k_{\rm e}(z)$, of $\Eheat$ in Eq.~\ref{Eq:EbhEhalo}.  An alternative approach is to move the evolving zero point to the SMHM relation in Eq.~\ref{Eq:MstarMvir} and remove it from $\Eheat$.  The expression that we have used for the SMHM relation in this case is derived in Appendix~\ref{SubSec:AppEvolvingSMHMZP}.  It makes the zero point of $\Mstar$ \vs $\Mvir$ larger by about 0.5~dex at $z = 0$ compared to $z = 2$.  This grows BHs more quickly, which allows them to overcome halos without resort to $k_{\rm e}(z)$.  Changes in the SMHM relation back in time that bear on this are described in Appendix~\ref{SubSec:AppEvolvingSMHMZP}.  There is some evidence for a change in the right direction, but the amount is smaller than required.

This model achieves the same match to the quenching boundaries as the empirical power-law model, but the growth of galaxy properties \vs galaxy mass is different, as shown by the dashed lines in Figure~\ref{Fig:TrajectoryM}.  BHs grow more rapidly at late times, so quenching starts too late, i.e., not until around $z = 1$ (see Figure~\ref{MEnergy_V2} in Appendix~\ref{SubSec:AppEvolvingSMHMZP}). Trying to move the onset of quenching earlier by increasing $\eta$ in Eq.~\ref{Eq:EbhEhalo} produces over-quenching at late times.  Also, the more rapid growth in $\Eheat$ with time no longer parallels the slower growth in halo-gas energy content, $\Eth$, as shown by the differing red and blue lines in Figure~\ref{MEnergy_V2} in Appendix~\ref{SubSec:AppEvolvingSMHMZP}, and so the appealing physical argument equating $\Eheat$ with halo-gas binding energy is therefore also broken.

The important lesson from Variant~2 is that changes in the zero point of the real SMHM relation back in time would require compensating changes in $\Eheat$, which could threaten its interpretation as halo-gas binding energy and also disturb the agreement on time of quenching.  Accurate measurements of the SMHM relation back in time are therefore needed to finalize the model.

\subsection{Variants~3 and 4: Mass Rates or Energy Totals?}\label{SubSec:RatesorTotals}

A major difference between the empirical  power-law model and other halo-heating models is that the latter often use heating and cooling \emph{rates} as opposed to cumulative energy totals as in Eq.~\ref{Eq:EbhEhalo}.  Use of rates is especially common in semi-analytic models \citep[e.g.,][]{Bower06,Croton06,Somerville08,Henriques15}.  Variants~3 and 4 explore whether Eq.~\ref{Eq:EbhEhalo} can be recast in terms of rates. In both variants, the halo energy quota, $\Eheat$, on the right side of Eq.~\ref{Eq:EbhEhalo} is replaced by the halo mass-cooling rate, $\mdotcool$, from \citet{Croton06} \citep[also][]{White91,Cole00,Somerville08,Guo11,Henriques15}.

To replace total BH energy on the left side of Eq.~\ref{Eq:EbhEhalo} by a BH heating rate, the BH mass accretion rate must also be known.  In Variant~3, this rate is obtained by differentiating the instantaneous $\Mbh$ values from the empirical power-law model.  In Variant~4, the BH accretion expression introduced by \citet{Croton06} for radio mode is used, but as this has no dependence on \Reff, this model is only a one-parameter family.  Equations are given in Appendix~\ref{SubSec:AppRatesorTotals}, and the accretion rates \vs time are compared in Figure~\ref{Fig:Trajectory_BH}.  It is seen that the Variant~3 BH accretion rates fall strongly at late times, the Variant~4 accretion rates grow super-exponentially at late times, while total BH masses in the empirical power-law model grow moderately at all times.

The consequences of these different rates are illustrated in Figures~\ref{Fig:MRate_V3}-\ref{Fig:SimuSize_V4}.  Variant~3 fails because the late turndown in BH accretion rates means that galaxies stop quenching at late times.  Variant~4 fails because the late turnup in BH accretion rates makes the  characteristic green valley mass fall by 1.0~dex from $z = 1.4$ to now, and because it is only a one-parameter family.

Collectively, Variants~3 and 4 illustrate two points: 1) BH and halo effects need to march closely together in time in order to feed galaxies smoothly into the green valley at constant rate and mass, and this is achieved more easily with cumulative energies than with rates; and 2) a successful quenching model needs two parameters working together to match the observed 2-D structural diagrams of galaxies, which automatically excludes one-parameter models that depend only on mass, like Variant 4.

\section{Discussion}\label{Sec:Discussion}

The Discussion has nine sections.  The first reviews and explains the successes that were  ascribed to the model in Table~\ref{Tb_1}.  From there, we consider possible explanations for the scatter in the effective radii of star-forming galaxies, which is a necessary ingredient in the model. Subsequent sections discuss quenching, compaction, BH growth through the green valley, and the different BH scaling laws found by different authors.  The latter open the way to thinking about the BH-halo connection, what is meant by  BH-galaxy ``co-evolution'', and the  implications of having \emph{three separate yet matching} quenching boundaries, in $\Sig1\text{-}\Mstar$,  $\Sig1\text{-}\Mbh$, and $\Mbh\text{-}\Mstar$.  We close with a comparison of our picture to the \citetalias{Lilly16} quenching model.

\subsection{Scorecard for the Binding-Energy Model}\label{SubSec:Successes}

This section reviews the ability of the model to match the observational data listed in Table~\ref{Tb_1}.  By ``model'' we henceforth mean the empirical power-law model from Section \ref{Sec:EvolPLModel} that includes the cartoon evolutionary tracks in Figure~\ref{Fig:Cartoon} but with the halo energy quota, $\Eheat$, replaced by 4$\times$ halo-gas binding energy, $\Eth$ -- i.e., \emph{the blue lines in Figure~\ref{Fig:Mstar_Sig1_SMA1}}.  This model will henceforth be called the \emph{``binding-energy model''}, and it is the major product of this paper.
The basic ingredients of this model are: 1) the known theoretical evolution of the halo properties $\Mvir$, $\Rvir$, and $\Vvir$; 2) the assumed links between halo properties and galaxy properties $\Mstar$, $\ReffMean$, and $\langle\Sig1\rangle$; 3) the scattering of the ratio $\Reff/\ReffMean$; 4) the criterion for entry into the green valley $\Ebh=f_B\times\Eth$; and 5) the relation $\Mbh = \rm{const.} \times \Sig1^{1.76}$ for star-forming galaxies.

Column~1 in Table~\ref{Tb_1} indicates the successes of this model by using the line number without brackets to indicate observations that we think are well explained by the model and the line number with brackets to indicate partial success.  We note that the binding-energy model has only one free parameter, and that is the constant $f_{\rm B}$ in Eq.~\ref{Eq:EbhEbind}, which has been set equal to 4 to make the zero point of the blue lines match that of the red lines in Figure~\ref{Fig:Mstar_Sig1_SMA1} on average.\footnote{The adopted value of $f_{\rm B}$ depends on the assumed value of the black hole efficiency factor $\eta$, which we have set to 0.01, and we are actually adjusting the ratio of these two quantities.}  Aside from the general insistence on power-laws, none of the observations in Table~\ref{Tb_1} were used to determine the five basic ingredients above, and hence all of the successes in Table~\ref{Tb_1} are new.

As noted, the starting point for the binding-energy model places star-forming central galaxies at the centers of dark halos with $\Mstar$ and $\ReffMean$ derived from halo $\Mvir$ and $\Rvir$.  As pointed out by \citetalias{vanderWel14} and noted in Section~\ref{SubSec:EvolTracks}, this halo-based approach matches rather well the ridgeline slope, zero-point evolution, and evolutionary track slopes of star-forming galaxies in $\Reff\text{-}\Mstar$ and, through the $p$,$q$ mapping, also in $\Sig1\text{-}\Mstar$.  These are lines~4, 5, 8, and 9 in Table~\ref{Tb_1}, which are indicated as full successes.  Moving down, the slopes of the quenching boundaries are constant with time in the binding-energy model, as observed, but the predicted slope is a bit too shallow in $\Sig1\text{-}\Mstar$ and a bit too steep in $\Reff\text{-}\Mstar$, and the zero-point motions with time are a bit too small (i.e., the blue lines do not quite match the red lines in Figure~\ref{Fig:Mstar_Sig1_SMA1}).  Because of these mismatches, lines~10, 11, and 12 for the quenching boundaries -- and their associated quenched ridgelines -- have been recorded as partial successes.

We pause here to comment on line 12, which says that the $\Mbh\text{-}\Mstar$ relation for quenched galaxies is offset to higher values at early times  \citep{Peng06,Treu07,Woo08,Gu09,Decarli10,Bennert10,Bennert11,Ding17,Ding20}.  This offset has often been interpreted to mean that BHs form \emph{before} the stars of their host galaxies \citep[e.g.,][]{Peng06,Treu07,Bennert11,Ding17}, making BHs appear larger back in time for their host stellar mass.  This interpretation is contrary to the binding-energy  model, which says that BHs experience their major growth in the green valley, and thus that black holes form \emph{after} most of the stars. The model has a different explanation for the offset, namely, that the quenching boundary in $\Mbh\text{-}\Mstar$ -- and therefore its associated ridgeline -- was higher in the past.  The predicted motion can be calculated from the observed motion of the $\Sig1\text{-}\Mstar$ ridgeline and is approximately $v \times 0.3 = 0.5$~dex back to $z = 3$, which is in good agreement with the observed shift in $\Mbh\text{-}\Mstar$  \citep{Bennert11,Ding17,Ding20}.  The point is that offsets of $\Mbh$ \vs $\Mstar$ with time are not necessarily due to growth-time differences between stars and black holes but may instead reflect motions of the quenching boundary.\footnote{The fact of a moving zero point of $\Mbh$ \vs $\Mstar$ does not contradict our  fundamental assumption that the zero point of  $\Mbh = \rm{const.}  \Sig1^{1.76}$  remains the same since $\Mstar$ and $\Sig1$ are not the same quantity.} Tnhe same point was made by \citet{Caplar18}, who explored a model similar to ours in which black holes grow rapidly just as galaxies enter the green valley.

Moving on, line 16 in Table~\ref{Tb_1} says that the observed zero point of the quenched ridgeline in $\Sig1\text{-}\Mstar$ evolves downward as $h(z)^{0.66}$ (Appendix~\ref{App:A2}), while the observed zero point of the star-forming ridgeline in the same plane evolves downward by the same amount (\citetalias{Barro17}).  This is the key equality that maintains star-forming galaxies \emph{in the same relationship} to quenched galaxies in $\Sig1\text{-}\Mstar$ and is the reason why the diagram looks the same at all redshifts (except for the bulk vertical shift).  This synchronized motion means that the quenching boundary cuts through the star-forming distribution \emph{over the same range of masses at all redshifts}. This is seen in the predicted model quenching loci of Figure~\ref{Fig:SimuCartoon}, where the smallest galaxies always enter quenching near $10^{9.5}~\Msol$ and are fully quenched by $10^{10}~\Msol$, which is the observed lower edge of the high-mass quenching channel at all redshifts \citepalias{Barro17}.  Finally, this same constant vertical offset between the quenching boundary and the star-forming cloud ensures a \emph{smooth, steady flow of galaxies across the boundary at all times} (line~3 in Table~\ref{Tb_1}). The delicacy of achieving this is illustrated by Variant~3, which fails to do this.

The synchronized downward motion of the quenched and star-forming clouds in $\Sig1\text{-}\Mstar$ is thus a fundamental feature of the data that demands an explanation, and the model provides estimates for the motions of both zero points separately.  The evolution of the star-forming ridgeline at fixed mass in $\Sig1\text{-}\Mstar$ goes as $\Reff^{-p} \sim \Reff^{-1} \sim \Rvir^{-1} \sim h(z)^{0.74}$ (this last is from Eq.~\ref{Eq:ReMstar} in Section~\ref{SubSec:EvolTracks}).  The zero-point motion of the quenching boundary (and thus the quenched ridgeline) varies as $\Eth^{1/v} \sim (\Mvir^2/\Rvir)^{1/v}$, which at fixed $\Mstar$ goes as $\Rvir^{-1/v} \sim h(z)^{0.74/v} \sim h(z)^{0.42}$.  The two exponents of $h(z)$ are therefore similar but not quite the same, differing by the factor $v = 1.76$.  This explains why the zero point of the blue line moves down more slowly than the red line in Figure~\ref{Fig:Mstar_Sig1_SMA1}, leaving a residual in $\Sig1$ of about 0.1~dex. Based on this partial success, line~16 is given a bracket.

It is helpful to rephrase the above mathematical language into something more intuitive. Simply put, black hole masses (at fixed $\Mstar$) grow smaller at late times because galaxies have larger $\Reff$.  However, halo binding energies (at fixed \Mvir) also become smaller with time because halos have larger \Rvir.  Together, these two factors cause the downward boundary motion.  This point is not new; it was made by \citet{Booth11}, who predicted that the quenching zero point of models in which BH energy output is balanced against halo-gas binding energy would decline with time. Our new insight is that the parallel evolution of black holes and halos is why the green-valley mass remains constant with time.  In more detail, the decline in halo binding energy is not quite fast enough to keep halos and black holes in perfect lockstep, but real halos might have additional factors, such as late-infall dry-up or the disappearance of cold streams, that might help them quench more easily at late times (see Section~\ref{SubSec:Quenching}).

A moving quenching boundary has another consequence.  If the evolution of individual galaxies \emph{after} quenching does not precisely follow the moving boundary, early-quenched galaxies will be  systematically offset from late-quenched ones, leading to age-stratification in the quenched ridgeline due to  ``progenitor bias'' \citep[e.g.,][]{Carollo13, Poggianti13}.

\citetalias{Barro17} listed three factors that help quenched galaxies stay close to the downward-evolving ridgeline in $\Sig1\text{-}\Mstar$. They are: 1) stellar mass-loss coupled with adiabatic expansion, 2) core scouring by BH mergers in merging galaxies, and 3) dry merging.  With three mechanisms helping, we expect that age stratification in $\Sig1\text{-}\Mstar$ will be only moderate.

In contrast, post-quenching evolution in $\Mbh\text{-}\Mstar$ ought not to track the moving boundary as closely. Stellar mass loss moves galaxies to the left, above the relation.  Also, if BH's merge along with their host galaxies, this would tend to evolve objects \emph{along the original locus} since the log slope of the relation is near unity, at best leaving their vertical residuals unchanged.  Finally, core scouring does not alter either $\Mbh$ or $\Mstar$.  The net result is that quenched objects are not expected to follow the moving ridgeline as closely in $\Mbh\text{-}\Mstar$ as in  $\Sig1\text{-}\Mstar$, and thus progenitor bias and age stratification should be larger in this ridgeline.

Age stratification in the quenched ridgeline has been measured in $\Mbh\text{-}\Mstar$ by \citet{Martin18} and in $\Sig1\text{-}\Mstar$ by \citet{Tacchella17} and \citet{Luo19}.  Early-quenched galaxies fail to fully keep up with the moving boundary in both spaces, but the mismatch in $\Sig1\text{-}\Mstar$ is significantly smaller, as the above logic predicts.  The magnitudes of both effects are generally consistent with the model, and line~15 has been marked as a success.

Line~2 notes the wide extent of the green valley in $\Mstar$.  As described in Section~\ref{SubSec:EvolModelResults}, this is explained by the highly acute ``angles of attack'' of the evolutionary tracks relative to the quenching boundary in $\Sig1$ \vs $\Mstar$, $\Reff$ \vs $\Mstar$, and $\Ebh$ \vs $\Eheat$ (and also $\Eth$), which enable star-forming galaxies with the same $\Mstar$ but only slightly different values of $\Reff$ and $\Sig1$ to evolve to very different final quenching masses.  These acute angles are a natural feature of the model, and line~2 is accordingly marked as a success.

Line~14 in Table~\ref{Tb_1} notes that the ridgelines of quenched galaxies are narrow in $\Sig1\text{-}\Mstar$ and $\Mbh\text{-}\Mstar$ but broader in $\Reff\text{-}\Mstar$, especially at high $z$.  The basic narrowness in $\Sig1$ and $\Mbh$ is due to the fact that galaxies stop evolving and pile up along these loci.  The broader width of $\Reff\text{-}\Mstar$ reflects the greater impacts on $\Reff$ of compaction (see Section~\ref{SubSec:Compaction}) and dry minor merging \citep[e.g.,][]{Daddi05,Trujillo06,vanDokkum08,Naab09,vanDokkum10,Szomoru12,Oser12}.  The evolution of the latter goes as $\Reff \sim \Mstar^2$, which is steeper than the quenched ridgeline itself ($\Reff \sim \Mstar^{0.75}$), thus broadening (and steepening) it significantly.  By invoking a little help from compaction and dry merging, we consider line~14 a success.

Finally, line~13 in Table~\ref{Tb_1} says that $\Mbh$ varies roughly as some power $\Mstar^w \Reff^{-w}$ (or as $\sigmae^{2w}$) for quenched and star-forming galaxies together but as a smaller power than $w$ for quenched galaxies alone (compare slopes from \citet{vandenBosch16} and \citetalias{Kormendy13}).  This difference turns out to be the same problem as the low value of $u = 3.52$ in Section~\ref{SubSec:InputPowerLaws}.  Both issues are discussed in Section~\ref{SubSec:PredictionsTests}, where we show that steep BH growth in the green valley can naturally explain both.  The same steep BH growth also explains the greater separation between quenched and star-forming galaxies in $\Mbh\text{-}\Mstar$ than in $\Sig1\text{-}\Mstar$, and both lines~13 and 18 are marked as successes.

The above discussion covers all the items without brackets in Table~\ref{Tb_1}.  Lines~6 and 7 (with brackets) involve the slope and zero point of the quenched ridgeline in $\Reff\text{-}\Mstar$.  As these depend on how $\Reff$ evolves after quenching, which is not included in the model, only partial credit is given for getting the locus in roughly the right place.  The different zero-point evolution of $\Reff\text{-}\Mstar$ \vs  $\Sig1\text{-}\Mstar$ (line~17) is attributable to the larger impact of post-quenching dry merging on $\Reff$ than on $\Sig1$, and so partial credit is given to that line as well.

This completes the review of all the items in Table~\ref{Tb_1}.  Yet another test would use X-ray luminosities to test whether star-forming galaxies with higher $\Sig1$ (and lower $\Reff$) at fixed $\Mstar$ have larger black holes. The findings of \citet{Ni19} suggest this, but their star-forming sample also includes green valley galaxies and thus is not quite pure enough.  A re-analysis of a pure main-sequence sample is underway (Kocevski et al., in prep.).  X-ray data for green valley galaxies are discussed in Section~\ref{SubSec:BoundaryCrossing}.

\subsection{The Origin of Scatter in Effective Radii}\label{SubSec:RadiusScatter}

A basic ingredient of the binding-energy model is that star-forming galaxies at a fixed stellar mass have different radii.  Radii differences correlate with central $\Sig1$ (Figure~\ref{Fig:Mstar_Sig1_SMA1}), and these induce differences in BH mass and BH feedback.  Where do radii differences come from?  Since the model assumes that all star-forming galaxies of the same stellar mass have the same dark-halo mass, $\Reff$ differences must come from some other parameter.

The connection between dark halos and galaxy radii was reviewed by \citet{Somerville18} and \citet{Wechsler18}.  \citet{Mo98} introduced the formula for the exponential disk scale length $r_{\rm d} = 1/\sqrt{2}\lambda\Rvir$, where $\lambda$ is the dimensionless halo spin parameter.  This suggested that $\lambda$-variations might cause scatter in $\Reff$.  However, the observed rms scatter in $\Reff$ is only 0.2~dex \citepalias{vanderWel14} compared to 0.25~dex for $\lambda$ in the models, which leaves no room for any other sources of scatter or measurement errors \citep{Desmond15}.  Hydro simulations also show that cooling gas gains or loses angular momentum unpredictably while falling onto galaxies  \citep{Pichon11,Stewart13,Danovich15,Stevens17,Jiang18}, blurring any connection between original and final angular momenta.  The net result is that galaxy radius is found to be almost uncorrelated with halo $\lambda$ in EAGLE \citep{Desmond17}, and  \citet{Jiang18} find at most poor correlations for VELA \citep{Ceverino14} and for NIHAO \citep{Wang15}.

An interesting alternative to $\lambda$ is halo concentration, $c$.  \citet{Jiang18} find that the radii of VELA and NIHAO galaxies obey $\Reff = 0.02 \Rvir(c/10)^{-0.7}$, with galaxies being smaller in higher-concentration halos.  This is plausible  since a greater fraction of the matter in a high-concentration halo falls in at earlier times, when the universe was denser.  The binding-energy model then says that such galaxies should have relatively larger black holes owing to their higher stellar densities.  For the purposes of this paper, the origin of radius scatter needs no explanation -- the $\Delta \log \Reff$ offset of a galaxy may simply be a random variable that is assigned at birth and is unconnected to halo properties -- it is only required that $\Delta \log \Reff$ stay constant during the star-forming phase so that galaxies evolve on parallel tracks in $\Reff\text{-}\Mstar$.  However, a picture in which the two leading structural parameters of star-forming galaxies, $\Mstar$ and $\Reff$, arise directly from two leading parameters of dark halos, \Mvir\ and $c$, would be both simple and elegant.

Recent analyses of EAGLE and IllustrisTNG see the predicted effect in which more concentrated halos form stars faster, have larger black holes, larger  values of $\Ebh/\Eth$, and are more quenched today \citep{Davies19a,Davies19b,Oppenheimer19}.  The effect in EAGLE is large -- halos in the top quartile of $\Ebh/\Eth$ were quenched several billion years ago while halos in the bottom quartile are still making stars today.  Furthermore, \citet{Davies19a} showed that it is the \emph{integral} of BH feedback, not the instantaneous BH feedback rate, that ultimately ejects baryons, and the best barometer for the degree of quenching is the ratio $\Ebh/\Eth$.

These properties support our assumption in the binding-energy model that quenching is determined by the ratio $\Ebh/\Eth$, not by instantaneous heating or cooling rates.  EAGLE quenching becomes perceptible when $\Ebh/\Eth \sim 2$ and is complete when $\Ebh/\Eth \sim 20$ whereas we predict the range 4-to-20 for this ratio (see Section~\ref{SubSec:BoundaryCrossing} below).  This is a promising level of agreement given the simplicity of our model.\footnote{It is important to note that the evacuation of gas from halos during quenching that is predicted by the binding-energy model cannot persist when halos later grow to cluster size.  Large clusters are seen to be full of gas at approximately the cosmic baryon-to-dark-matter ratio \citep[e.g.,][]{Tumlinson17}, and so they must refill.  EAGLE and IllustrisTNG clusters in fact do refill \citep{Oppenheimer19,Davies19b}.  A different form of BH feedback must take over to to maintain quenching at the centers of gas-rich groups and clusters. This later evolutionary stage is not covered by the model.}

Guided by EAGLE and IllustrisTNG, it is tempting to create a revised binding-energy model by adding halo concentration to $\Mvir$ as a secondary parameter.  This addition of a second halo parameter in addition to $M_{\rm halo}$ would be an example of \emph{assembly bias} \citep{Wechsler18}.  However, doing so would raise important questions.  For one thing, N-body simulations show that mass accretion rates onto halos at fixed time vary with halo concentration \citep{Lin19}, which, if unopposed, might disturb the observed absence of any trend in star-formation rate \vs galaxy size at fixed $\Mstar$ (\citealp{Omand14,Whitaker17}; \citetalias{Fang18}; \citealp{Lin19}).  Indeed, star-formation rates \emph{do} vary with size in EAGLE \citep{Oppenheimer19}, in disagreement with the observations.  A dodge around this could posit weaker galactic winds in small galaxies on account of their higher escape velocities \citep{Dutton10,Lin19}.  But even if \emph{current} star-formation can be evened out in this way, high-concentration halos still need to form stars earlier in order to give their galaxies smaller radii.  This means that their star-formation rates had to be different in the past, i.e., that mass cohorts \emph{cannot} stay together over long times, which violates another of our key  assumptions.  Finally, it is not clear that halos in a given mass cohort would  maintain their \emph{relative} concentration rankings at different epochs, which would be needed to keep their evolutionary tracks parallel.

In summary, adding halo concentration (and perhaps also time of formation) as a second parameter to the binding-energy model is an intriguing suggestion, but important issues remain unresolved.

\subsection{Quenching in the Binding-Energy Model Compared to Other Models}\label{SubSec:Quenching}

It is instructive to compare the BH accretion and quenching processes in the binding-energy model to analogous mechanisms in current SAM and hydro models.  The following summarizes the kinds of feedback mechanisms in these models:

\emph{Stellar feedback}. This is the most basic feedback mechanism, present in all models.  Energy/momentum from young massive stars and SNae drive a wind that expels gas from a galaxy during the star-forming phase.

\emph{Quasar mode} \citep{Croton06,  Somerville08, Henriques15}, also called ``bright mode'' or ``high mode'' \citep{Somerville08,Pillepich18} or ``radiative mode'' \citep{Somerville15}. This is the  principal mode of BH growth in many models, triggered by mergers \citep{Hopkins06}, disk instabilities \citep{Cattaneo06}, or other  unspecified secular processes \citep{Martin18}. Key features of quasar mode (in most models) are a high black-hole accretion rate, formation of a thin accretion disk, and high radiative efficiency from the accretion disk (see review by \citealt*{King15}).  It is generally quasar-mode winds that quench star formation in semi-analytic models and in many hydro models.

\emph{Radio mode} \citep{Croton06,  Somerville08, Henriques15}, also called ``kinetic mode'' or ``low mode'' \citep{Weinberger17,Pillepich18}, or ``jet mode'' \citep{Somerville15}.  Hot gas is accreted slowly onto a central BH, producing high-velocity winds and/or radio jets that escape into the halo.  Key features of radio mode (in most models) are a low black-hole accretion rate, formation of a thick accretion disk, and low radiative efficiency (see review by \citealt*{Yuan14}).  Radio mode may quench, as in IllustrisTNG \citep{Weinberger17,Davies19b,Terrazas19,Nelson19}, or it may maintain a low star-formation rate in gas-rich halos after other mechanisms have quenched.

\emph{Halo quenching} \citep{Rees77, Blumenthal84,Silk86,Birnboim03,Keres05,Dekel06,Dekel19}. As halos grow in mass, their gas becomes hotter and more dilute, which retards cooling.  A critical halo mass has been proposed, $\Mcrit \sim 10^{11.5\text{-}12}~\Msol$ \citep{Dekel06,Dekel19}, above which halos fill with hot gas that cannot easily cool, and star formation shuts down.  This transition is also accompanied by a change from ``cold streams'' to ``hot flows'' \citep{Keres05,Dekel06,Keres09}, which are less able to penetrate to the central galaxy.

\emph{Late infall dry-up}. The specific accretion rate of gas onto halos declines after $z \sim 1$ in all cosmologies, regardless of $\Omega_{\rm m}$ or $\Omega_{\Lambda}$ \citep{Dekel14}.  Though not a feedback process per se, it lowers the amount of gas available for star formation \citep{Bouche10, Dave11, Dave12, Krumholz12, Dekel13, Lilly13, Dekel14, Forbes14, Feldmann15, Mitra15, Rodriguez16_1} and retards halo cooling by keeping gas densities low at late times.  Along with the transition from cold streams to hot streams \citep{Keres05,Dekel06}, this makes late halos easier to quench and would help to lower the zero point in the quenching boundary relation (red lines) in $\Sig1\text{-}\Mstar$, as seen in Figure~\ref{Fig:Mstar_Sig1_SMA1}.

Three features of the binding-energy model collectively make it unique compared to all other models.  The first is that the power of AGN feedback to quench is based on the \emph{rate} of BH gas accretion \vs the \emph{rate} of halo-gas cooling in some models \citep[e.g.,][]{Croton06,Bower06,Somerville08,Henriques15} whereas the binding-energy model integrates \emph{total} BH energy over time and compares it to the \emph{total} binding energy of the halo gas.  The impact of the difference was explored in Variants~3 and 4 in Section~\ref{Sec:Variants}, where it was shown that rate formulations do not feed gas to black holes at the right rate.  As noted above, two recent hydro simulations also seem to exhibit the total-energy criterion (EAGLE: \citet{Davies19a,Oppenheimer19} and IllustrisTNG: \citet{Davies19b,Terrazas19}).

The second difference is that multiple mechanisms typically operate together in SAM and hydro models to initiate quenching, and the decisive mechanism in any one galaxy can be hard to identify.  Such is the case, for example, in the \citet{Somerville08} SAM \citep{Pandya17} and in TNG50 \citep{Nelson19}. In contrast, entry into the green valley in the binding-energy model is the well-defined point when total AGN output surpasses $f_{\rm B} \times$halo-gas binding energy, where $f_{\rm B} \simeq 4$.

Finally, the third major difference between the binding-energy model and  most other models is the mechanism that maintains the constant value of the mean green valley mass, $\langle\MstarGV\rangle$, over time. A plausible value for this mass, cited in many works, is the value of $\Mstar$ at the critical halo cooling mass $\Mcrit$ \citep[e.g.,][]{Birnboim03,Keres05,Dekel06,Dekel19}.  A second characteristic mass has also emerged in recent hydro models from balancing supernovae wind energy against galaxy or halo potential well depth \citep{Dubois15,Angles17,Bower17,Habouzit17,Davies19a,Dekel19,Oppenheimer19,Henriques19}. SNae wind failure allows central gas to accumulate \citep[e.g.,][]{Dekel86}, which fuels rapid BH growth and causes quenching. \citet{Dekel19} even argue that these two masses converge to create a single ``golden mass'' at early times.

In contrast, the constancy of $\langle\MstarGV\rangle$ in our model has nothing to do with $\Mcrit$ or SNae winds but comes instead from the fact that halos and galaxies evolve so as to keep the quenching boundary \emph{located in the same place relative to the blue cloud} over time.  This means that quenching galaxies always exit the blue cloud with the \emph{same characteristic mass}.  Granted, it is fortunate that $\Mcrit$ for hot halos is near $10^{11.5\text{-}12.0}~\Msol$ and not larger, since having a hot halo may facilitate BH quenching.  Central SNae wind failure might also do this. However, we equate quenching with cross-over into the green valley, and, in the binding-energy model, this point is not set by the ability of halo gas to cool, or by the failure of SNae winds to keep galaxy centers gas free, but by the \emph{cumulative effect of black hole feedback on the halo gas}.  The characteristic masses set by halo cooling and by SNae winds are not fundamental -- they merely need to be small enough to set the stage for the halo binding-energy criterion, which is determinative.

It may be asked if there \emph{is} a characteristic halo mass in the binding-energy model, and the answer is yes.  This is the halo mass, $M_{\rm vir,cross}$ where the growing energy from the BH in a typical galaxy (with \Reff = \ReffMean) during the star-forming phase crosses the binding energy of halo gas.  Scaling laws in the Appendix may be combined with the quenching condition in Eq.~\ref{Eq:EbhEbind} to yield:
\begin{eqnarray}\label{Eq:MvirCross}
\left(\frac{M_{\rm vir,cross}}{10^{12}~\Msol}\right)^{1.60v-1.67} & = & 9.5  k_{\rm{Re}}^{v}  k_{\rm{BH}}^{-1}   k_{\rm{\Sig1}}^{-v}   k_{\rm{Rvir}}^{v-1}   k_{\rm m}^{-1.1v} \frac{f_{\rm B}}{4} \times \nonumber\\
 & & \frac{f_{\rm hot}}{0.1} \times \left(\frac{\eta}{0.01}\right)^{-1} \times  h(z)^{0.74(v-1)},
\end{eqnarray}
where the constants $k$ are the zero points of the five relevant scaling laws from Appendix~\ref{App:A3} and we have inserted our best values for $p$, $q$, $s$, and $t$.   $M_{\rm {vir,cross}}$ will be a fundamental mass if the constants $k$ are fundamental numbers. The first two of the constants are already known, and it will be possible to write down the values of the other three  once galaxy radii and black-hole building are fully understood. Two more free parameters at the present time are $f_{\rm B}$ and $\eta$. These also should be expressible (someday) using fundamental constants.

For reference, we insert $k$-values from the Appendix, use $v = 1.76$, and substitute $\Mstar$ for $\Mvir$ to get the crossing mass in terms of $\Mstar$, which should equal to $\Mstar$ in the middle of the green valley.  We obtain:
\begin{eqnarray}\label{Eq:MstarCross}
\frac{M_{\rm *,cross}}{10^{10}~\Msol} & = & 4.0 \left( \frac{f_{\rm B}}{4} \right)^{1.54} \times \left( \frac{f_{\rm hot}}{0.1}  \right)^{1.54} \times \nonumber\\
  & & \left( \frac{\eta}{0.01} \right)^{-1.54} \times \left( \frac{h(z)}{2.3} \right)^{-0.86},
\end{eqnarray}
where $h(z)$ has been normalized by its value at $z = 1.5$, where $k_{\rm {Rvir}}$ is calculated.  $M_{\rm *,cross}$ therefore equals $10^{10.6}~\Msol$ at $z = 1.5$, which is a good match to the green valley stellar mass.  However, note that $M_{\rm *,cross}$ is not exactly constant with time due to the presence of the term $h(z)^{-0.86}$.  This term arises from the same zero-point mismatch between the blue lines and the red lines in Figure~\ref{Fig:Mstar_Sig1_SMA1} that was pointed out in Section~\ref{SubSec:Successes}.  However, the residuals relative to $z = 1.5$ in mass are $-0.2$~dex at $z = 3$ and $+0.2$~dex at $z = 0.5$, which are not large.

To summarize, the binding energy-model, like other models, depends on SNae feedback to keep galaxies on the SMHM relation while star-forming.  For quenching, some form of AGN feedback is required, which could be either high mode or low mode.  Since BH energy overcomes halo-gas binding energy in this model, preventive feedback is probably the dominant mode though ejective feedback may also play a role.  Halo quenching may be a prerequisite for the binding-energy criterion to work, but it does not set the fundamental green valley mass.  Late infall dry-up (and the transition from cold streams to hot streams) may make late halos easier to quench than early halos, thereby contributing to the faster lowering of the quenching boundary zero point (compared to the model prediction) that is observed in $\Sig1\text{-}\Mstar$ in Figure~\ref{Fig:Mstar_Sig1_SMA1}.

\subsection{Compaction in the Binding-Energy Model}\label{SubSec:Compaction}

Implicit in the binding-energy picture is that galaxies, halos, and black holes evolve smoothly and steadily together during the star-forming phase and that crossing into the green valley is a quiet event with no notable fireworks.  That seems a good model for quenching today, particularly for S0 galaxies, which have not experienced recent major mergers because their disks are intact.  However, it will be noticed that many dense star-forming galaxies are visible on the \emph{wrong} side of the boundary at high redshift but then disappear at low redshift.  This is especially true using $\Reff$ as a measure of density as opposed to $\Sig1$ (cf. red points below the boundary in the top row of Figure~\ref{Fig:Mstar_Sig1_SMA1}a).  The topic of this section is how these dense, discrepant galaxies might be accommodated in the binding-energy model.

These small, dense galaxies are the so-called ``blue nuggets'' at $z = 2$  (\citealp{Barro13,Barro14,Nelson14,Williams14,vanDokkum15}; \citetalias{Barro17}).  They are thought to be the star-forming precursors of the similarly dense, quenched ``red nugget'' galaxies \citep{Daddi05, Trujillo06, vanDokkum08, Damjanov11}, which are visible in Figure~\ref{Fig:Mstar_Sig1_SMA1} at the same redshifts. VELA simulations \citep{Ceverino14} suggest that both populations originated from a ``compaction'' event in which gas was driven strongly to the center, forming a burst of stars that consumed the central gas and drove gas from the rest of the galaxy \citep{Dekel14,Zolotov15,Tacchella16b}. This scenario says that star-forming galaxies compactified to form blue nuggets, which then faded to become red nuggets. Simulations by \citet{Choi18} and by the IllustrisTNG team \citep{Habouzit19} produce similar objects.

Actually, VELA galaxies never fully quench owing to their lack of central black holes.  But X-ray data show a large population of X-ray sources near the quenching boundary at $z \sim 2$ \citep{Kocevski17}, right where maximum compaction occurs \citep{Dekel19}. If these accreting black holes were added to VELA, their feedback might achieve quenching.  In contrast to VELA, BHs \emph{are} present in IllustrisTNG and in the \citet{Choi18} models, and in these simulations compact blue nuggets quench.

A variety of factors can trigger compaction events in central galaxies, including mergers \citep{Barnes91, Hernquist91, Hopkins06}, violent-disk-instability-driven inflows \citep{Dekel14}, counter-rotating streams \citep{Danovich15}, a tri-axial halo core \citep{Tomassetti16}, and the return of recycled low-angular-momentum gas \citep{Elmegreen14}.  However, most of these events are stochastic in nature and uncorrelated with the underlying galaxy structural parameters.  Galaxies therefore tend to evolve along their native evolutionary trajectories in these scenarios, their parameters bobbing up and down due to occasional compaction events until a final, major drop in star-formation occurs. Such behavior is seen in VELA, where the final drop is driven by late infall dry-up \citep{Tacchella16a}.

We take this as the basic picture for compaction in the binding-energy model, to which we add black holes.  As blue nuggets compactify, their central densities rise rapidly and their BHs grow according to the green valley vectors in Figure~\ref{Fig:Cartoon}. Compaction thus sweeps galaxies \emph{rapidly} over the quenching boundary at high redshift, at which point their growing BHs begin to impede star formation according to Eq.~\ref{Eq:EbhEbind}.

Adding black holes in this way solves four problems.  First, the inclusion of black holes makes galaxies quench.  Second, quenching according to the binding-energy equation produces \emph{sloping} quenching boundaries, which are seen at high redshift as well as at low redshift (Figure~\ref{Fig:Mstar_Sig1_SMA1}).  Third, the model sharpens the rather fuzzy boundary that might otherwise result from compaction alone due to its stochastic nature.  Finally, the disappearance of blue nuggets at low redshift is natural since compaction is strong in gas-rich galaxies but should tend to taper off as galaxies become gas-poor.

The one remaining issue is timing: why does star-formation continue \emph{well after} the BH mass has crossed the quenching boundary?  We theorize that this occurs because compaction delivers an unusually large amount of fresh gas to the center of a galaxy in a short space of time.  In order for the galaxy to quench, this gas must either be consumed or ejected, which imposes a delay.  During that delay, the galaxy sits on the wrong side of the boundary, being compact yet still star-forming.  After the central gas has cleared, the recently enlarged BH can transmit energy and/or momentum all the way to the halo, as required by the binding-energy picture.

These last remarks echo the ``fast track-slow track'' dichotomy for galaxy quenching that has been advanced by several authors (e.g., \citealt{Yesuf14,Barro17,Woo15,Wu18}; \citealt*{Woo19}).  Fast-track galaxies, sometimes called ``post-starburst galaxies'' \citep{Dressler83, Balogh99, Dressler99}, exhibit strong Balmer absorption lines and small radii, which indicate rapid quenching and strong gaseous dissipation \citep{Whitaker12,Yesuf14,Yano16,Almaini17,Wu18}.  The binding-energy model attributes both characteristics to the rapid arrival of gas at the center.  A starburst ensues while radii are small, which places them temporarily on the wrong side of the quenching boundary until black holes can grow and quench. Slow-track galaxies in contrast slide over the boundary gradually, and their star-formation rates never get strongly out of synch with their structural properties and BH masses.  Since the fast-track scenario requires lots of gas, the fraction of fast-track to slow-track galaxies should decrease with time, as observed (\citealp{Whitaker12,Yano16}; \citetalias{Barro17}).  In conclusion, although fast-track galaxies are not explicitly included in the binding-energy model, modest extensions to it seem able to accommodate them.

\subsection{Evolution Through the Green Valley}\label{SubSec:BoundaryCrossing}

The manner in which galaxies evolve through the green valley is important in the model because the slope of \emph{fully-quenched galaxies} is assumed for BH growth in \emph{star-forming galaxies} (Section~\ref{SubSec:InputPowerLaws}), and these two populations are connected by the green valley.  This section seeks to understand why green valley BH growth vectors might be both steep and uniform over time (cf.~Figure~\ref{Fig:Cartoon}).

Brief evidence to support the steepness and uniformity of GV tracks was presented in Section~\ref{SubSec:InputPowerLaws}, which we expand on here.  The first fact is the sharp ``elbow'' seen in $\Delta \log SSFR$ \vs $\Delta \log \Sig1$ by \citetalias{Fang13} and again in Figure~6b of \citetalias{Barro17}, in which $SSFR$ declines steeply through the GV while $\Sig1$ hardly changes (see cartoon version of this in Figure~\ref{Fig:Elbow}a here).  It is this pile-up at near-constant $\Sig1$ that creates the narrow ridgeline of quenched galaxies in $\Sig1\text{-}\Mstar$ (cf.~Figure~\ref{Fig:Mstar_Sig1_SMA1}).  The constant morphology of the elbow (\citetalias{Fang13,Barro17}; \citealp{Lee18,Mosleh18}) sets the growth of $\log \Sig1$ through the GV at roughly 0.2~dex at all masses and redshifts $z \leq 3$.  Consistent with this, the GV entry locus also lies a constant 0.2~dex below the quenched ridgeline for SDSS galaxies at all masses today \citep{Luo19}.  In summary, there seems to be a universal $\Sig1$ ``elbow diagram'' such that the growth of $\Sig1$ through the GV is small, only about $\sim0.2$~dex, while $SSFR$ falls by much more than this  (\citetalias{Fang13,Barro17}; \citealt{Luo19}).

The evolution of black-hole mass through the green valley is different.  \citet[][Figure 3]{Terrazas17} have shown that BH mass \emph{continues to grow} through the GV even as star formation declines.  That is, $\Delta \log SSFR$ \vs $\Delta \log \Mbh$ does not show a sharp elbow but rather a smooth, monotonic decline, with $\Mbh$ increasing across the green valley by about 1.5~dex (cf.~Figure~\ref{Fig:Elbow}b here).  The same difference is also seen in  Figure~\ref{Fig:ThreePanel}c, where star-forming and quenched galaxies are well separated, in contrast to $\Sig1$ in Figure~\ref{Fig:ThreePanel}b, where the two populations overlap.  Further evidence for large BH growth in the green valley comes from the enhanced X-ray luminosities of galaxies as they evolve off the star-forming ridgeline and into the green valley; these show a remarkably similar pattern at all redshifts in which $L_{\rm X}$ rises by a factor of 2\text{-}3 compared to the star-forming main sequence \citep[Figure 10c]{Aird19}. If X-ray luminosity accurately tracks BH accretion (but see below), these data suggest high BH growth rates in the green valley for all quenching galaxies since $z = 4$. Rapid BH growth in the GV based on X-ray fluxes has also been deduced by \citet{Delvecchio19}, and a similar picture in which black holes accrete most of their mass just before quenching has been modeled by \citet{Caplar18}.

To summarize, in the binding-energy model, $\Sig1$ virtually stops growing as galaxies enter the green valley whereas $\Mbh$ grows considerably.  This is the behavior captured by the steep green tracks in Figure~\ref{Fig:Cartoon}.  The different tracks from left to right represent differences in $\Reff$ during the star-forming phase: galaxies with smaller initial $\Reff$ enter the green valley sooner at lower values of $\Sig1$ and $\Mbh$ and vice versa.  As the quenching boundary moves down with time, the range of green vectors being populated at any redshift also moves down, as shown by the moving green parallelogram.

The late rapid BH growth depicted in the green valley in Figure~\ref{Fig:Cartoon} resembles BH growth in the following hydro models: RAMSES/Horizon-AGN \citep{Dubois15,Habouzit17,Dekel19}, EAGLE \citep{Bower17}, FIRE \citep{Angles17}, IllustrisTNG \citep{Habouzit19}, and MARVEL \citep{Bellovary19}.  BH growth in these models is retarded during star formation due to SNae feedback, wandering motions of BHs in shallow central potential wells, and low central gas densities. Retardation ends when SNae are no longer able to evacuate the central regions (for various reasons) and when growing central stellar surface density captures the black hole and concentrates the gas around it.  These conditions will also occur in the binding-energy model, since declining star formation at entry to the GV means less SNae feedback, and higher $\Sig1$ means a deepening central potential well.  In short, the steep BH growth rates in the green valley in the binding-energy model find some support from similarly-growing black holes near the end of star formation in hydro models.

The late growth of black holes implies that BH masses lie \emph{below} the blue line in Figure~\ref{Fig:Cartoon} for most of the star-forming phase, crossing it only later and approaching it from below.  Possible tracks are schematically indicated by the steep dashed green lines below the blue line in that figure.   This scenario has important implications, as it changes the blue line from an evolutionary track into a \emph{boundary}  that is crossed by galaxies that are moving upward in Figure~\ref{Fig:Cartoon} into the green valley.  This would  profoundly alter the meaning of the $\Mbh \sim \Sig1^v$ scaling law, as discussed further in Section~\ref{SubSec:BHGrowth}.

In the above, we have interpreted both loci in Figure~\ref{Fig:Elbow} as evolutionary tracks that galaxies traverse while evolving from the blue cloud through the green valley and onto the red sequence.  Panel b, showing black-hole growth, is a re-rendering of data from \citet{Terrazas17}.   However, these authors favored a different interpretation in which green-valley galaxies can remain for long periods of time in a quasi-equilibrium state, with BH feedback regulating their star-formation at a fairly constant level. We favor the more rapid evolutionary track interpretation because of the observed net flow of galaxies from the blue cloud to the red sequence over cosmic time (\citealt{Bell04, Faber07, Huang13, Muzzin13, Ilbert13, Tomczak14,Straatman15}; \citetalias{Fang18}). On the other hand, the Terrazas sample probably does contain galaxies that quenched at different times, and hence this locus is not the track of any single object.  If the quenching boundary has been moving down over time, earlier-quenched galaxies would lie too far right in this diagram, and the amount of BH growth would look too large.  The total amount of the boundary motion was estimated to be $\sim$0.5 dex in \Mbh\text{-}\Mstar\ from $z = 3$ to now in  Section~\ref{SubSec:Successes}. Even if this full correction were applied, we would still have a net BH growth factor across the GV of 1.0~ dex, or $\times10$.  We take this as our estimate of the  \emph{minimum} amount of BH growth in the GV in what follows.

We can now use this growth factor to estimate how much energy BHs emit during their stay in the GV.  Galaxies enter the green valley when their effective black hole energy equals $f_{\rm B}$ $\times$ their halo-gas binding energy, where $f_{\rm B}$ is set to 4 to match the zero points of the blue lines to the red lines in Figure~\ref{Fig:Mstar_Sig1_SMA1}.  We assume for the sake of argument that $\Mstar$ grows like $\Sig1$ through the GV (0.2~dex) and that $\Mvir$ grows similarly, galaxies being near the knee of the SMHM relation at this point, where the log slope is near unity  (Figure~\ref{Fig:HMSM}).  Then $\Eth \sim \Mvir^{5/3}$ grows by $5/3 \times 0.2 = 0.3$~dex (Eq.~\ref{Eq:GasBindingEnergy}), and the ratio $\Ebh/\Eth$ grows by at least $1.0-0.3 = 0.7$~dex, which is a factor of 5.  This together with $\Ebh/\Eth  = 4$ on entry to the green valley then implies that $\Ebh/\Eth  \gtrsim 20$ when galaxies come to a halt on the quenched side.  The factor $\times$20 has margin to accommodate as-yet-unknown radiative losses and other inefficiencies and is plausibly large enough to complete quenching.  If green valley vectors are indeed universal, this new quantity $\Ebh/\Eth \sim 20$ at arrival on the quenched ridgeline is yet another constant of quenching.

The foregoing discussion shows that postponing major BH growth to the green valley has important consequences for how BHs accrete their gas, generate AGNs, produce feedback, and impact their environment.  We therefore need to review the evidence carefully, especially since the \citet{Terrazas17} sample of black holes is rather small, highly biased to massive galaxies (as shown in Figure~\ref{Fig:ThreePanel}c), and reflects only BHs today.  We start by listing seven different pieces of evidence that do favor uniform and steep black hole growth in the green valley over extended cosmic times:

1) Rodgr\'iguez-Puebla et al.~(in prep.) have assembled a sample of reliable BH masses that is considerably larger than the sample in \citet{Terrazas17}.  Using $\Delta\log SSFR$ to locate a galaxy's position in the green valley, as predicted by Figure~\ref{Fig:Cartoon}, they derive a single, unified scaling law that fits all BH masses.  The coefficient of  $\Delta\log SSFR$ in the new law is consistent with the slope and length of the GV vectors in Figure~\ref{Fig:Cartoon}.

2) The zero points of the $\Mbh\text{-}\sigmae$ and $\Mbh\text{-}\Mstar$ relations for star-forming galaxies are roughly 1~dex below those of quenched galaxies (\citetalias{Kormendy13}; \citealp{Ho14,Reines15,Savorgnan16,Terrazas16,Davis18}).  These offsets make sense if it is assumed that black holes grow steeply in the green valley as galaxies pass between the two relations.

3) A long-standing puzzle has been the very different black hole masses of the Milky Way ($4\times10^6~\Msol$) and M31 ($1.5\times10^8~\Msol$) \citepalias{Kormendy13} despite the small difference in their stellar masses of only 0.2-0.3~dex \citep{Sick14,Licquia15}.  No standard BH scaling law has been able to fit both galaxies.  The two objects are plotted in Figure~\ref{Fig:Cartoon}.  M31 lies close to the quenched locus while the Milky Way lies slightly below the star-forming locus, and the slope of the line connecting them is roughly parallel to the adopted growth vector in the green valley.  Thus, most of the discrepancy between the two galaxies  would be resolved if the Milky Way is just entering the green valley and M31 is just exiting it.  This is consistent with the fact that \citetalias{Kormendy13} classify the Milky Way as a pseudo bulge (i.e., gas-rich and star-forming) but classify M31 as a classical bulge.

4) At fixed $\Mstar$, AGN frequency rises strongly in the green valley.  This is true for SDSS galaxies \citep[e.g.,][]{Schawinski07,Schawinski10,Guo19} and also for CANDELS galaxies \citep[e.g.,][]{Nandra07,Gu18,Aird19}.

5) \citetalias{Rodriguez17} used abundance matching to measure evolutionary tracks for $\Sig1$ and $\Mbh$ back in time.  Rodr\'iguez et al.~(in prep.) now find, using similar methods, that black holes in quenched galaxies follow evolutionary tracks that look much like the broken power-laws in Figure~\ref{Fig:Cartoon}.

6) In Section~\ref{SubSec:PredictionsTests}, we show that steep BH growth through the green valley (like that in Figure~\ref{Fig:Cartoon}) can explain the different BH scaling-laws measured by different authors for different galaxy samples, including the very steep slope of the BH ``fundamental plane'' found by \citet{vandenBosch16} when fitting all galaxies.

7) Lastly, there is the theoretical justification from the hydro simulations cited above, which predict that black hole growth is unleashed when bulges reach a threshold size, resulting in steep growth trajectories similar to the green vectors in Figure~\ref{Fig:Cartoon}.

On the other hand, not all evidence supports sufficiently rapid BH growth in the green valley to satisfy the model.  Most troubling are the X-ray AGNs analyzed by \citet{Aird19}, which show only a modest increase in X-ray luminosity as galaxies enter the GV ($\times$2-3).  If black holes grow by at least $\times$10 in the GV, they accrete 10\% of their mass on the main sequence and 90\% of their mass in the GV (assuming no further growth after quenching).  If main sequence and GV lifetimes are similar (which if anything probably overestimates time in the GV), average X-ray fluxes should be roughly an order of magnitude higher in the GV, not the factor of $2\text{-}3$ observed.  In short, AGNs should not only be prevalent in the GV, they should also be \emph{very luminous X-ray sources}, which is not seen.

There appear to be only two ways out.  First, the bulk of the X-ray flux might be obscured.  Indeed, the sample studied by \citet{Aird19} has not been corrected for Compton-thick sources with $N_{\rm H} > 10^{24}$~cm$^{-2}$ \citep{Aird18}. Obscured AGN are reviewed by \citet{Hickox18}, who conclude that their number could be large.  The second option is that BH mass accretion in the GV is radiatively inefficient, i.e., that black holes can accrete much mass without shining in X-rays.  This is thought to occur in ADAFs \citep{Yuan14}, but their mass accretion rates are low whereas our green valley sources would be accreting fast.  The point is that a \emph{switch in mode} would need to occur as galaxies enter the green valley, whereby the ratio of radiative energy per unit BH mass accreted would need to decline. Whether such a change, or obscuration, or both together can solve the problem of missing X-ray flux in the GV is unknown.  Right now, missing X-ray flux in the green valley appears to be the biggest challenge to the binding-energy model.

Finally, we note that the tight relationship between black hole mass and halo mass that exists at the end of quenching, where $\Ebh \sim 20\times\Eth$, must break down later as the masses of central galaxies and their BHs stagnate while halos continue to grow.  The BH-halo relationship in different stages of a galaxy's lifetime is discussed in the next section.

\subsection{The Black-Hole/Dark Halo Connection}\label{SubSec:BHsandHalos}

The nature of the link between black holes and halos is controversial.  A tight correlation between $\Mbh$ and $\Mvir$ was claimed by \citet{Ferrarese02}, \citet{Baes03}, and \citet{Bandara09} but was disputed by \citetalias{Kormendy13} because neither very large nor very small halos fit their relations. The binding-energy model sheds further light.

The first point is that the different scaling laws in Figure~\ref{Fig:Cartoon} clearly imply different correlations for different kinds of galaxies.  The equation $\Sig1 \sim \Mstar^{0.88}$ is the observed ridgeline slope of star-forming galaxies at fixed epoch in $\Sig1\text{-}\Mstar$ \citepalias{Barro17}.  Converting this to BH mass using $\Mbh \sim \Sig1^{1.76}$ yields $\Mbh \sim \Mstar^{1.55}$.  This is steeper than the ridgeline for quenched galaxies ($\Mbh \sim \Mstar^{1.16}$), lies below it, and should show more scatter because star-forming galaxies have different  $\Reff/\ReffMean$, and therefore different BH masses at fixed $\Mstar$.  These predictions agree tolerably well with observations (\citetalias{Kormendy13}; \citealp{Reines15,Savorgnan16,Terrazas16,Davis18,Delvecchio19}).  Finally, using $\Mstar \sim \Mvir^{1.75}$ for star-forming galaxies predicts $\Mbh \sim \Mvir^{2.71}$, but with large scatter due to $\Reff$ variations.

For quenched galaxies, the derived slope depends on whether the quenched galaxies are in halos above or below the knee of the SMHM relation, which is at $\Mstar = 10^{10.7}~\Msol$ (Figure~\ref{Fig:HMSM}).  Since most quenched ridgeline galaxies are more massive than this \citepalias{Kormendy13}, we start with the quenched ridgeline slope $\Mbh \sim \Mstar^{1.16}$ and use the SMHM relation $\Mstar \sim \Mvir^{0.35}$ above the knee to obtain $\Mbh \sim \Mvir^{0.40}$ for quenched galaxies, i.e., dramatically shallower than the star-forming slope of 2.17.  The main reason for the difference is the flattening of the SMHM relation above the knee at $\Mvir \sim 10^{12.5}~\Msol$.  The net result is that BH mass is predicted to rise strongly in star-forming galaxies with halo mass below $\Mvir = 10^{12.5}~\Msol$ but much more slowly above that, as BHs stagnate but halos continue to grow.  Hence, no single power-law relation is expected between BHs and their halos, and the scatter among star-forming galaxies (and in the green valley) should be large.

In contrast, \citet{Ferrarese02} found a single tight relation for all galaxies with slope 1.65-1.82 by estimating $\Mvir$ from the outer rotation curves of spirals in the range $\log \Mvir = 11.5\text{-}12.5$ and the outer kinematic profiles of ellipticals \citep{Kronawitter00} in the range $\log \Mvir = 12.5\text{-}13.5$.  \citet{Baes03} found a single slope of 1.27 from a larger sample of spirals but the same ellipticals from \citet{Kronawitter00}.  Finally, \citet{Bandara09} found a slope of 1.55 using strong lensing to estimate halo masses of mainly ellipticals in the range $\log \Mvir = 13.0\text{-}13.8$.  None of these results agrees with the predictions, the slopes being too uniform and equal to some compromise value between the two predicted extremes of 0.40 to 2.71.

The steep slopes found for the massive end in these papers clearly disagree with elementary estimates.  Large spirals like M31 have $\Mbh$'s a few times $10^8~\Msol$ and dark halos a few times $10^{12}~\Msol$, while large clusters like Coma have central $\Mbh$'s a few times $10^9~\Msol$ but dark halos a few times $10^{15}~\Msol$.  The differential log slope is about 1/3, in good agreement with the predicted binding-energy slope of 0.40 but vastly smaller than the measured slopes of 1.27-1.82 from these authors.  Similar points were made by \citetalias{Kormendy13} and \citet{Delvecchio19}.  The problem may be trying to estimate the total mass of huge dark halos from data that do not extend far enough from the center -- the resulting values of $\Mvir$ are too small, and the calculated slopes of $\Mbh$ \vs $\Mvir$ are too steep.

To summarize, it is not true that there is \emph{no} relation between the mass of a halo and its central BH.  Indeed, it is a fundamental tenet of the binding-energy model that halos build galaxies, which in turn predictably build BHs.  However, the resulting slopes and scatter are very different for quenched and star-forming galaxies.  This chain of reasoning answers the fundamental question posed by \citetalias{Fang13}, who guessed that a close coordination between halo and BH is needed in order to produce a narrow quenched $\Sig1\text{-}\Mstar$ ridgeline but wondered how the halo ``knows'' what the BH is doing. The answer is that the halo \emph{makes} the black hole during the star-forming phase and then is quenched by it when $\Ebh \sim 20\times \Eth$ (Section~\ref{SubSec:BoundaryCrossing}), closing the loop.  Black-hole mass differences due to $\Delta \log \Reff$ cause extra scatter about the $\Mbh\text{-}\Mvir$ relation for star-forming and green valley galaxies but do not fundamentally change this basic picture.

\subsection{Black Hole Scaling Laws and Fundamental Planes}\label{SubSec:PredictionsTests}

This section returns to the problem mentioned in Section~\ref{SubSec:QuenchSlope} that the value of $v$ in the model seems too small.  Specifically, our value $v = 1.76$ implies that $\Mbh \sim \sigmae^{3.52\pm0.26}$ (if $\sigma_1 \sim \sigmae$) whereas the observed relation (for classical bulges and ellipticals in \citetalias{Kormendy13}) is $\Mbh \sim \sigmae^{4.4\pm0.3}$.  This suggests that $v$ is too small by the factor $4.4/3.52 \simeq 1.25 \pm 0.12$.  \citet{vandenBosch16F} fitted all galaxies together and found an even steeper law $\Mbh \sim \sigmae^{5.35\pm0.23}$, for a difference with us of $5.35/3.52 = 1.5\pm 0.13$.  We don't have measurements of $\sigmae$ in the CANDELS database, but we do have measurements of $\Mstar$ and $\Reff$, which were also fitted by \citet{vandenBosch16}.  He found $\Mbh \sim \Mstar^{2.91\pm0.14}\Reff^{-2.77\pm0.22}$ from fitting all galaxies, which is close to $\Mbh \sim \Sig1^{2.7\pm0.17}$ using the $p$,$q$ mapping.  This implies that $v$ is too small by the factor $2.7/1.76 \simeq 1.5\pm0.14$, i.e., the same as $\sigmae$. (Note: this last also implies that $\Mstar/\Reff \sim \sigmae^2.0$.)  The goal of this section is to understand these slope differences.

We have taken advantage of these rough correspondences between $\Sig1$, $\sigma_1$, $\sigmae$, and $\Mstar/\Reff$ to label the X-axis in Figure~\ref{Fig:Cartoon} with all four quantities, which allows us to crudely compare slope results from different data sets.  We stress that $v$ is our slope for star-forming galaxies only (blue line in Figure~\ref{Fig:Cartoon}), whereas \citet{vandenBosch16} lumps all galaxies together and fits them with a single power law.  Being an average over both the shallow and steep portions of the tracks, his slope should be steeper than the blue track in Figure~\ref{Fig:Cartoon}.  A rough estimate is shown by the black dashed line in Figure~\ref{Fig:Cartoon}, which is schematically fit to  all galaxies.  The slope of the black line is 2.6, very close to the \citet{vandenBosch16} value of 2.7 above.  The construction is crude but illustrates clearly how major BH mass growth through the GV, accompanied by little or no growth in $\Sig1$, can produce a big increase in the average slope of $\Mbh$ \vs $\Sig1$ (or $\sigma$, or $\Mstar/\Reff$) when all galaxies are fitted together.

The difference between our slope and that of \citetalias{Kormendy13} for E's and classical bulges is only 1.25 but should be 1.00 if their sample is pure.  However, at the low-mass end, some of the classical bulges in \citetalias{Kormendy13} may actually be in the green valley, and at the high-mass end, \citet{Lauer07} have argued that $\Mbh\text{-}\sigma$ has been steepened due to galaxy merging.  Contamination by both of these effects would steepen the slope in the direction seen. (We note that the derivation of $t$ and $v$ in Section~\ref{SubSec:QuenchSlope} deliberately chose the  $\Mbh\text{-}\Mstar$ relation, not $\Mbh\text{-}\sigma$, for quenched galaxies, precisely because it is less sensitive to merging.)

The discussion above has focused on scaling laws \vs one variable, typically $\Sig1$ or  $\sigma$.  However, the close relation of these quantities to $\Mstar/\Reff$ has led some authors to fit using two variables from among $\Mstar$, $\Reff$, and $\sigmae$.  These expanded correlations are often referred to as the BH fundamental plane (BHFP).  Van den Bosch's planar fit to all galaxies yielding $\Mbh \sim \Mstar^{2.91}\Reff^{-2.77}$ is an example.  However, he noted his opinion that the fundamental variable is really $\sigmae$ and that $\Mstar$ and $\Reff$ are stand-ins for it.  In that sense, the BH ``fundamental'' plane is not as fundamental as the relationship with $\Sig1$ (or with $\sigma$).

Separately, \citet{Hopkins07a} fitted a plane to quenched ridgeline galaxies only and found $\Mbh \sim \Mstar^{1.78 \pm 0.40} \Reff^{-1.05 \pm 0.37}$.  As mentioned in Section~\ref{SubSec:InputPowerLaws}, the  $p$, $q$ mapping for quenched galaxies is empirically $\Sig1 \sim \Mstar^{1.0} \Reff^{-0.5}$.  Our model then predicts $\Mbh \sim \Sig1^{1.76} \sim \Mstar^{1.76} \Reff^{-0.88}$ on the quenched ridgeline, which agrees with the Hopkins exponents within their errors.  We therefore propose that the BHFP for quenched galaxies, like the plane of \citet{vandenBosch16} for all galaxies, is really a stand-in for $\Sig1$ (or $\sigmae$).  Similar calculations (not shown) can explain the slopes of BHFPs fitted to $\sigmae\text{-}\Reff$ and $\sigmae\text{-}\Mstar$ by \citet{Hopkins07a} and by \citet{Aller07}.

Yet another expression offered for the quenched BHFP is $\Mbh \sim (\Mstar \sigmae^2)^{\alpha}$, where \citet{Hopkins07a,Hopkins07b} find $\alpha \simeq 0.7$ and \citet{Aller07} find $\alpha \simeq 0.6$. The quantity $\Mstar \sigmae^2$ is proportional to \emph{galaxy} binding energy per particle, which these authors proposed as the property that regulates BH mass.  This was motivated by the merger scenario of \citet{Hopkins07b}, in which gas that is brought to the center of the galaxy builds the black hole until the BH energy unbinds the surrounding gas.  In spirit, this is similar to the unbinding criterion in Eq.~\ref{Eq:EbhEbind}, but we use gas-binding energy to the halo, $\Mvir\Vvir^2 \sim \Mvir^{5/3}$, whereas Hopkins-Aller use gas-binding energy to the galaxy.

A problem noted in both works is that $\alpha$ ought to be 1.0 in a simple binding energy argument, not 0.6-0.7 as seen.  However, the binding-energy model again sheds light.  From  Section~\ref{Sec:ThermalModel} halo gas-binding energy $\Eth \sim  \Mvir\Vvir^2 \sim \Mvir^{5/3}$.  Setting $\sigmae^2 \sim \Mstar/\Reff$, noting that $\Reff \sim \Mstar^{0.44}$ along the quenching boundary, and substituting $\Mstar \sim \Mvir^{1.75}$ on the SMHM relation yields $\Mvir^{5/3} \sim (\Mstar \sigmae^2)^{0.61} \sim \Eth$.  In other words, the unexpected exponent with the value 0.6-0.7 in the model of \citet{Hopkins07b} and \citet{Aller07} is simply that needed to convert halo gas-binding energy into galaxy gas-binding energy!

In summary, fitting BH scaling relations to different samples has historically yielded slopes that differ by more than their formal errors.  This is true whether single-variable or planar fits are used.  The broken power-law model in Figure~\ref{Fig:Cartoon} helps to explain this by positing different scaling laws (and $\Reff$ scatter) for different evolutionary phases.  Fitting any single law to the entire population gives an average.  Better fits can be obtained by fitting each evolutionary stage separately or by including a new term in $\Delta\log SSFR$ to reflect a galaxy's location in the green valley (see Item 1 in  Section~\ref{SubSec:BoundaryCrossing}).  Finally, it appears that planar fits to two variables are really stand-ins for fits to $\Sig1$ (or $\sigmae$), which are the fundamental quantities that describe $\Mbh$.

\subsection{Black Hole-Galaxy ``Co-Evolution''; the Origin and Meaning of BH Scaling Laws}\label{SubSec:BHGrowth}

Because of its highly transparent nature, the binding-energy model is a good tool for  probing the nature of so-called BH-galaxy ``co-evolution''.  In our opinion, this word has been used too loosely by astronomers simply to signify that the properties of two objects (e.g., black holes and galaxies) are correlated.  What we really want to know is the nature and arrow of causality: does A cause B, does B cause A, are A and B both caused separately by C, or do A and B reciprocally influence one other throughout time?  The dictionary defines co-evolution as ``the \emph{reciprocal influence} of one evolving entity on another evolving entity'', and we adopt that standard here.

It is then illuminating to compare the BH growth processes in the binding-energy model to those in other models, and detailed descriptions of BH growth mechanisms exist in the literature for the following hydro simulations: RAMSES/Horizon-AGN \citep{Dubois15, Habouzit17, Dekel19}, EAGLE \citep{Bower17,Davies19a,Oppenheimer19}, and FIRE \citep{Angles17}.  We do not include IllustrisTNG because it uses two different BH feedback modes (thermal-quasar mode and kinetic-radio mode;  \citet{Weinberger18,Habouzit19,Nelson19}), which would complicate the discussion.

Three different evolutionary phases are considered:

1) \emph{BH growth on the star-forming main sequence:} In the \emph{binding-energy model}, all star-forming galaxies evolve along the BH growth law $\Mbh = \rm{const.}\times\Sig1^{1.76}$ (blue line in Figure~\ref{Fig:Cartoon}), which is an evolutionary track. Growth is steady on average but may be episodic on short timescales. Since $\Mstar$ and $\Reff$ determine \Sig1\ through the $p$,$q$ mapping,  BHs are rigidly linked to their host galaxies during this phase. If halos also determine $\Reff$ (through halo mass and concentration, for example), black holes are rigidly linked to their halos as well.  In short, the properties of  galaxies are determined by halos and the properties of black holes are in turn determined by galaxies, but there is no reciprocity at either level and so, strictly speaking, there is no co-evolution.

In the above-cited \emph{hydro models}, BH growth during the star-forming phase is severely limited by bursty stellar feedback, which periodically ejects gas from the center.  Growth overall is episodic and inefficient. Instantaneous BH growth takes place by Bondi accretion (RAMSES, Horizon-AGN, EAGLE) or is torque-limited (FIRE), but the time-averaged rate is more likely to be limited by the rate at which gas refills in the central regions.  The central stellar density is low, the central potential is not very deep or well-focused, and BH wandering is strong \citep{Dekel19,Bellovary19}, which may further reduce BH gas accretion.  BHs are typically started with seed masses of around $10^5~\Msol$, and growth beyond that in this phase is typically only $\times3$ even though galaxy masses increase by up to $\times100$.  Owing to the different physical processes governing central BHs and host galaxies, their properties during the star-forming phase are not closely linked, and we conclude that black holes and galaxies do {not} co-evolve -- they are not even closely correlated.

2) \emph{Entry to the green valley; transition to rapid BH growth:} In the \emph{binding-energy model}, both of these phenomena occur together, when the BH energy output passes $4 \times \Eth$.  The BH has been growing steadily and is just beginning to perceptibly reduce the ability of gas in the halo to fall onto the galaxy.  Keying the location of this transition to a fixed fraction of halo binding energy closely matches the observed, sloping quenching boundaries (blue vs. red lines in Figure~\ref{Fig:Mstar_Sig1_SMA1}). Galaxy, halo, and black hole properties are perfectly correlated at the transition, as in the previous star-forming phase, but now black holes are back-reacting on their halos and galaxies. This reciprocal influence satisfies the definition of co-evolution, and so black holes, galaxies, and their halos \emph{are co-evolving} as galaxies enter the green valley.

In the cited \emph{hydro models}, the transition to rapid BH growth occurs \emph{before} entry into the green valley, when the stellar-feedback-driven wind fails.  The stated reason is different in the different models (failure to reach escape velocity \citep{Dekel19}, winds no longer buoyant \citep{Bower17,McAlpine18}, major mergers \citep{McAlpine18}, deepening central potential well \citep{Angles17,Dekel17}, but the net result is that gas starts to collect in the center of the galaxy, unleashing BH growth.  This transition tends to occur in a small range of bulge mass near $10^{9\text{-}10}~\Msol$, a halo mass near $10^{12}~\Msol$,  or a fixed halo virial temperature $\Tvir$ near $10^{6{\text-}7}$~K, all of which are plausibly related to the kinetic energy content of supernovae winds.  Since this is a fixed number \citep{Dekel86}, the theory tends to pick out a single stellar or halo mass where BH growth is launched, thus creating a momentary correspondence between the black hole, galaxy, and halo but not true co-evolution since there is no reciprocity a this point.

3) \emph{BH growth in the green valley; final transition to fully quenched:} In the \emph{binding-energy model}, the BH growth trajectory \vs $\Sig1$ in the green valley steepens.  The log slope of the blue track in Figure~\ref{Fig:Cartoon} is 1.76, whereas the log slope of the green GV vectors is 1.5 dex/0.2 dex = 7.5 (for these numbers, see Figure~\ref{Fig:Elbow}; with minimum BH growth of 1.0 dex in the green valley, the ratio would be 5.0).  The steep tracks needed by the model closely resemble the steep growth tracks in the hydro simulations, but the detailed physics shaping those tracks is not understood.  Meanwhile, global gas content and star formation are falling due to AGN feedback.  A reasonable inference, then, is that black hole, galaxy, and halo are continuing to ``talk'' to one another, and \emph{this co-evolution is plausibly what makes all green valley vectors parallel and of the same length}.  Ultimately, galaxies arrive on the fully quenched ridgeline when $\Ebh/\Eth \simeq 20$, which enforces absolute correspondence between halos and black holes again at that point.  After that, halos continue to gain mass and drift away from both black holes and galaxies, and co-evolution ends (except possibly for ``maintenance mode'').

In \emph{hydro models}, it is less certain what sets the rate of BH growth in the green valley or the moment of final quenching.  The picture is perhaps clearest for EAGLE \citep{Davies19a,Oppenheimer19}, in which fractional halo gas mass, $f_{\rm CGM}$, appears to be a barometer of the degree of quenching.  According to this parameter, entry into the green valley corresponds to $\Ebh/\Eth \simeq  2$, and complete loss of halo gas corresponds to $\Ebh/\Eth \simeq  20$ \citep[][Figure 2b]{Oppenheimer19}.  These numbers are not far from the binding-energy model, which gives 4 and 20 respectively.  On that basis, it is plausible that co-evolution proceeds in the green valley for EAGLE much as it does in the binding-energy model.  A similar case can be made for IllustrisTNG, where $\Ebh/\Eth$ again tends to be a fixed factor at full quenching \citep{Terrazas19}.

Based on the above, we see that real co-evolution does occur in the binding-energy model, but only in the green valley, where it plausibly is the agent that produces growth vectors of similar length and slope for all galaxies (cf.~Figure~\ref{Fig:Cartoon}).  But the ultimate uniformity of quenched galaxies also depends on the uniformity of galaxies \emph{as they enter the green valley}, which is impressed in our model by the initial scaling law $\Mbh = \rm{const.}\times\Sig1^{1.76}$ for star-forming galaxies.  Thus, in the binding-energy model, the final black-hole regularities of galaxies on the quenched ridgeline arise from the combination of a reciprocal, co-evolving phase in the green valley that builds on a uni-directional, causal phase for star-forming galaxies.

This insight motivates a new look, in our model, at the nature of the two $\Mbh \sim \Sig1$ scaling laws, for star-forming and quenched galaxies separately.  In the star-forming phase, $\Sig1$ determines $\Mbh$ single-handedly according to $\Mbh = \rm{const.}\times\Sig1^{1.76}$, and the scaling relation is an \emph{evolutionary track} that all galaxies must follow (blue line in Figure~\ref{Fig:Cartoon}). Eventually, galaxies arrive on the quenched ridgeline, and a second scaling law for $\Mbh$ \vs $\Sig1$ emerges having the same slope as the blue line but higher zero point (red line in Figure~\ref{Fig:Cartoon}).  However, despite the mathematical similarity of the second law to the first, its interpretation is entirely different -- galaxies do not evolve along this second line, and it is not an evolutionary track.   Rather, it is the \emph{boundary} where $\Ebh/\Eth \simeq 20$, and $\Sig1$ along this line \emph{labels the paths that galaxies took through the green valley to get there}.

In summary, some scaling laws may represent unique evolutionary tracks but others may be boundaries linking points on \emph{different} tracks where a third quantity takes on a common value.  The scaling law for quenched galaxies in our model is of the second sort.  It is not a track like the blue line, and galaxies do not evolve along it;  rather, it connects galaxies where $\Ebh/\Eth$ has the common value $\simeq20$.

With that as background, we now explore briefly whether the blue line might itself be a boundary rather than an evolutionary track.  Up to now, we have regarded the blue line as a track that all star-forming galaxies must evolve along while building their black holes.  But actually, BHs need not follow this track for their entire lives -- they only need to obey the requisite relations \emph{as they enter the green valley}.  The blue line would then express the relationship between $\Mbh$ and $\Sig1$ that needs to obtain only there.

This change of perspective allows us to address a discrepancy that we have thus far glossed over, which is a disagreement between the binding-energy model and state-of-the-art theory.  As stressed in Section~\ref{SubSec:InputPowerLaws}, a generally accepted \emph{analytical} explanation of the BH scaling laws is thus far lacking, which means that the best theory comes from hydro models.  However, these models disagree with the binding-energy model in that their black holes in star-forming galaxies lie ``dormant'' and evolve rapidly only later, approaching the blue line \emph{from below} (see the dashed green lines in Figure~\ref{Fig:Cartoon}). Galaxies no longer evolve along the blue line -- instead, it becomes a \emph{boundary scaling law} linking together different entry points to the green valley along different evolutionary tracks.

This new interpretation has several implications.  First, it means that there is no relation like $\Mbh = \rm{const.}\times\Sig1^{1.76}$ for star-forming galaxies or, if there is, it is much broader and fuzzier than what we've assumed.  The zero point of the blue line was chosen to fit nearby star-forming galaxies with \emph{detected} black holes.  The new interpretation says that many star-forming galaxies with the same $\Mstar$ and $\Sig1$ have even smaller black holes.  In other words, the locus of measured black holes in star-forming galaxies, as in \citet{Ho14} and \citet{Reines15}, is only the \emph{upper envelope} of a much larger population with smaller black holes below this locus.  This conclusion agrees with estimates that the measured BH relation is biased high for star-forming galaxies, having missed a population of small black holes \citep{Shankar16}.  It would also explain the large scatter in the $\Mbh \sim \sigmae^4$ law that is seen among star-forming  galaxies  (\citetalias{Kormendy13}, \citealt{Ho14}) and the anomalously small BHs found in some of them \citep[e.g., M33,][]{Gebhardt00}.

With the abandonment of the blue line as an evolutionary track, a new theory for the nature of the boundary to the green valley is needed.  One possibility is the model of AGN winds by \citet{King03} and \citet{King15}, which says that the structure of quasar outflows fundamentally changes at a certain black hole mass.  According to this theory, winds change from being momentum-driven to energy-driven when
\begin{equation}\label{Eq:KingPounds1}
\frac{M_{\rm BH}}{10^8~\rm{M_{\odot}}}
\simeq 3.2 \times \left(\frac{f_{\rm g}}{0.16}\right) \sigma_{200}^4.
\end{equation}
Here 0.16 is the cosmic baryon-to-dark matter ratio and $f_{\rm g}$ is the total gas mass fraction in the central galaxy potential well (which is assumed to be isothermal with velocity dispersion $\sigma_{200}$).  This change makes winds much more efficient at driving gas out of the galaxy and halo.  Combining this with Eq.~\ref{Eq:Sig1_sig1} from Appendix~\ref{App:A3} gives
\begin{equation}\label{Eq:KingPounds2}
\frac{M_{\rm BH}}{10^8~\rm{M_{\odot}}}
\simeq 0.09\times \left(\frac{f_g}{0.16}\right) \left(\frac{\Sig1}{10^9~\Msol~{\rm kpc^{-2}}}\right)^{2.0}.
\end{equation}
This can be compared to the equation of the blue line in Figure~\ref{Fig:Cartoon}, given by Eq.~\ref{Eq:Mbh_Sig1} in Appendix~\ref{App:A3}, which is:
\begin{equation}\label{Eq:KingPounds3}
\frac{M_{\rm BH}}{10^8~\rm{M_{\odot}}} = 0.025\times\left(\frac{\Sig1}{10^9~\Msol~{\rm kpc^{-2}}}\right)^{1.76}.
\end{equation}
If the small difference between the two exponents can be ignored, we see that the King-Pounds expression will lie on or near the blue line provided
\begin{equation}\label{Eq:KingPounds4}
\frac{f_g}{0.16} = \frac{0.025}{0.09} \sim 0.3,
\end{equation}
or
\begin{equation}\label{Eq:KingPounds5}
{f_g}\sim 0.04,
\end{equation}
which is a reasonable value for the gas mass fraction at the center of a galaxy that is building a black hole.

The King-Pounds theory therefore predicts a wind-enhancement boundary in the vicinity of the blue line.\footnote{This point was also made by \citet{King03} and \citet{King15}.}  Hydro models approaching this boundary from below would experience a large increase in wind efficiency at this point such that feedback energy from the BH is suddenly transferred more efficiently to the halo gas.  With higher feedback, quenching becomes perceptible, and galaxies enter the green valley along the steep green tracks in Figure~\ref{Fig:Cartoon}.

We therefore have an \emph{analytic} theory that predicts a boundary in the right place at the entrance to the green valley, together with a plausible way of feeding that boundary (from below) as in the hydro simulations.  However, a major ingredient is still missing: the theory predicts a boundary in $\Mbh\text{-}\Sig1$ only, whereas matching quenching boundaries are also seen in $\Mbh\text{-}\Mstar$  and $\Sig1\text{-}\Mstar$ (cf.~Figure~\ref{Fig:ThreePanel}bc).\footnote{The boundary in $\Reff\text{-}\Mstar$ in Figure~\ref{Fig:ThreePanel}a is not fundamental, being a mapping of the $\Sig1\text{-}\Mstar$ boundary into $\Reff\text{-}\Mstar$ via the $p$ ,$q$ relation.}  In other words, to match all the data, a galaxy needs to cross over the King-Pounds boundary in $\Mbh\text{-}\Sig1$ \emph{with the right $\Mstar$}.

In summary, the existence of a quenching boundary involving \emph{three} separate variables ($\Mbh$, $\Sig1$, and $\Mstar$) is a major constraint whose significance has not been appreciated until now.  The original binding-energy model met this constraint by making the $\Mbh\text{-}\Sig1$ relation a tight evolutionary track obeyed by \emph{all} star-forming galaxies, which removed one degree of freedom.  If the $\Mbh\text{-}\Sig1$ relation is instead a boundary that can be crossed in different places by galaxies with different $\Mstar$, a separate process is needed to align $\Sig1$ and $\Mbh$ with the proper value of $\Mstar$ at the crossing point.  Meeting this second constraint is a new test for hydro models.

\subsection{Comparison to the Model of Lilly et al. (2016)}\label{SubSec:L16}

We close this Discussion by returning to the model of Lilly et al.~(2016; L16), which was mentioned in the Introduction. This model is distinguished by the assumption that the quenching probability is a function of $\Mstar$ only, not of, say, $\Mstar$ and $\Reff$. Despite this, the model still manages to reproduce the sloping contours of equal quenched fraction in $\Reff\text{-}\Mstar$ by exploiting two factors: 1) galaxies that quench earlier have smaller $\Reff$ and thus lower the quenched ridgeline zero point at fixed $\Mstar$ (progenitor bias), and 2) $\Reff$ values are light-weighted and shrink upon fading in this model owing to lower-$M/L$ stars in their outskirts (also mentioned by \citetalias{Fang13}). These two effects together increase the fraction of quenched galaxies at smaller $\Reff$ at fixed $\Mstar$.

Despite this success, the model fails to explain other data:

1) The difference in $\Reff$ between quenched \vs star-forming galaxies in the model is smaller at high redshift than at low redshift (see \citetalias{Lilly16}, Figure 3), whereas the reverse is true in real data: quenched galaxies, particularly lower-mass ones, are \emph{much} smaller than star-forming galaxies at $z \sim 2$ than at later times (cf.~Figure~\ref{Fig:Mstar_Sig1_SMA1}). This is naturally accommodated in the binding-energy model because $\Reff$ is free to shrink more at early times due to compaction (Section~\ref{SubSec:Compaction}), whereas the \citetalias{Lilly16} model does not have this extra degree of freedom.

2) The \citetalias{Lilly16} model predicts a $\times10$ radial increase in SSFR from 1 to 10~kpc within star-forming galaxies of mass $10^{10.75}~\Msol$ at all redshifts, which is generated by the assumption of strong inside-out star formation. A gradient of this magnitude is needed in order to produce the required amount of $\Reff$ shrinkage upon fading (see above). Current data on distant galaxies do not support gradients of this magnitude within 10~kpc -- the SSFR profiles of most galaxies are flat out to this radius \citep{Wang17, Tacchella18,Woo19}.

3) The \citetalias{Lilly16} model does not include BHs and therefore does not explain the $\Mbh\text{-}\Mstar$ scaling relation of quenched galaxies. Quenching is assumed to occur with uniform probability per unit stellar mass growth, but no mechanism is offered to achieve this and the role of BHs in quenching is unaddressed.

4) Most important, the \citetalias{Lilly16} model has not yet attempted to match the $\Sig1\text{-}\Mstar$ diagram.  $\Sig1$ is likely to remain rather constant through and after quenching in the \citetalias{Lilly16} picture. But $\Sig1$ is set by the $p$, $q$ mapping for star-forming galaxies, and hence by $\Reff$ and $\Mstar$.  If quenching is equally probable for all galaxies at a given mass regardless of $\Reff$, the $\Sig1$ distributions of green valley and star-forming galaxies at a given $\Mstar$ should be the same.  Instead, the $\Sig1$ distributions of real green-valley galaxies are \emph{shifted to much higher values and are much narrower} than those of star-forming galaxies at all redshifts (\citetalias{Fang13,vanDokkum15,Barro17}; \citealp{Woo17,Lee18,Luo19}).  The key point is that GV galaxies are observed \emph{close to the beginning of quenching}, so that the evolutionary time difference between them and their star-forming progenitor galaxies should be small.  Even so, progenitor bias might induce a small offset in \emph{average} $\Sig1$, but it cannot explain the large difference in \emph{widths}.  Moreover, as a mass density, $\Sig1$ is not affected by fading. Thus, $\Sig1$ sidesteps both effects that the \citetalias{Lilly16} model uses to convert quenching contours that are intrinsically vertical at fixed mass into sloping contours in $\Reff$ \vs $\Mstar$.  The sloping behavior of $\Sig1$ \vs $\Mstar$ signals that a second physical variable is involved in quenching beyond $\Mstar$, which we have suggested is $\Reff/\ReffMean$.

Parenthetically, we note that \emph{all} quenching models that rely largely on stellar (or halo) mass meet the same objection, of which the obvious examples are the pure ``halo-quenching'' models \citep[e.g.,][]{Ostriker77,Blumenthal84,Cattaneo06,Dekel06,Zu15,Zu18}.  Being one-dimensional theories in which the only variable is mass, they inherently fail to reproduce the observed diversity of properties of galaxies at a given stellar mass.

\section{Summary}\label{Sec:Summary}

A schematic model is presented for the growth of BHs in \emph{central} star-forming galaxies and their role in causing central galaxies to quench. The simplicity of the model is useful for appreciating how various factors in galaxy evolution -- such as structural evolution, black-hole growth rules, and quenching criteria -- interact and how results depend on these inputs.  The impetus for the new model is the belief that the observed scaling laws of galaxies contain important clues to quenching that can be elucidated from a proper model with the right degrees of freedom.  The paper starts with a list of 18 separate observations for quenched and star-forming galaxies that a successful model should explain.  The final binding-energy model presented here explains 11 of these items fairly well and sheds useful light on the remaining seven.

The main feature that enables this success is a new degree of freedom in which larger star-forming galaxies at fixed mass have smaller BHs owing to their lower central densities, $\Sig1$.  A second key assumption is that galaxies enter the green valley when the total effective energy emitted by their central black holes equals some fixed multiple, $f_{\rm B}$, of the binding energy of the gas in their dark halos.  Together, these assumptions imply that large-$\Reff$ star-forming galaxies must evolve to higher stellar masses before AGN feedback terminates the gaseous infall from their halos, and they accordingly have higher values of $\Mbh$, $\Sig1$, and $\Reff$ once quenched.  In other words, the simple addition of an $\Reff$ dependence to BH mass explains the otherwise mysterious \emph{sloped} quenching boundaries in $\Sig1\text{-}\Mstar$, $\Reff\text{-}\Mstar$, and $\Mbh\text{-}\Mstar$ whereby more massive quenched galaxies have larger central densities, larger effective radii, and larger BHs. For star-forming galaxies, these three spaces are different projections of the same 4-D parent space, and mappings are presented to transform among them: $\Sig1 \sim \Reff^{-p}\Mstar^q$ is the transformation between $\Sig1$ and $\Reff$, where $p = 1$ and $q = 1.1$, while $\Mbh \sim \Sig1^v$ is the transformation between $\Sig1$ and $\Mbh$, where $v = 1.76$.

As long as galaxies are star-forming, they are assumed to evolve along log-linear parallel tracks in this 4-D space.  Each track is labeled with two numbers: the starting redshift, $z_{\rm s}$, when the galaxy passes $\Mstar = 10^9~\Msol$, and the factor by which its radius $\Reff$ differs from the mean $\ReffMean$ at that mass and time.  These two numbers make central star-forming galaxies a two-parameter family, which can also be labeled by $\Mstar$ (or equivalently, $\Mvir$) and $\Reff/\ReffMean$ today.  The physical origin of the scatter in $\Reff$ is not known, but halo concentration and/or halo formation time are promising candidates. If further tests support this, then $\Mvir$ and $c$ (or formation time) would be the two principal halo parameters that predict the evolutionary paths of central galaxies.

All star-forming galaxies in a given slice of $\Mstar$ are assumed to have the same halo mass, $\Mvir$, at all times, and the mean effective $\ReffMean$ in that slice is always $0.02 \times \Rvir$. The star-forming branch of the stellar-mass-halo-mass relation (SMHM) is represented by a power-law $\Mstar = {\rm const.} \times \Mvir^s$ with zero scatter, where $s = 1.75$ and the zero point is constant with time. Both slope and zero point are taken from abundance matching \citepalias{Rodriguez17}.  Another key assumption is that all galaxies in a single mass slice make stars at the same rate regardless of $\Reff/\ReffMean$, in agreement with observational data.   These assumptions, together with the physics of evolving halos, yield mean evolutionary track slopes and zero points for star-forming galaxies in $\Reff\text{-}\Mstar$ and $\Sig1\text{-}\Mstar$ that generally agree with observations.

Galaxies are assumed to enter the green valley when they pass over a boundary in $\Sig1\text{-}\Mstar$ (and matching boundaries in $\Reff\text{-}\Mstar$ or  $\Mbh\text{-}\Mstar$). The empirically observed boundary, shown as the red lines in Figure~\ref{Fig:Mstar_Sig1_SMA1}, is based on the well-measured quenched ridgeline in $\Sig1\text{-}\Mstar$, which is offset downwards by 0.2~dex in $\Sig1$ to represent entry into the green valley.  In the evolving empirical power-law model in Section~\ref{Sec:EvolPLModel}, this boundary is defined as the point where the total energy deposited in the halo by the black hole, $\eta \Mbh c^2$, equals some empirical energy quota needed to impact the halo and stop gas infall. To fit the observed quenching boundary slopes in $\Sig1\text{-}\Mstar$ and $\Mbh\text{-}\Mstar$ forces the energy quota to have the form $\Eheat = {\rm const.}\times h(z)^{0.66v} \Mvir^t$, where $h(z)$ is the normalized Hubble constant $H(z)/H(0)$ and $t = 2.03$. This prescription fits the red lines precisely, and the log-linear slope of the boundary in $\Sig1\text{-}\Mstar$ is $t/vs = 0.66$, in $\Reff-\Mstar$  it is $q-pt/vs = 0.44$, and in $\Mbh\text{-}\Mstar$ it is $t/s = 1.16$.

The above empirical expression for $\Eheat$ is then compared to $4\times$ the total binding-energy content of the hot halo gas, \Eth, and it is shown that their behaviors \vs mass and time are similar (red lines \vs blue lines in Figure~\ref{Fig:Mstar_Sig1_SMA1}).  Hence, nearly the same quenching boundary is predicted using this more physically-based expression for halo gas-binding energy.

Four variations on this basic picture are considered in Section~\ref{Sec:Variants}, including two that express the black hole-halo balance in terms of an energy-\emph{rate} balance instead of total-energy equality. Such rate prescriptions have been employed in SAM models in the past but fail badly (as implemented here) to match the observed width of the green valley, the sloping power-law nature of the quenching boundaries, or the steady flow of galaxies from the blue cloud to the red sequence since $z \sim 3$.

The formula for BH mass growth, $\Mbh = \rm{const.}\times\Sig1^{1.76}$ with constant zeropoint, applies only to galaxies on the star-forming main sequence (i.e., down to $-0.45$~dex below the ridgeline). Evidence is given that the BH growth relation steepens as galaxies enter the green valley, indicating an acceleration in BH growth relative to $\Sig1$ at that point (Figure~\ref{Fig:Cartoon}).  Increased BH accretion in the green valley is also seen in X-ray luminosities and in hydro models.  Total BH mass growth while galaxies are in the green valley is at least $\times$10.  The star-forming slope, $\Mbh = \rm{const.}\times\Sig1^{1.76}$, is reacquired when galaxies come to rest on the quenched ridgeline, but now with a different zero point that is $\sim$10 times larger than the star-forming zero point.  This scenario for BH growth is summarized in Figure~\ref{Fig:Cartoon}.

Rapid black hole growth in the green valley implies that AGN activity should be high there.  AGNs are indeed frequent in the green valley, but integrated X-ray luminosity is not as high as expected.  Perhaps AGNs are obscured, or accretion may be radiatively inefficient.  This conflict with the X-ray data is the biggest unanswered challenge for the binding-energy model.

A surprise from the model is that the derived galaxy evolutionary tracks in $\Sig1\text{-}\Mstar$ and $\Reff\text{-}\Mstar$ are nearly parallel to the quenching boundary and most galaxies would not cross it if it were stationary.  In practice, it is the decline in halo gas-binding energy (at fixed $\Mvir$) that brings the quenching boundary down and across the tracks.  The angle between the tracks and boundary remains acute, however, and the small value of this angle is responsible for the 1$^+$-dex width of the green valley.

An important requirement at the heart of any successful model is that the quenching boundary and the star-forming galaxy cloud must maintain a constant relationship to one another over time in $\Sig1\text{-}\Mstar$, i.e., their zero points must evolve together in lockstep \citepalias{Barro17}. This constancy is observed and is responsible for the unchanged appearance of the $\Sig1\text{-}\Mstar$ diagram over time (except for the downward evolution in both zero points) and is also why the green valley quenching channel has nearly the same mass at all redshifts.  An exact match between the boundary and the star-forming cloud is imposed by fiat in the empirically derived quenching boundary, $\Eheat$, in Eq.~\ref{Eq:EbhEhalo}.  The more physically motivated quenching boundary based on halo gas-binding energy, $\Eth$ in Eq.~\ref{Eq:EbhEbind}, also moves down but not quite fast enough, and the average green-valley mass would vary slightly on that account from $-0.2$~dex at $z = 3$ to $+0.2$ dex at $z=0.5$.

For most of the paper, we assume that star-forming galaxies build their black holes by evolving \emph{along} the relation $\Mbh = \rm{const.}\times\Sig1^{1.76}$ (blue line in Figure~\ref{Fig:Cartoon}).  If this is true, then BH masses in star-forming galaxies can be read off directly from the $\Sig1\text{-}\Mstar$ diagram using $\Mbh = \rm{const.}\times\Sig1^{1.76}$, or from the $\Reff\text{-}\Mstar$ diagram using $\Mbh = \rm{const.}\times\Reff^{-1.76}\Mstar^{1.94}$.

However, inspired by hydro models, Section~\ref{SubSec:BHGrowth} explores an alternate picture in which black holes in star-forming galaxies spend most of their lives \emph{below} the blue line and grow rapidly only later. The wind theory of \citet{King03} and \citet{King15} predicts that AGN winds should suddenly become more efficient as galaxies cross a relation very near the blue line, setting the stage for entry into the green valley.  The blue line would then no longer be an evolutionary track but rather the boundary where the King-Pounds relation is satisfied.  This picture  is still incomplete, however, because galaxies must cross the boundary with the right combination of $\Sig1$, $\Mbh$, \emph{and $\Mstar$}, and the tight connection to $\Mstar$ is still theoretically unexplained.

The major takeaway messages from the binding-energy model are:

1) Central galaxies perceptibly quench when the total effective energy emitted by their black holes equals some constant multiple, $f_{\rm B}$, of the binding energy of gas in their dark halos.  Galaxies enter the green valley when $f_{\rm B}$ is near 4 (Section~\ref{Sec:ThermalModel}) and are fully quenched when $f_{\rm B}$ is near 20 (Section~\ref{SubSec:BoundaryCrossing}).

2) Effective radius is an important second parameter in shaping the life histories of galaxies and the masses of their black holes.  Promising candidates for the origin of radius scatter (at fixed stellar mass) are halo concentration and/or halo formation time (Section~\ref{SubSec:RadiusScatter}).

3) In the simple version of the theory, star-forming galaxies build their black holes \emph{along} the relation $\Mbh = \rm{const.}\times\Sig1^{1.76}$ (blue line in Figure~\ref{Fig:Cartoon}).  The positive dependence on central stellar density means that black holes are smaller in galaxies with larger $\Reff$ at fixed mass (Section~\ref{SubSec:InputPowerLaws}).  Points (1) and (2) then explain the sloping boundaries between quenched and star-forming galaxies in key structural diagrams (Section~\ref{SubSec:EvoTracksQuenchBound}). In this interpretation, the blue line is an evolutionary track.

4) An alternate version of the theory says that black holes in star-forming galaxies spend most of their time \emph{below} the relation $\Mbh = \rm{const.}\times\Sig1^{1.76}$ and grow only later when SNae winds fail, as in hydro models.  The King-Pounds wind theory then predicts a jump in wind feedback efficiency when black holes reach the blue line, which would propel host galaxies into the green valley.  In this interpretation, the blue line is a boundary not an evolutionary track, and many small black holes should exist below the blue line in star-forming galaxies that have not yet been detected (Section~\ref{SubSec:BHGrowth}).

5) The quenching boundary moves with time in structural diagrams because the gas binding energies of halos at fixed $\Mvir$ are lower at late times (Section~\ref{SubSec:Successes}). The evolutionary tracks of star-forming galaxies are intrinsically parallel to the boundary, and it is mainly the sideways motion of the boundary that causes them to quench (Sections~\ref{SubSec:EvoTracksQuenchBound} and \ref{Sec:ThermalModel}).

6) The downward motion of the quenching boundary in $\Sig1\text{-}\Mstar$ is matched by a similar motion of the star-forming cloud. This holds the quenching and star-forming populations in the same relationship to one another, which keeps the green valley mass nearly constant. In physical terms, black hole masses are getting smaller with time owing to the growth in galaxy radii (at fixed $\Mstar$) while dark halos are getting easier to quench with time owing to the decline in halo-gas binding energy (at fixed $\Mvir$) (Section~\ref{SubSec:Successes}).  This approximate equality is why quenching has looked so similar since $z \sim 3$ (Section~\ref{SubSec:Successes}).

7) Black-hole scaling laws and their relationships to dark halos are well defined but are different for star-forming galaxies, the green valley, and quenched galaxies (Section~\ref{SubSec:PredictionsTests}). At least 90\% of black hole mass is gained in the green valley (Sections~\ref{SubSec:InputPowerLaws} and  \ref{SubSec:BoundaryCrossing}).

8) The quenching process is highly uniform from halo to halo.  The rms raw scatter in the factor $f_{\rm B}$ is at most 0.28~dex based on the observed scatter in the $\Sig1\text{-}\Mstar$ quenched ridgeline and will probably be considerably less than this once progenitor bias is fully allowed for (Section~\ref{Sec:ThermalModel}).

9) Despite the successes of the binding-energy model, the physical basis of the model remains incomplete.  The foundation of the model is the way that black holes grow in star-forming galaxies, and the default version assumes that a tight evolutionary track, $\Mbh = \rm{const.} \times \Sig1^{1.76}$, is followed by all  star-forming galaxies.  This assumption has empirical support but disagrees with recent hydro models, which predict more scatter and generally smaller black holes in star-forming galaxies.  An alternative theory predicts that AGN wind efficiency should increase strongly with $\Sig1$ in star-forming galaxies, but this creates a boundary in $\Mbh\text{-}\Sig1$, not an evolutionary track.  It explains the boundary in $\Mbh\text{-}\Sig1$ but not the associated boundaries in $\Sig1\text{-}\Mstar$ or $\Mbh\text{-}\Mstar$ since $\Mstar$ is not linked to these other two quantities.  The existence of \emph{three separate, matched quenching boundaries} -- in $\Sig1\text{-}\Mbh$, $\Sig1\text{-}\Mstar$, and $\Mbh\text{-}\Mstar$ -- is  currently the biggest challenge to theoretical quenching models.

\begin{acknowledgments}
The major impetus for the model in this paper was the structural and star-forming data on galaxies from the CANDELS program on the \textit{Hubble Space Telescope}.  We thank the many workers who helped to create this unique database.  The main CANDELS observations were supported under program number HST-GO-12060, provided by NASA through a grant from the Space Telescope Science Institute, which is operated by the Association of Universities for Research in Astronomy, Incorporated, under NASA contract NAS5-26555.  C. Z. acknowledges support from NFSC grants 11403016,11433003,11673018, Innovation Program 2019-01-07-00-02-E00032 of SMEC. C. Z. thanks the support by Science and Technology Commission of Shanghai Municipality to the Key Lab for Astrophysics, International Joint Lab and Project No. 18590780100.  Members of the CANDELS team at UCSC acknowledge support from NASA \textit{HST} grant GO-12060.10-A and from NSF grants AST-0808133 and AST-1615730.  We thank the anonymous referee for raising important questions that stimulated several significant and beneficial improvements.
\end{acknowledgments}

\appendix
\section{A.\ The $p$,$q$ Mapping in $\Sig1\text{-}\Reff\text{-}\Mstar$ for Star-Forming Galaxies}\label{App:A1}

\begin{figure*}[htbp]
\begin{minipage}[t]{\linewidth}
\centering
\includegraphics[scale=0.6]{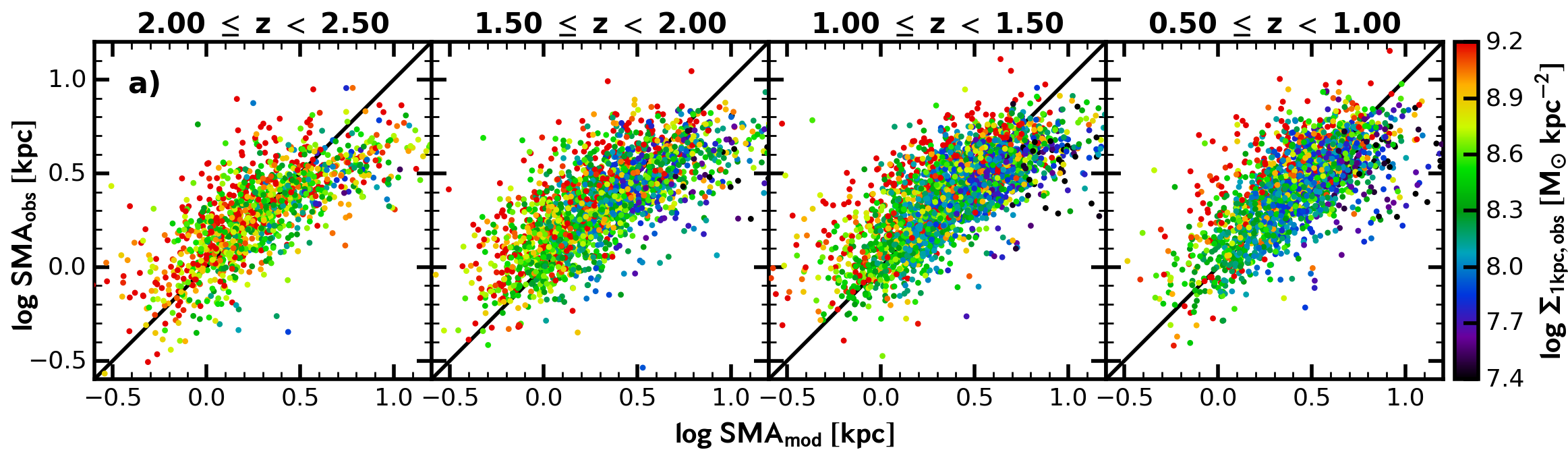}
\begin{center} (a) Observed \vs model values of $\Reff$ using the $p$,$q$ mapping from $\Sig1\text{-}\Mstar$ with $p=1.00$ and $q=1.10$, color-coded by $\Sig1$. (SMA stands for effective half-light semi-major axis.)\end{center}
\end{minipage}
\begin{minipage}[t]{\linewidth}
\centering
\includegraphics[scale=0.6]{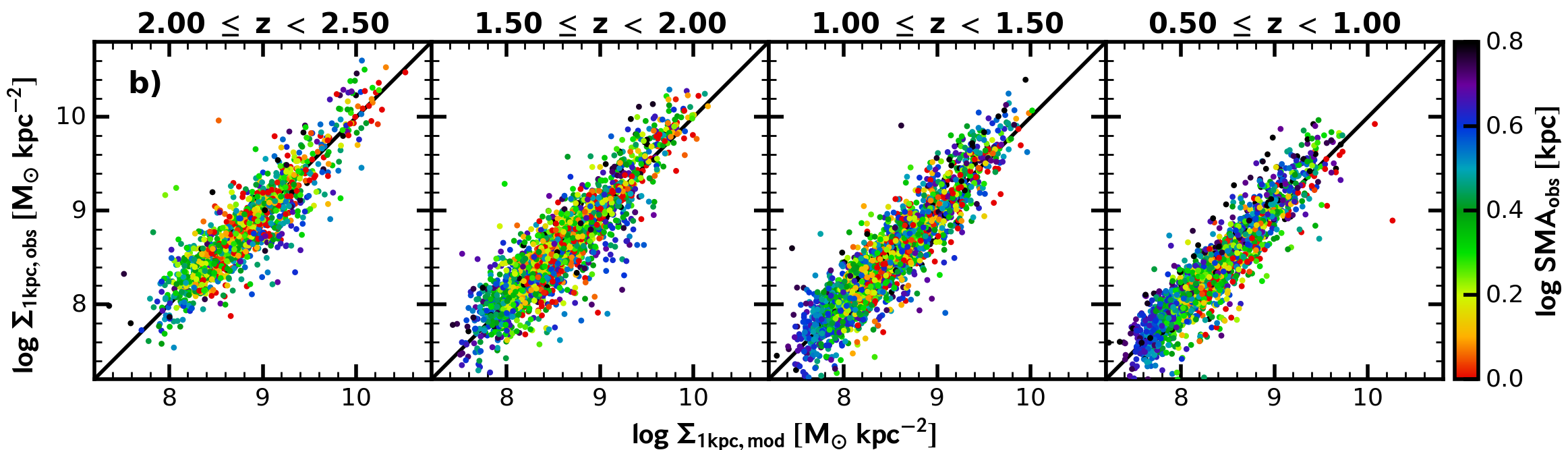}
\begin{center} (b) Observed \vs model values of $\Sig1$ using the same $p$,$q$ mapping from $\Reff\text{-}\Mstar$, color-coded by $\Reff$.\end{center}
\end{minipage}
\caption{\label{Fig:A1} This figure shows the results of mapping the two spaces $\Reff\text{-}\Mstar$ and $\Sig1\text{-}\Mstar$ onto one another using the $p$,$q$ power-law expression $\Sig1 = k_{\Sig1} \Reff^{-p} \Mstar^q$ with $p=1.00$ and $q=1.10$.  Predicted values are on the X-axes, and observed values are on the Y-axes.  The sample consists of the same face-on star-forming galaxies shown in Figure~\ref{Fig:Mstar_Sig1_SMA2}. All five CANDELS fields are used, and ``star-forming'' is defined as main-sequence galaxies lying above the line that lies $-0.45$~dex below the main sequence ridgeline.  There is no systematic shift in zero point with time, and the rms scatter in both relations is about 0.25~dex. Some systematic residuals are seen, as discussed in the text.}
\end{figure*}

Figures~\ref{Fig:A1}ab show observed \vs model values of $\Sig1$ and $\Reff$ predicted by our adopted power-law $p$,$q$ mapping $\Sig1 \sim \Reff^{-p} \Mstar^q$.  The sample shown is described in \citetalias{Fang18}; basic parameters are $H \leq 24.5$ mag, $\Mstar \geq 10^{9.0}~\Msol$, and axis ratio $b/a \geq 0.5$. $\Reff$ (SMA) comes from GALFIT fits by \citet{vanderWel12}.  Galaxies with peculiar or poorly measured structure are omitted, as indicated by the GALFIT quality flag PhotFlag = 0, which excludes 13\% of the selected objects lying within our mass and redshift limits. Visual inspection shows that $\sim$75\% of these objects appear to suffer from contamination from nearby objects, some fraction of which are mergers and disturbed (images are shown in \citetalias{Fang18}). The remaining $\sim$25\% appear normal, but many of these latter have small angular sizes, which may preclude reliable fits \citep{vanderWel12}. In this paper, we are not in general using absolute counts of objects, and so the loss of these objects per se is not an issue.

Completeness as a function of mass is described in \citetalias{Fang18}.  To summarize, quiescent galaxies down to $10^{10}~\Msol$ are included at all redshifts, which captures nearly all of them.  Star-forming galaxies are likewise complete at most redshifts but are only $\sim$50\% complete for $\Mstar \leq 10^{9.5}~\Msol$ at $z \sim 2$. However, all of the missing galaxies are captured if the sample is extended to $H = 25.5$ mag, so we redid the $p$,$q$ fitting formula and replotted  Figure~\ref{Fig:A1} to this lower limit. Neither the figure nor the formula changed perceptibly.

Figures~\ref{Fig:A1}ab show that points are distributed in a plane, and fitted values are $p=1.00$, $q = 1.10$, and  $k_{\Sig1}=1.38$.  The formal error in both $p$ and $q$ is 0.01. There is no systematic shift in the zero point with time, and the rms scatter in both panels of Figure~\ref{Fig:A1} is about 0.25~dex.  However, some curvature of the plane is indicated by the curvature of the $\Reff$ residuals at high redshift and by systematic errors as a function of $\Sig1$ at all redshifts.  We rotated the space by hand to verify that the fitted values are optimum; further investigation is postponed to future work.

\section{B.\ The Evolving Zero Point of the $\Sig1\text{-}\Mstar$ Ridgeline}\label{App:A2}

\begin{figure}[htbp]
%\centering
\begin{center}
\includegraphics[scale=0.65]{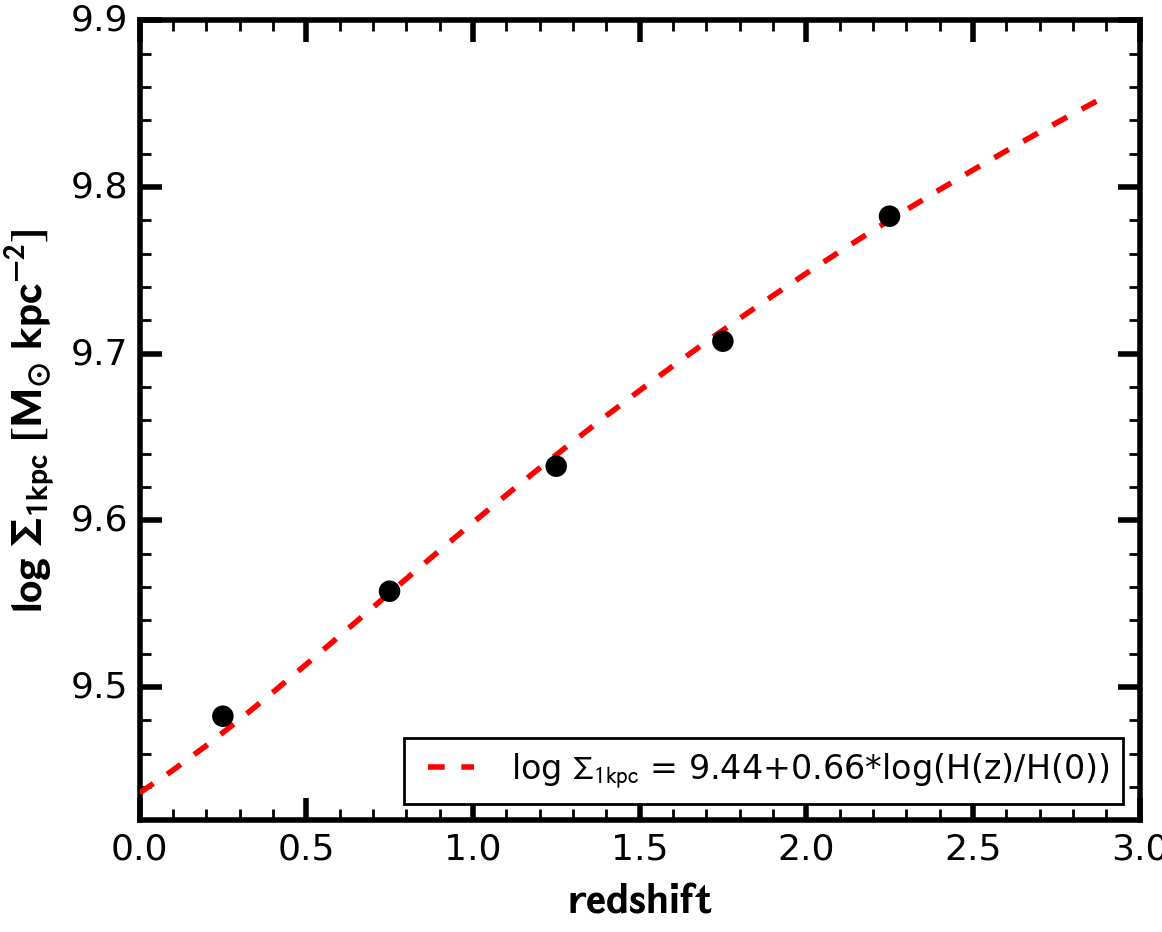}
\caption{\label{Fig:A2}The zero point evolution of the $\Sig1\text{-}\Mstar$ relation from \citetalias{Barro17}, fitted as a function of $H(z)/H(0)$.}
\end{center}
\end{figure}

The ridgeline of quenched galaxies in $\Sig1\text{-}\Mstar$ determines the location of the quenching boundary, which is offset downwards by 0.2~dex in $\Sig1$ from the ridgeline at all redshifts and masses.  The zero point of the ridgeline was measured in redshift bins by \citetalias{Barro17}.  Figure~\ref{Fig:A2} fits a smooth curve to the \citetalias{Barro17} zero points as a function of the normalized Hubble constant $h(z) = H(z)/H(0)$.  It is seen that the zero point evolves as $h(z)^{0.66}$.   This functional form is used to calculate the evolving zero point in the quenching equation Eq.~\ref{Eq:EbhEhalo}.

\section{C.\ Power-law expressions in the empirical power-law model}\label{App:A3}

This section collects together all the exponents and zero points of the various power laws used in the empirical power-law model (Section~\ref{Sec:EvolPLModel}) and its variants (Section~\ref{Sec:Variants}).  This model has a fixed zero point in the stellar-mass-to-halo-mass relation (SMHM, Eq.~\ref{Eq:MstarMvir}) and an evolving zero point in the BH-halo energy quota equation (Eq.~\ref{Eq:EbhEhalo}).
\vspace{\baselineskip}

1. \emph{The $\Sig1\text{-}\Reff\text{-}\Mstar$ relation ($p$,$q$ mapping):}
From Appendix~\ref{App:A1}, the $\Sig1\text{-}\Reff\text{-}\Mstar$ relation is:
\begin{equation}\label{Eq:Sig1_Re_Mstar}
\frac{\Sig1}{10^9~\Msol~{\rm kpc^{-2}}} = k_{\Sig1}\times\left( \frac{\Mstar}{10^{10}~\Msol}\right)^q\times\left(\frac{\Reff}{\rm 1~kpc}\right)^{-p},
\end{equation}
where $p=1.00$, $q=1.10$, and $k_{\Sig1}=1.38$.
\vspace{\baselineskip}

2. \emph{The SMHM relation:}
This is the power-law fit by eye to the average stellar-mass-to-halo-mass relation (SMHM) for star-forming galaxies from \citetalias{Rodriguez17}.  It is assumed to hold with constant zero point in the range $z = 0\text{-}4$. It also forms the lower branch of the broken power-law model (Variant~1) and as such is plotted in Figure~\ref{Fig:HMSM}.  The adopted relation is:
\begin{equation}\label{Eq:SMHM_lin}
\frac{\Mstar}{10^{10}~\Msol} = k_{\rm m}\times\left(\frac{\Mvir}{10^{12}~\Msol}\right)^s,
\end{equation}
where $s=1.75$ and $k_{\rm m}=2.51$.
\vspace{\baselineskip}

3. \emph{$\Rvir$ as a function of $\Mvir$:}
This is the relation between $\Rvir$ and $\Mvir$ based on the definition of halo overdensity $\Delta_{\rm vir}$ from \citet{Bryan98}.  We solved Eq.~\ref{Eq:RvirMvir} for $\Rvir$ and fitted $\Delta_{\rm vir}^{-1/3}\rho_{\rm m}^{-1/3}$ as a power of $h(z) \equiv H(z)/H(0)$ between $z = 0.5$ and $z = 3$ to find:
\begin{equation}\label{Eq:RvirMvirApp}
\frac{\Rvir}{\rm 1~kpc} = k_{\rm {Rvir}} \left(\frac{\Mvir}{10^{12}~\Msol}\right)^{1/3} \left( \frac{h(z)}{2.3} \right)^{-0.74},
\end{equation}
where the constant $k_{\rm Rvir} = 215$ is evaluated at $z = 1.5$ and $h(z)/2.3 = h(z)/h(1.5)$.
\vspace{\baselineskip}

4. \emph{$\Reff$ \vs $\Rvir$:}
This is the assumed relation between the effective radius of a typical star-forming galaxy, $\ReffMean$, and the radius of its halo:
\begin{equation}
\ReffMean = k_{\rm{Re}} \times \Rvir,
\end{equation}
where  $k_{\rm Re} = 0.02$.
\vspace{\baselineskip}

5. \emph{The $\Sig1\text{-}\sigma_1$ relation:}
$\Sig1$ is the projected stellar surface density within a radius of 1 kpc from the center of the galaxy, and  $\sigma_1$ is the projected velocity dispersion along the line of sight within the same aperture.  The relation we use is from \citetalias{Fang13}:
\begin{equation}\label{Eq:Sig1_sig1}
\frac{\Sigma_{\rm {1~kpc}}}{10^9~\rm{M_{\odot}~kpc^{-2}}} = k_{\rm{\Sig1,\sigma_1}} \times \left(\frac{\sigma_1}{200~\rm{km s^{-1}}}\right)^{1.99},
\end{equation}
where $k_{\rm{\Sig1}} = 6.0$.
It is assumed to hold for all galaxies at all redshifts.
\vspace{\baselineskip}

6. \emph{The $\Mbh\text{-}\Sig1$ relation for star-forming galaxies:}
This is the equation of the blue line in Figure ~\ref{Fig:Cartoon}. In Section~\ref{SubSec:InputPowerLaws}, we motivated the use of the power law $\Mbh \sim \Sig1^v$ to express black-hole mass in star-forming galaxies, and we found the value of the exponent $v = 1.76$ by taking the ratio of the quenched ridgeline slopes in $\Sig1\text{-}\Mstar$ and $\Mbh\text{-}\Mstar$. Here we set the zero point of this relation by equating our strongly star-forming ridgeline galaxies to the pseudo-bulge galaxies in Figure~26 of \citetalias{Kormendy13} and noting that $\Mbh/\Mstar = 10^{-3.7}$ for such objects on average (with scatter).  Normalizing $\Sig1$ at $10^{9.0}~\Msol~\rm kpc^{-2}$ corresponds to $\Mstar = 10^{10.1}~\Msol$ for star-forming SDSS galaxies in \citetalias{Fang13}.  We therefore have:
\begin{equation}\label{Eq:Mbh_Sig1}
\frac{M_{\rm BH}}{10^8~\rm{M_{\odot}}} = k_{\rm BH}\times\left(\frac{\Sig1}{10^9~\Msol~{\rm kpc^{-2}}}\right)^{v},
\end{equation}
where $v = 1.76$ and $k_{\rm BH}=0.025$.  This relation is assumed to be valid for all star-forming galaxies at all redshifts. It says that BH mass is purely a function of central stellar density within 1~kpc in star-forming galaxies.
\vspace{\baselineskip}

7. \emph{The $\Sig1\text{-}\Mstar$ quenched ridgeline relation:}
The equation of the quenched ridgeline in $\Sig1\text{-}\Mstar$ is \citepalias{Barro17,Fang13}:
\begin{equation}\label{Eq:Sig1_Mstar_ridge}
\frac{\Sig1}{10^9~\Msol~{\rm kpc^{-2}}} = 1.26\times\left(\frac{\Mstar}{10^{10}~\Msol}\right)^{0.66} \times\ h(z)^{0.66}.
\end{equation}
The location of the quenching boundary is taken to be 0.2~dex below this ridgeline based on the location of the ``structural valley'' in \Sig1\text{-}\Mstar\ identified by \citet{Luo19} for SDSS galaxies today.
\vspace{\baselineskip}

8. \emph{The $\Mbh\text{-}\Mstar$ quenched ridgeline relation:}
The basic scaling relation for BHs in quenched galaxies is the red dashed line in Figure~\ref{Fig:Cartoon}, which plots $\Mbh$ \vs $\Sig1$.  The slope of this line comes from the quenched ridgeline in $\Mbh\text{-}\Mstar$ today from \citetalias{Kormendy13}:
\begin{equation}\label{Eq:Mbh_Mstar_ridge}
\frac{\Mbh}{10^8~\Msol} = 0.36\times\left(\frac{\Mstar}{10^{10}~\Msol}\right)^{1.16} \times\  h(z)^{0.66v},
\end{equation}
which is converted to $\Sig1$ using $\Sig1 \sim \Mstar^{0.66}$ for quenched ridgeline galaxies (Item 7 above).  However, the zero point of the red dashed line comes from the zero point for star-forming galaxies in Eq.~\ref{Eq:Mbh_Sig1} incremented by the length and slope of the green valley vectors in Figure~\ref{Fig:Cartoon}.  These in turn come from the growth of $\Mbh$ in the green valley from \citet{Terrazas17} and the growth of $\Sig1$ in the GV from \citetalias{Fang13,Barro17} and are (coincidently) such that the zero point of the red line is offset from the blue line by almost  precisely a factor of 10.  Using this number and combining Eqs.~\ref{Eq:Mbh_Sig1} and \ref{Eq:Sig1_Mstar_ridge} above, we can therefore perform a sanity check to see if the zero point for quenched galaxies in Figure~\ref{Fig:Cartoon} is consistent with the measured \citetalias{Kormendy13} zero point for quenched galaxies in $\Mbh\text{-}\Mstar$.  The predicted zero point using this route is 0.37, which compares well with the observed zero point of 0.36 used in Eq.~\ref{Eq:Mbh_Mstar_ridge}.
\vspace{\baselineskip}

9. \emph{The empirical power-law expression for \Eheat:}
$\Eheat$ is the empirically derived value for the halo quenching-energy quota defined in Eq.~\ref{Eq:EbhEhalo}. By fitting to the red lines in Figure~\ref{Fig:Mstar_Sig1_SMA1}, we find the evolving constant in the following equation:
\begin{equation}\label{Eq:Eheat_Ebh}
\Ebh = \Eheat =
\eta \frac{\Mbh}{10^8~\Msol} c^2 = k_{\rm e}(z) \times \left(\frac{\Mvir}{10^{12}~\Msol}\right)^t,
\end{equation}
where
\begin{equation}
k_{\rm e}(z) =  k_{\rm e}(0) \times\ h(z)^{0.66v},
\end{equation}
the term $h(z)^{0.66v}$ is based on the zero point fit in Figure~\ref{Fig:A2},
and $k_{\rm e}(0)= 10^{17.66}$.
\vspace{\baselineskip}

10. \emph{The zero point $k_{\rm SMHM}(0)$ of the SMHM relation in Eq.~\ref{Eq:Variant2}:}
This is $k_{\rm SMHM}(0)$ in the Variant~2 model:
\begin{equation}
k_{\rm SMHM}(z) = k_{\rm SMHM}(0)\times\ h(z)^{-1.00},
\end{equation}
where $k_{\rm SMHM}(0)=2.51$.
\vspace{\baselineskip}

11. \emph{The $\Mbh\text{-}\Reff\text{-}\Mstar$ relation}:
For convenience, we include the power law that makes it possible to read off BH mass for star-forming galaxies from $\Reff$ and $\Mstar$.  From Eq.~\ref{Eq:Sig1_Re_Mstar} and Eq.~\ref{Eq:Mbh_Sig1}, the relation is:
\begin{equation}\label{Eq:Mbh_Mstar_Re}
\frac{\Mbh}{10^8~\Msol} = k_{\rm BH1}\times\left(\frac{\Mstar}{10^{10}~\Msol}\right)^{qv}\times\left(
      \frac{\Reff}{\rm 1~kpc}\right)^{-pv},
\end{equation}
where $qv = 1.94$, $pv = 1.76$, and $k_{\rm BH1}=0.044$.  This is the value of BH mass predicted by the blue line in Figure~\ref{Fig:Cartoon}.  Note that the alternative picture of BH growth in star-forming galaxies that is presented in Section~\ref{SubSec:BHGrowth} says that BHs in star-forming galaxies actually spend most of their lives \emph{below} this value, growing only later as galaxies near the green valley.
\vspace{\baselineskip}

We conclude this section with some useful power laws that approximate the evolutionary tracks of three quantities \vs $\Mstar$ for galaxies with $\Reff/\ReffMean =1$ \emph{during the star-forming phase}.  These are evolutionary tracks, not quenched ridgelines. The slopes are taken from Figure~\ref{Fig:TrajectoryM}, and the zero points are set by eye to fit the middle curves in the redshift range $z = 0.5$-3. The actual model calculations use the real, curved tracks in Figure~\ref{Fig:TrajectoryM}, not these power laws.
\vspace{\baselineskip}

12. \emph{The average radius-mass evolutionary track:}
  \begin{equation}\label{Eq:Re_Mstar}
\frac{\Reff}{\rm 1~kpc} = \kappa_{\rm R}\times\left(\frac{\Mstar}{10^{10}~\Msol}\right)^R,
\end{equation}
where $R=0.40$ and $\kappa_{\rm R} \sim 3$.
\vspace{\baselineskip}

13. \emph{The average $\Sig1$-mass evolutionary track:}
 \begin{equation}\label{Eq:Sig1_Mstar}
\frac{\Sig1}{10^9~\rm{M_{\odot}~kpc^{-2}}} = \kappa_{\rm \Sig1}\times\left(\frac{\Mstar}{10^{10}~\Msol}\right)^S,
\end{equation}
where $S=0.70$ and $\kappa_{\rm \Sig1} \sim 0.6$.
\vspace{\baselineskip}

14. \emph{The average $\Mbh$-mass evolutionary track:}
\begin{equation}\label{Eq:Mbh_Mstar}
\frac{\Mbh}{10^8~\Msol} = \kappa_{\rm \Mbh}\times\left(\frac{\Mstar}{10^{10}~\Msol}\right)^B,
\end{equation}
where $B=1.23$ and $\kappa_{\rm \Mbh} \sim 0.008$.  This assumes that star-forming galaxies evolve along the blue line in Figure~\ref{Fig:Cartoon}.  The alternative picture for BH growth that is presented at the end of  Section~\ref{SubSec:BHGrowth} says that black holes in star-forming galaxies are smaller than this mass until their host galaxies near the green valley.

\section{D.\ Four Variants on the Empirical Power-Law Model}\label{App:A4}

This section provides more details on the four variants of the empirical power-law model that are discussed in Section~\ref{Sec:Variants}.

\subsection{Variant 1: A Broken Power-Law Model}\label{SubSec:AppBrokenPLModel}

A feature of the empirical power-law model that does not quite match observed data is the long ``tongue'' of star-forming galaxies extending up to $\Mstar > 10^{12}~\Msol$ at late redshifts in Figure~\ref{Fig:SimuCartoon}. This tongue is due to the fact that we have used the curved HMSM relation from \citetalias{Rodriguez17} at $z = 0$ to find the starting redshifts, $z_{\rm s}$, but have calculated stellar masses going forward using the power-law SMHM relation of Eq.~\ref{Eq:MstarMvir}, $\Mstar = k_{\rm m} \Mvir^s$. Since the latter keeps climbing with $\Mvir$, massive galaxies that start near the knee of the real SMHM relation today can wind up with higher stellar mass than their assumed initial mass. Galaxies with $\Mstar > 10^{10.7}~\Msol$ and large $\Reff$ (and small BHs) are the objects that populate the massive tongue. We tried to limit this by keeping starting stellar masses close to the knee, which is at approximately $10^{10.7}~\Msol$ in \citetalias{Rodriguez17}. However, that is not high enough to populate the full mass range of observed star-forming galaxies, which reaches to $\sim 10^{11.2}~\Msol$ in quantity at all redshifts \citepalias{Fang13, vanderWel14, Barro17}.

\begin{figure}[htbp]
\centering
\includegraphics[scale=0.75]{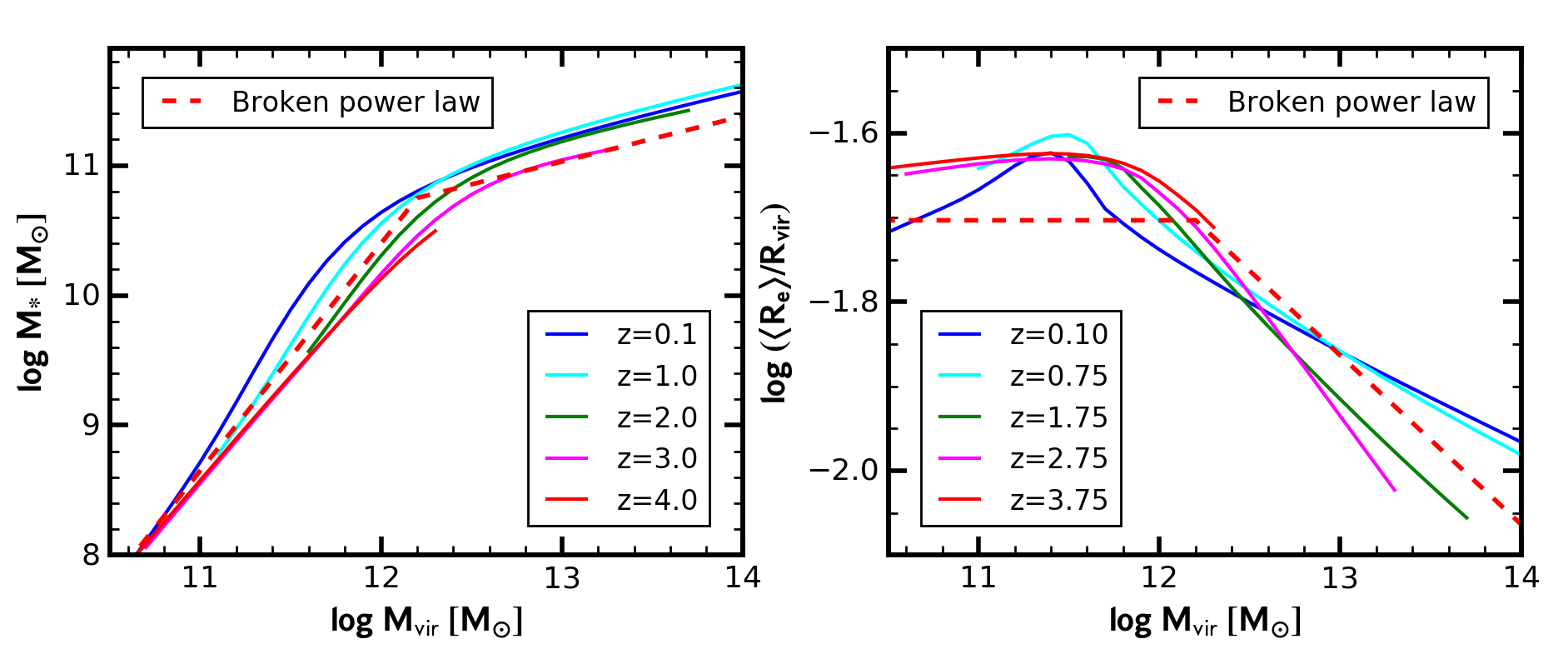}
\caption{\label{Fig:HMSM}  Broken power laws used for the Variant~1 evolving model in Section~\ref{SubSec:BrokenPLModel}.  Left panel: $\Mstar$ \vs $\Mvir$. The power-law slope of the dashed line below $\Mvir = 10^{12.2}~\Msol$ is 1.75, which is also the value of $s$ used in the empirical power-law model.  Above $\Mvir = 10^{12.2}~\Msol$, the slope is 0.35.  The overplotted colored curves for comparison are the \citetalias{Rodriguez17} SMHM relations from abundance matching.   Right panel:  $\ReffMean/\Rvir$ \vs $\Mvir$ for the corresponding lines in the left panel.  The colored curves are calculated from equations in \citetalias{Rodriguez17}. The constant ratio for the dashed line below  $\Mvir = 10^{12.2}~\Msol$ is 0.02, which is used in the empirical power-law model and agrees fairly well in the mean with \citetalias{Rodriguez17} at those halo masses.  Above $\Mvir = 10^{12.2}~\Msol$ the broken power-law slope is $-0.2$. }
\end{figure}

To attempt to cure this problem, the Variant~1 broken-power-law model uses the same power-law SMHM relation for star-forming galaxies up to $10^{10.7}~\Msol$ but switches to a flatter power-law for SMHM above that.  This retards stellar mass growth and also black hole growth, since $\Sig1$ is not increasing.
We also take into account the related fact that $\ReffMean$ now also lags behind $\Rvir$ above the knee and the ratio $\ReffMean/\Rvir$ therefore declines. The latter effect has been estimated using the abundance-matching results of \citetalias{Rodriguez17}, which give $\ReffMean/\Rvir$ \vs $\Rvir$ for massive halos. The adopted new broken power-law relations for $\Mstar$ and $\Reff$ are compared to the \citetalias{Rodriguez17} SMHM relations in Figure~\ref{Fig:HMSM}.

\begin{figure*}[htbp]
\centering
\includegraphics[scale=0.65]{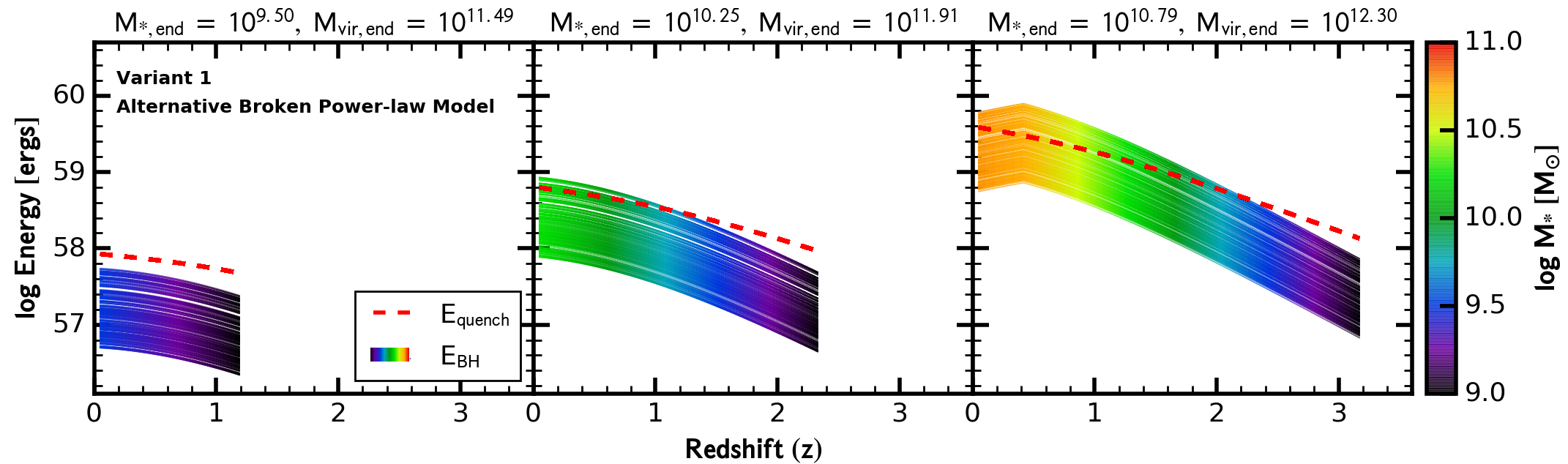}
\caption{\label{MEnergy_V1} Evolutionary plots of $\Ebh$ and $\Eheat$ for the Variant~1 broken power-law model in the style of Figure~\ref{Fig:BHEnergy}.   The single power-law SMHM and $\ReffMean/\Rvir$ relations of the empirical power-law model are replaced with the broken power laws shown in Figure~\ref{Fig:HMSM}.  The evolution of galaxies below the break point at $10^{10.7}~\Msol$ is unaffected.  Above this, BHs fail to grow fast enough to keep up with the heat requirements of their halos, and the majority of the most massive galaxies never quench.}
\end{figure*}

Inserting these new relations produces a difference between high-mass and low-mass galaxies. The evolution is identical for galaxies whose stellar masses stay below the knee today, and so the $\Reff\text{-}\Mstar$ and $\Sig1\text{-}\Mstar$ diagrams remain the same below $10^{10.7}~\Msol$.  Above this mass, $\Mstar$ tends to saturate, and BH growth is retarded. This is evident in Figure~\ref{MEnergy_V1}, which replots $\Ebh$ and $\Eheat$ as in Figure~\ref{Fig:BHEnergy} but with the new broken power laws.  With the new relation, high-mass galaxies now stop growing but also never quench. This solves the too-massive galaxy problem, but these galaxies are now permanently stuck in a star-forming state because their BHs never grow large enough. We were not able to find a broken power-law scaling relation that both prevents the formation of too-massive star-forming galaxies and at the same time allows typical slightly less-massive star-forming galaxies to quench.

On the other hand, more recent information about the shape of the quenching boundary from a large sample of SDSS galaxies \citep{Luo19} offers a different solution. The quenched ridgeline in $\Sig1\text{-}\Mstar$ in this new sample is seen to bend over to lower $\Sig1$ at high stellar mass, which would create a matching bend in the quenching boundary using our assumptions.  Our assumed straight power law lies about 0.2~dex above the new SDSS ridgeline at $10^{11.4}~\Msol$, and it is seen from Figure~\ref{Fig:SimuCartoon} that bending the ridgeline down by this amount would remove all of the problem massive galaxies.

We conclude that the existence of too-massive star-forming galaxies in the model may be an artifact of assuming a perfectly linear power-law quenching boundary, whereas the real boundary may curve slightly down in $\Sig1$ at high mass.  This curvature could be due either to larger BH mass growth per galaxy than assumed in the model or to a change in the structure of massive halos that makes them easier to quench.  Either way, the lesson of Variant~1 is that the fraction of quenched \vs star-forming galaxies at high masses is extremely sensitive to the shape of the quenching boundary, and a detailed physical understanding is needed to predict it accurately. A related consequence is that the shape of the mass function of quenched galaxies at high values of $\Mstar$ is also very difficult to predict.

\subsection{Variant 2: Move the Evolving Zero Point to the SMHM Relation}\label{SubSec:AppEvolvingSMHMZP}

This variant moves the evolving zero point from the halo energy quota in Eq.~\ref{Eq:EbhEhalo} to the SMHM relation.  The expression used for $\Mstar$ is then:

\begin{equation}\label{Eq:Variant2}
\frac{\Mstar}{10^{10}~\Msol} = k_{\rm SMHM}(0) \times\left(\frac{\Mvir}{10^{12}~\Msol}\right)^s \times\ h(z)^{-1.00},
\end{equation}
where $k_{\rm SMHM}(0)=2.51$.

This equation was derived by substituting $\Mstar$ for $\Mvir$ in Eq.~\ref{Eq:EbhEhalo} and demanding that galaxies quench with their previous values of $\Mbh$ and $\Mstar$ at the same $z$.  The exponent is fortuitously $-1.00$ owing to the zero point evolution of the $\Sig1\text{-}\Mstar$ ridgeline as $h(z)^{-0.66}$ in Figure~\ref{Fig:A2} and the relationship among $s$, $t$, and $v$ from Eq.~\ref{Eq:Sig1Mstar}.

The new law makes the zero point of $\Mstar$ larger at late times, by about 0.5~dex at $z = 0$ compared to $z = 2$.  The growth of $\Reff$ \vs $\Mvir$ is unaffected, but $\Mstar$ grows more steeply \vs $\Mvir$ than before. This change effectively grows BHs more strongly at late times, allowing them to overcome halos without having to reduce $\Eheat$.  The evolutionary tracks for this model are shown as the dashed lines in Figure~\ref{Fig:TrajectoryM}.  The main change is a reduction of the log-log slope in $\Reff$ \vs $\Mstar$ from 0.40 to 0.33, which actually agrees better with the slope of 0.3 estimated by \citetalias{vanDokkum15}.

\begin{figure*}[htbp]
\centering
\includegraphics[scale=0.65]{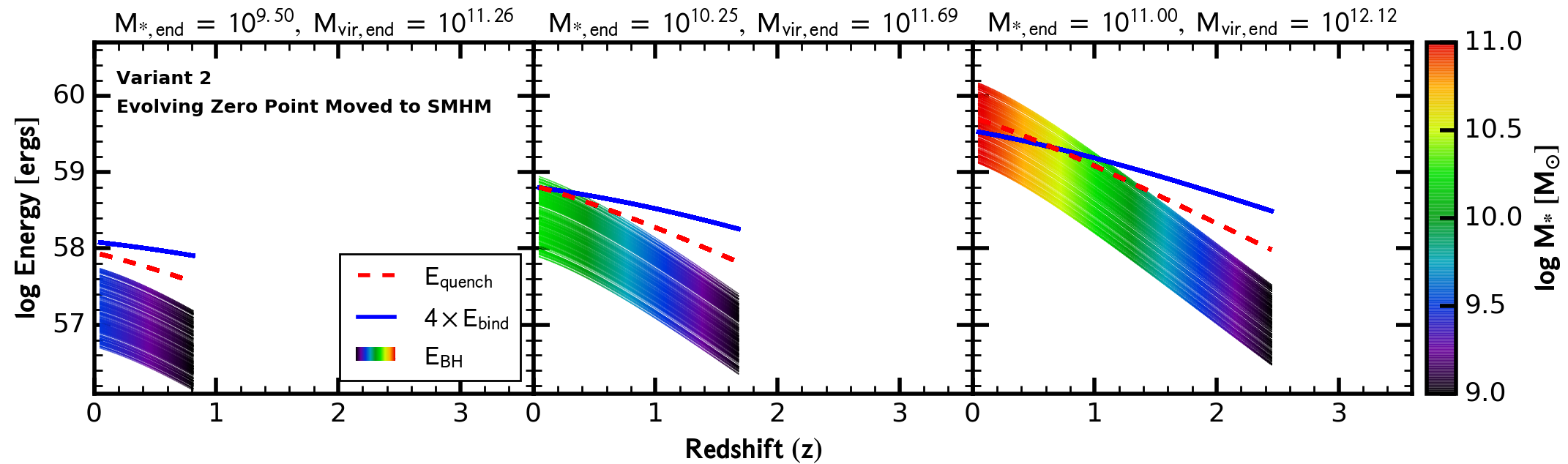}
\caption{\label{MEnergy_V2} The BH energy output and the halo quenching energy quota are plotted in the style of Figure~\ref{Fig:BHEnergy} for the Variant~2 model, which puts the evolving zero point into the SMHM relation rather than $\Eheat$. The red dashed lines are the new halo quenching energy $\Eheat$, and the blue solid curves repeat the energy content of the halo gas, $\Eth$, from Figure~\ref{Fig:BHEnergy}.  $\Eheat$ no longer follows $\Eth$, which means that the energy needed to quench a halo is not closely equal to the gas binding energy in this model (see Section~\ref{SubSec:EvolvingSMHMZP}). Also, quenching for both curves begins too late at $z \sim 1$ rather than at $z \sim 2.5$ as observed.}
\end{figure*}

The shift of the evolving zero point affects stellar mass, black-hole growth, and $\Eheat$, but $\Eth$ remains unchanged. Figure~\ref{MEnergy_V2} plots these quantities for the Variant~2 model in the style of Figure~\ref{Fig:BHEnergy}.  \Eheat\ is now simply proportional to $\Mvir^t = \Mvir^2$, which climbs strongly with time (red dashed lines). However, it still roughly parallels $\Ebh$, which also climbs rapidly, and this maintains a wide green valley.  On the other hand, close equality between $\Eheat$ and halo-gas binding energy, $\Eth$, has been lost. This is shown by the blue curves, which are now much shallower than $\Eheat$.  This has the consequence that quenching using $\Eth$, which began for massive galaxies at the plausible redshift $z \sim 2.5$ in Figure~\ref{Fig:BHEnergy}, does not start here until $z = 1$.  Quenching using $\Eheat$ also starts very late.  Thus, even though Variant~2 can match the moving quenching boundary, its $\Eheat$ no longer has physical justification, and the late growth of BH masses makes quenching too recent under all scenarios.

A related question is whether the measured changes in the zero point of the SMHM relation back in time are as large as needed by this variant.  Historically, measurements of SMHM back in time have scattered.  Several authors have reported finding similar values of SMHM at $\Mvir \sim 10^{10}~\Msol$ but lower values back in time near the knee at $\Mvir \sim 10^{12}~\Msol$ (e.g., \citealp{Moster13,Skibba15}; \citetalias{Rodriguez17}), implying shallower slopes at higher $z$.  However, some works \citep{Yang12,Leauthaud12,Behroozi13, Behroozi19} find roughly parallel slopes at all redshifts but a shift to a lower zero point in $\Mstar$ at early times.  This is the right direction needed for Variant~2.  However, the zero point shift from $z = 0$ to $z = 1$ in \citet{Behroozi19} is 0.25~dex, with little change  beyond that to $z = 3$.  This is only half the value of 0.5~dex needed by this model.

\subsection{Variants 3 and 4: Energy Rates or Energy Totals?}\label{SubSec:AppRatesorTotals}

Variants~3 and 4 replace the total energies on both sides of Eq.~\ref{Eq:EbhEhalo} with mass rates. Entry into the green valley then occurs when the halo mass-heating rate from the black hole equals the rate at which mass cools in the halo.  For both, the halo energy term $\Eheat$ on the right side is replaced by the standard halo mass-cooling rate, $\mdotcool$, from \citet{Croton06} \citep[see also][]{Somerville08,Henriques15}:

\begin{equation}\label{Eq:MdotCool}
\dot{m}_{\rm cool} = M_{\rm hot}\frac{r_{\rm cool}}{R_{\rm vir}}\frac{1}{t_{\rm dyn,h}},
\end{equation}
where
\begin{equation}\label{Eq:RCool}
r_{\rm cool} = \biggl[\frac{t_{\rm dyn,h}M_{\rm hot}\Lambda (T_{\rm hot},Z_{\rm hot})}{6\pi \mu m_{\rm H}\kappa T_{\rm hot}R_{\rm vir}}\biggr]^{\frac{1}{2}},
\end{equation}
and where $\mu$ is the mean molecular weight, set here to 0.59, $\kappa$ is Bolzmann's constant, and $t_{\rm dyn,h}$ is the halo dynamical time $= \Rvir/\Vvir$ (\citealt{Henriques15}). $M_{\rm hot}$ is the mass of hot gas defined as:
\begin{equation}\label{Eq:f_hot}
M_{\rm hot} = f_{\rm hot} \times M_{\rm vir},
\end{equation}
where $f_{\rm hot}$ is the hot-gas fraction, set here to 0.1. $\Lambda$ is the equilibrium cooling function for collisional processes, which depends both on the metallicity and temperature but ignores radiative ionization effects (\citealt{Sutherland93}). For $Z_{\rm hot}$ we use one-third solar metallicity at all redshifts. $T_{\rm hot}$ is the post-shock temperature of the infalling gas, which is assumed to be the virial temperature of the dark halo (\citealt{Henriques15}), that is:
\begin{equation}
T_{\rm hot}=\Tvir=35.9\left(\frac{\Vvir}{\rm km~s^{-1}}\right)^2K.
\end{equation}
(Note that this $\Tvir$ is half the virial temperature of \citet{Bryan98}.)

To replace total energy from the BH on the left side of Eq.~\ref{Eq:EbhEhalo} by an energy input rate, two formulations for the BH mass accretion rate are used. Variant~3 uses the rate obtained by differentiating the instantaneous $\Mbh$ values given in our empirical power-law model, while Variant~4 uses the BH accretion expression introduced by \citet{Croton06} (and modified by \citealp{Henriques15}) to supply the heating for radio mode:
\begin{equation}\label{Eq:MbhAccretion}
\mdotbh = k_{\rm AGN}\left(\frac{\Mbh}{10^8~\Msol}\right)\left(\frac{f_{\rm hot}}{0.1} \right)\left(\frac{V_{\rm vir}}{200~{\rm km~s^{-1}}}\right)^3,
\end{equation}
where $\Mbh$ is the black hole mass, $f_{\rm hot}$ is the hot-gas mass fraction, and $k_{\rm AGN}=6\times10^{-6}$ is a free parameter with units of $\Msol~{\rm yr^{-1}}$. We note that $\Vvir^3$ is approximately $\Mvir$ and thus the BH grows super-exponentially with time in this formulation. This has important consequences for the shape and location of the quenching boundary, as discussed below.

Both mass accretion rates are converted into a halo mass-heating rate, $\mdotheat$, in the standard way:
\begin{equation}\label{Eq:MdotHeat}
\mdotheat = \frac{2\dot{E}_{\rm radio}}{V_{\rm vir}^2}
          = \frac{2\eta\mdotbh c^2}{V_{\rm vir}^2},
\end{equation}
where $\eta = 0.1$ is an efficiency parameter and $c$ is the speed of light.

To use the \citet{Croton06} expression in Eq.~\ref{Eq:MbhAccretion} for Variant~4 requires two more inputs. The initial values of $\Mbh$ in \citet{Croton06} (and in \citealt{Henriques15} and \citealt{Somerville08}) come from growing the BH masses previously in quasar mode. We do not have access to those masses, so we start each galaxy at $z_{\rm s}$ with $\Mbh$ given by the empirical power-law model using $\Sig1$ from $\Reff$ and $\Mstar$ and grow the BH thereafter using Croton's $\mdotbh$.  Meanwhile, host galaxy properties $\Sig1$, $\Reff$, and $\Mstar$ are tracked using the empirical power-law model.  The coefficient $k_{\rm AGN}$ out in front of the accretion rate in Eq.~\ref{Eq:MbhAccretion} also needs to be adjusted. This was originally set low in \citet{Croton06} since the purpose was to keep an already-hot halo hot, but we are using it to stop accretion in a cold halo and need to maximize it within reason. We therefore adjust it so that the total BH mass growth in radio mode approximately equals the total mass growth in the empirical model (Eq.~\ref{Eq:Sig1Mbh}), which has the benefit of making the final BH masses satisfy the $\Mbh\text{-}\Mstar$ relation today.  It turns out that a wide range of BH masses can be approximately matched today by adjusting this one zero point (compare the BH masses in panel c with those in panel a  at $z = 0$ in Figure~\ref{Fig:Trajectory_BH} below).

In order to compute Variants~3 and 4 in a fashion that is exactly parallel to the empirical power-law model, we would need to take into account any effect that radius variations, $\Reff/\ReffMean$, would have on $\mdotbh$.  This is automatically taken care of for Variant~3 because $\mdotbh$ is based on the time derivative of $\Mbh$ from the empirical model, which itself is scattered by $\Reff/\ReffMean$.  In contrast, as written, $\mdotbh$ in Variant~4 depends only on $\Mbh$ and halo $\Vvir$, neither of which explicitly has a theory to relate it to  $\Reff/\ReffMean$ in this alternate picture.  For Variant 4, we have therefore simply used the \emph{average galaxy} with $\Reff = \ReffMean$ in the empirical power-law model as a starting point and evolved its black hole mass from there.  Variant 4 is therefore effectively a one-parameter model at each epoch.

\begin{figure}[htbp]
\centering
\includegraphics[scale=0.7]{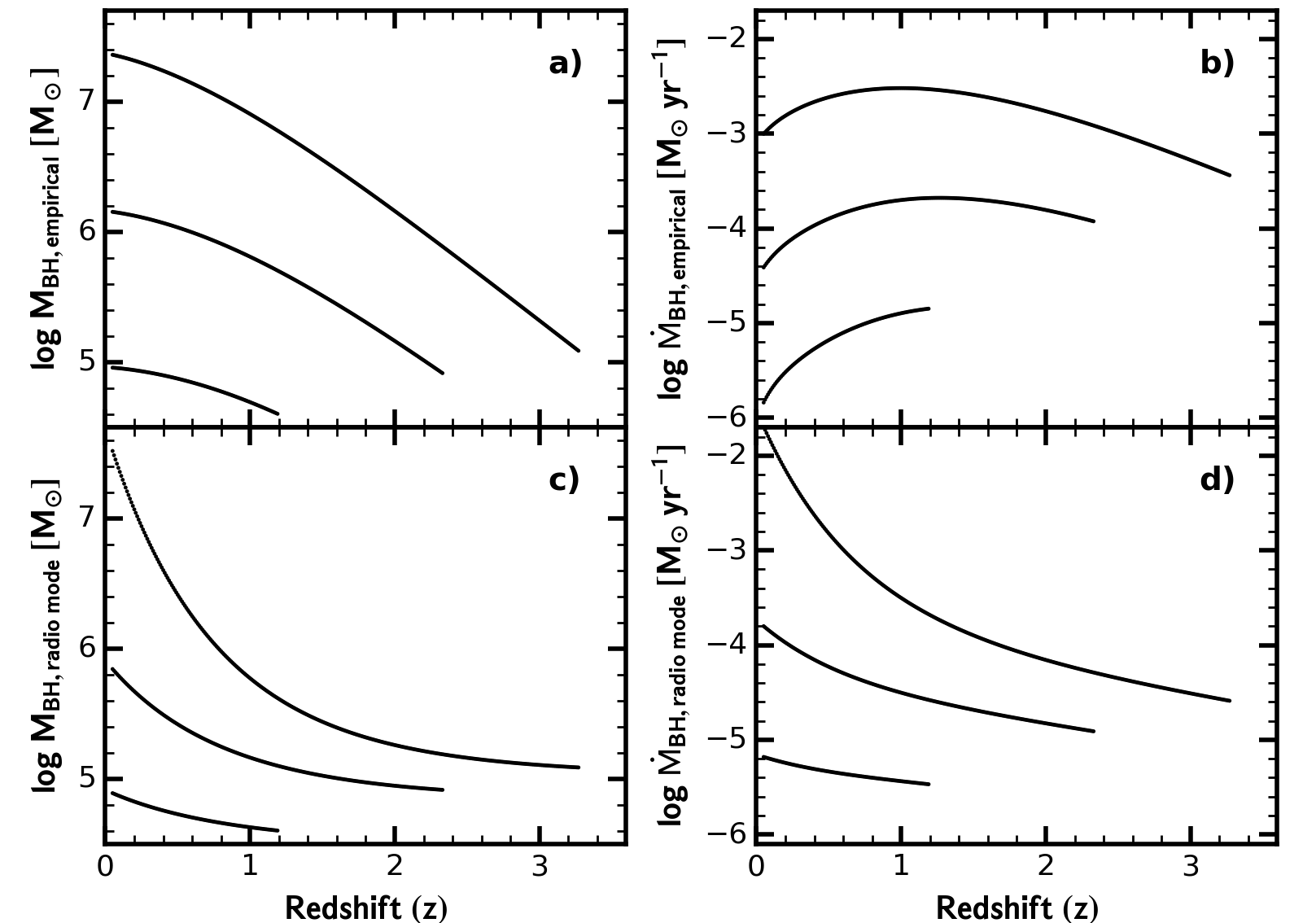}
\caption{\label{Fig:Trajectory_BH} BH-related quantities used in Variants~3 and 4.  The three lines correspond to the final halo masses of Figure~\ref{Fig:Trajectoryz}.  Panels a) and b): The original BH masses from the empirical power-law model, $\Mbh \sim \Sig1^v$, are plotted at upper left, and the BH accretion rates obtained by differentiating those masses \vs time are plotted at upper right. Panels a) and b) are used in Variant~3.  Panel c): BH masses in the \citet{Croton06} radio-mode model, obtained by integrating the BH accretion rates in that model, which are shown in panel d).  They have been normalized by forcing them to match, on average, the BH masses from the empirical model in panel a) at $z = 0$.  Since the ranges of these masses are roughly similar, this proves easy to do just by adjusting the zero point $k_{\rm AGN}$ in Eq.~\ref{Eq:MbhAccretion}.   Panels c) and d) are used in Variant~4. The different behaviors of the accretion rates \vs time predict very different shapes and evolution of the quenching boundaries, as shown in Figures~\ref{Fig:MRate_V3} and \ref{Fig:MRate_V4}.}
\end{figure}

The resulting $\Mbh$ and $\mdotbh$ curves for Variants~3 and 4 are shown in Figure~\ref{Fig:Trajectory_BH}. The $\Mbh$ growth curve from the empirical power-law model is shown at upper left, and its time derivative, $\mdotbh$, is shown at upper right.  This latter is the mass accretion rate used in Variant~3.  The lower panels show $\Mbh$ and $\mdotbh$ from the \citet{Croton06} radio mode prescription.   These have been normalized by forcing the final BH masses today to roughly equal the values in the empirical model by adjusting $k_{\rm AGN}$ in Eq.~\ref{Eq:MbhAccretion}.

Comparing the four panels, it is seen that the $\Mbh$ curves from the empirical power-law model (panel a) curve gently down and are very different from the exponentially growing, upwardly curved $\Mbh$ curves in the radio-mode formulation (panel c).  The mass accretion rates on the right side also differ.  $\mdotbh$ trends down with time using the empirical-model accretion rate (panel b).  This is because $\Mbh \sim \Mstar^{1.23}$ in this model (Figure~\ref{Fig:TrajectoryM}), and thus $\mdotbh \sim \dot{M}_{*}$ very approximately.  Since the star-formation rate is falling at late times, $ \dot{M}_{*}$ is falling and $\mdotbh$ falls, too.  In contrast, $\mdotbh$ in the exponentially growing radio-mode trends up super-exponentially (panel d).

To summarize, Figure~\ref{Fig:Trajectory_BH} compares three separate models for the energy coming from the BH: 1) empirical power-law model, where the relevant energy quantity is total BH mass (panel a); 2) BH accretion rates from the empirical model (Variant~3; panel b); and 3) the radio-mode accretion  rates from \citet{Croton06} (Variant~4; panel d).  The first is rather log-linear \vs redshift, the second turns steeply down, and the third turns steeply up.  The three models therefore quite fortuitously exemplify three very different BH heating histories.

\begin{figure*}[htbp]
\centering
\includegraphics[scale=0.65]{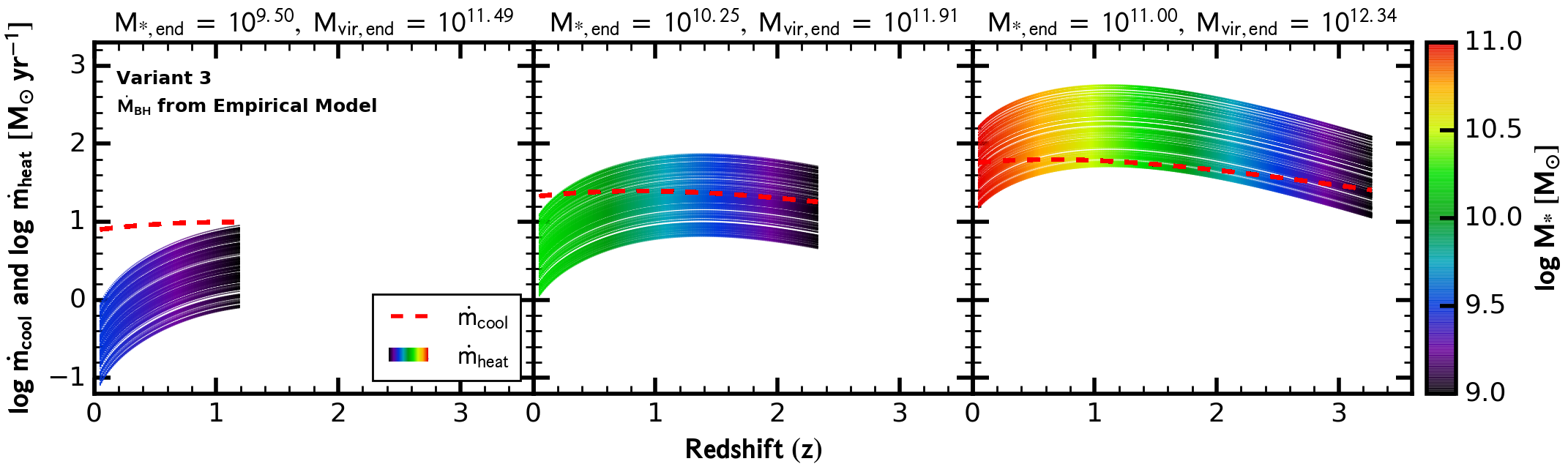}
\caption{\label{Fig:MRate_V3} $\mdotheat$ and $\mdotcool$ \vs redshift for the Variant~3 model. This figure is analogous to Figure~\ref{Fig:BHEnergy} for the empirical power-law model but uses heating and cooling \emph{rates} instead of total energies. (Note that, as in  Figure~\ref{Fig:BHEnergy}, galaxies are plotted ignoring the fact that objects above the dashed lines would have quenched, so that actual values of $\Mbh$ and $\Ebh$ for these objects would be smaller than shown.)   $\mdotcool$ for the halo cooling rate is given by Eq.~\ref{Eq:MdotCool}, while BH $\mdotheat$ uses the BH mass accretion rates from the empirical model, shown in panel b of Figure~\ref{Fig:Trajectory_BH}.  Note how the BH energy output rate turns down at late times and also how halo $\mdotcool$ stays very constant with mass while $\mdotheat$ trends sharply down with time.  These features generate very different quenching boundaries with time from the empirical model, as shown in Figure~\ref{Fig:SimuSize_V3}.}
\end{figure*}

Figure~\ref{Fig:MRate_V3} plots the resulting values of  $\mdotheat$ and $\mdotcool$ \vs redshift for Variant~3. Compared to the analogous plot for the total-energy-based empirical model in Figure~\ref{Fig:BHEnergy}, the halo term $\mdotcool$ is much more constant with redshift and mass than the halo term $\Eheat$ is in Figure~\ref{Fig:BHEnergy}, whereas $\mdotheat$ from the BH varies more.  (Note that values of $\Mbh$ and $\Ebh$ are again plotted as though quenching did not occur, so that actual values for galaxies above the dashed line would be smaller than shown. See description for Figure~\ref{Fig:BHEnergy}).   The two terms move roughly together in Figure~\ref{Fig:BHEnergy}, which mean that all masses are never far from quenching. In Variant~3 by contrast, the highest halo mass is quenched early at high redshifts, while the lowest halo mass never quenches at all. The rate of BH heating, $\mdotheat$, also tends to turn down at late times. That is because BH masses grow as $\Mstar^{1.23} \sim \Mstar$ (Figure~\ref{Fig:TrajectoryM}), and \mdotstar, and therefore also \mdotbh, are falling. Since $\mdotcool$ stays flat, this means that BHs are generally less able to quench at late times in this model.\footnote{Once a galaxy is quenched, it is counted as quenched forever even if its $\mdotheat$ falls below $\mdotcool$ later.} Finally, $\mdotheat$ and $\mdotcool$ are roughly parallel at middle halo masses. This means that some galaxies at that mass will quench very early while others never will.

\begin{figure*}[htbp]
\centering
\includegraphics[scale=0.6]{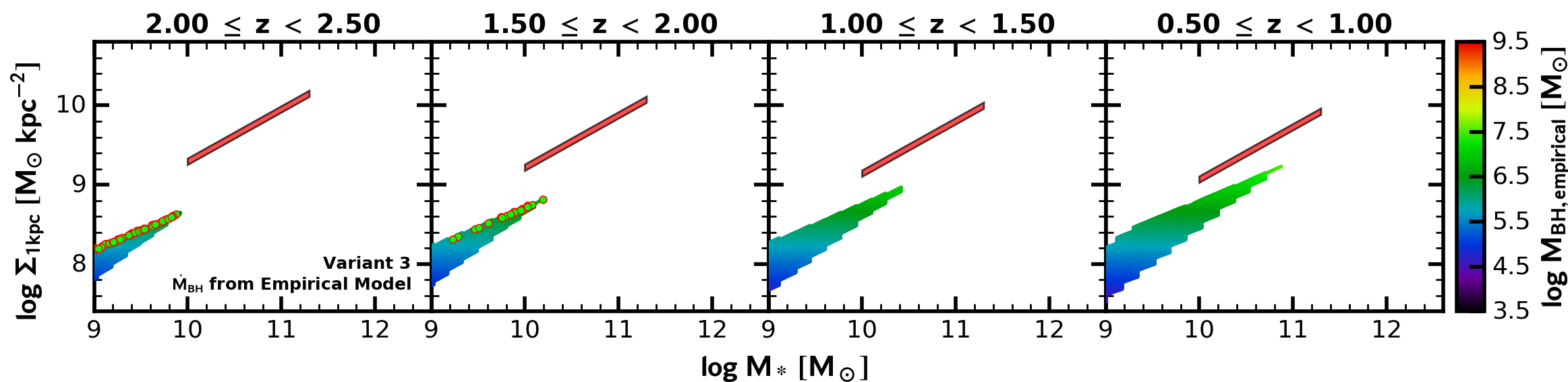}
\caption{\label{Fig:SimuSize_V3} $\Sig1$ \vs $\Mstar$ for the Variant~3 model.  This figure is analogous to Figure~\ref{Fig:SimuCartoon} for the empirical power-law model but uses the $\mdotheat$ and $\mdotcool$ rates from Figure~\ref{Fig:MRate_V3}.  The observed quenching boundary that we want to match at each redshift is given by the straight red power law.  Model galaxies that are actually quenching at that redshift are the pale green points.  The colored regions represent objects that are still star-forming.  No galaxies quench at late times in this model, which stems from the turndown in $\mdotheat$ at late times.}
\end{figure*}

These circumstances produce a very uneven feeding of galaxies through the quenching boundary \vs time for Variant 3. This is shown in Figure~\ref{Fig:SimuSize_V3}, which is similar to Figure~\ref{Fig:SimuCartoon} showing plots of $\Sig1$ \vs $\Mstar$ in four redshift bins. The locations of actually-quenching galaxies are shown as the pale green points, and the target quenching boundary at that redshift is the red line.  Galaxies that are still star-forming fill the colored regions. The quenching loci fall very far from the target locations in this model. Galaxies that quench do so too soon and at too low a mass.  They come primarily from the easily quenched high-mass halos. In addition, there is a general lack of quenching at late times due to the late  turndown in $\mdotheat$. This is reflected in the complete disappearance of the pale green points in the lowest redshift panels, which does not match the observed continued growth in the number of quenched galaxies after $z \sim 1$ \citep[e.g.,][]{Bell04, Faber07, Muzzin13, Tomczak14, Straatman15}. Finally, the zero point of the predicted quenching boundary hardly falls with time, unlike the observations, and that is because $\mdotcool$ is nearly constant with time.

Variant~3 illustrates the point that the BH and halo heating and cooling terms need to march closely together in zero point \vs time in order to feed galaxies smoothly across the entire width of the green valley with the same average mass at all redshifts and at a uniform rate. BH energy \emph{rates} vary a lot at late times, whereas $\mdotcool$ for halos (Eq.~\ref{Eq:MdotCool}) hardly varies, and this produces a mismatch. More generally, matching a power-law-shaped quenching boundary is most naturally accomplished using BH and halo expressions that are logarithmically featureless. This favors the use of galaxy and halo structural parameters that vary smoothly with time and mass and disfavors the use of rates, which can vary irregularly (cf.~Figure~\ref{Fig:Trajectory_BH}).

We conclude that using the mass-rate formulation for halo cooling together with black-hole accretion rates based on the empirical power-law model, as in Variant 3, fails to match the observed quenching boundary and its evolution with time.

\begin{figure*}[htb]
\centering
\includegraphics[scale=0.65]{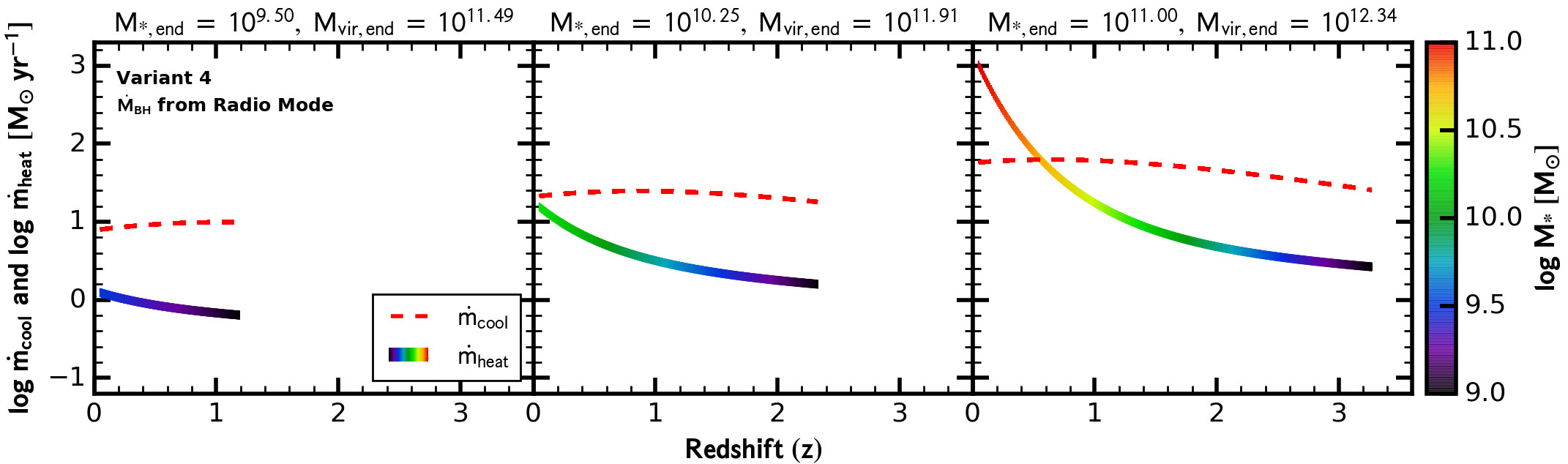}
\caption{\label{Fig:MRate_V4}  $\mdotheat$ and $\mdotcool$ \vs redshift for the Variant~4 model. This figure is analogous to Figure~\ref{Fig:MRate_V3} but uses BH mass accretion rates based on the radio-mode model of \citet{Croton06} (Figure~\ref{Fig:Trajectory_BH}, panel d).  No scatter is applied since there is no explicit dependence of black-hole accretion rate on \Reff\ (Eq.~\ref{Eq:MbhAccretion}); tracks are widened slightly to show \Mstar. Note how the BH energy output rate turns up sharply at late times due to super-exponential growth of the black hole, while $\mdotcool$ stays roughly constant with mass.  This together with zero scatter generate yet different quenching boundaries with time, as shown in Figure~\ref{Fig:SimuSize_V4}. }
\end{figure*}

\begin{figure*}[htbp]
\centering
\includegraphics[scale=0.6]{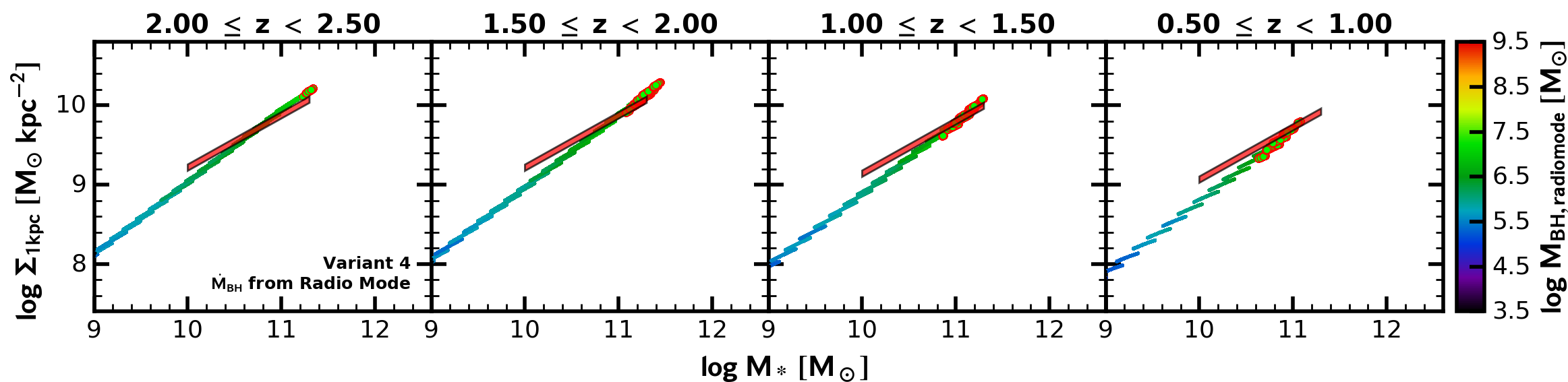}
\caption{\label{Fig:SimuSize_V4} $\Sig1$ \vs $\Mstar$ for the Variant~4 model.  Similar to Figure~\ref{Fig:SimuSize_V3} but uses the $\mdotheat$ and $\mdotcool$ rates from Figure~\ref{Fig:MRate_V4}.  The observed quenching boundary at each redshift is given by the straight red power law.  Model galaxies that are actually quenching at that redshift are the pale green points, and the colored regions indicate galaxies that are still star-forming.  No scatter is applied since there is no explicit dependence of black-hole accretion rate on \Reff\ (Eq.~\ref{Eq:MbhAccretion}); tracks are widened slightly simply to show \Mbh. This model matches the location of the quenching boundary fairly well but predicts that only a single mass should be quenching at any epoch. }
\end{figure*}

Figures~\ref{Fig:MRate_V4} and \ref{Fig:SimuSize_V4} are the analogous plots for the Variant~4 model, which uses the BH accretion rate $\mdotbh$ from radio mode, as shown in Figure \ref{Fig:Trajectory_BH}d. The halo term $\mdotcool$ is again rather constant whereas $\mdotheat$ turns up steeply with time on account of the super-exponential BH mass growth in Eq.~\ref{Eq:MbhAccretion}.  This model does a better job of matching the observed quenching boundary locations, but its characteristic quenching mass drifts down by 1.1 dex from $z = 1.4$ to now, in disagreement with the observed constant green valley mass.  This is because the super-exponential growth sets in earlier for more massive BHs (Figure \ref{Fig:Trajectory_BH}d), which makes them quench at higher redshift. A second deficit is the lack of any theory to predict how BH mass would scale with galaxy radius (or any other potential second parameter), which means that neither the finite width of the green valley nor the slopes of the quenching boundaries through it (in \Sig1\text{-}\Mstar, \Reff\text{-}\Mstar, and \Mbh\text{-}\Mstar) are explained.

Variant 4 proves the inverse of Variant 3, that accretion rates cannot rise too rapidly with time.  It also illustrates the point made in the Introduction that the observed properties of quenching galaxies are at least a two-parameter family at each epoch whereas models based just on halo cooling (like Variant 4) are only a one-parameter family.
\vspace{\baselineskip}
\vspace{\baselineskip}

\end{document}